\documentclass[fleqn,usenatbib]{mnras}
\usepackage{newtxtext,newtxmath}
\usepackage[T1]{fontenc}
\usepackage{ae,aecompl}

\usepackage{graphicx}	
\usepackage{amsmath}	
\DeclareGraphicsExtensions{.pdf,.jpeg,.png,.eps,.jpg}
\graphicspath{{figs}}
\usepackage{algpseudocode}
\usepackage{algorithmicx}
\usepackage{algorithm}
\usepackage{hyperref}
\usepackage{pdflscape}
\usepackage{subcaption}

\usepackage{filecontents}

\newcommand{\alg}[1]{\textsc{#1}}
\newcommand{\ds}{\displaystyle}
\newcommand{\Conv}{\mathop{\scalebox{1.5}{\raisebox{-0.2ex}{$\ast$}}}}%

\newcommand{\bb}{\ensuremath{\boldsymbol{b}}}

\newcommand{\db}{\ensuremath{\boldsymbol{d}}}
\newcommand{\kb}{\ensuremath{\boldsymbol{k}}}
\newcommand{\lb}{\ensuremath{\boldsymbol{\ell}}}

\newcommand{\xb}{\ensuremath{\boldsymbol{x}}}
\newcommand{\yb}{\ensuremath{\boldsymbol{y}}}
\newcommand{\zb}{\ensuremath{\boldsymbol{z}}}
\newcommand{\nb}{\ensuremath{\boldsymbol{n}}}
\newcommand{\gb}{\ensuremath{\boldsymbol{g}}}
\newcommand{\rb}{\ensuremath{\boldsymbol{r}}}
\newcommand{\Db}{\ensuremath{\boldsymbol{\mathsf{D}}}}
\newcommand{\Fb}{\ensuremath{\boldsymbol{\mathsf{F}}}}
\newcommand{\Gb}{\ensuremath{\boldsymbol{\mathsf{G}}}}
\newcommand{\Ib}{\ensuremath{\boldsymbol{\mathsf{I}}}}

\newcommand{\Ub}{\ensuremath{\boldsymbol{\mathsf{U}}}}

\newcommand{\Yb}{\ensuremath{\boldsymbol{\mathsf{Y}}}}
\newcommand{\Zb}{\ensuremath{\boldsymbol{\mathsf{Z}}}}
\newcommand{\Bb}{\ensuremath{\boldsymbol{\mathsf{B}}}}

\newcommand{\unb}{\ensuremath{\boldsymbol{1}}}

\newcommand{\Psib}{\ensuremath{\boldsymbol{\Psi}}}
\newcommand{\Phib}{\ensuremath{\boldsymbol{\Phi}}}

\newcommand{\Gc}{\ensuremath{\mathcal{G}}}

\newcommand{\Pc}{\ensuremath{\mathcal{P}}}
\newcommand{\Hc}{\ensuremath{\mathcal{H}}}
\newcommand{\eC}{\mathbb{C}}
\newcommand{\eD}{\mathbb{D}}

\newcommand{\eN}{\mathbb{N}}
\newcommand{\eR}{\mathbb{R}}

\newcommand{\oN}{{N^\prime}}

\newcommand{\prox}{\ensuremath{\operatorname{prox}}}

\algblock[Block]{Block}{EndBlock}
\algblockdefx[Block]{Block}{EndBlock}%
[1]{{#1}}%
[1]{{#1}}
\newcommand{\Given}[1]{\State{\bf given} {#1}}

\newcommand{\RepeatFor}[1]{\Repeat {\bf~for} {#1}}

\usepackage[table,xcdraw]{xcolor}
\selectcolormodel{cmyk}

\definecolor{darkgreen}{cmyk}{0.8,0,0.8,0.45}

\newcommand{\algc}{\color{darkgreen}}

\title[]{Cygnus A jointly calibrated and imaged via non-convex optimisation from {{VLA}} data}
\author[A. Dabbech et al.]{
A. Dabbech,$^{1}$\thanks{E-mail: a.dabbech@hw.ac.uk}
A. Repetti,$^{1,2,3}$
R.~A. Perley,$^{4}$
O.~{M}. Smirnov$^{5,6}$
and Y. Wiaux$^{1}$
\\
$^{1}$Institute of Sensors, Signals and Systems, Heriot-Watt University, Edinburgh EH14 4AS, UK\\
$^{2}$Department of Actuarial Mathematics \& Statistics, Heriot-Watt University, Edinburgh EH14 4AS, UK\\
$^{3}$Maxwell Institute for Mathematical Sciences, Bayes Centre, Edinburgh, UK\\
$^{4}$National Radio Astronomy Observatory, P.O. Box 0, Soccoro, NM 87801, USA\\
$^{5}$Department of Physics and Electronics, Rhodes University, PO Box 94, {{Makhanda}} (Grahamstown), 6140, South Africa,\\
$^{6}${{South African Radio Astronomy Observatory, 2 Fir Street, Black River Park, Observatory, Cape Town 7925, South Africa}}
\\
}

\date{Accepted XXX. Received YYY; in original form ZZZ}

\pubyear{2020}

\begin{document}

\label{firstpage}
\pagerange{\pageref{firstpage}--\pageref{lastpage}}
\maketitle
\begin{abstract}


Radio interferometric (RI) data are noisy under-sampled spatial Fourier components of the unknown radio sky affected by direction-dependent antenna gains. Failure to model these antenna gains accurately results in a radio sky estimate with limited fidelity and resolution. The RI inverse problem has been recently addressed via a joint calibration and imaging approach which consists in solving a non-convex minimisation task, involving suitable priors for the DDEs, namely temporal and spatial smoothness, and sparsity for the unknown radio map via an $\ell_1$-norm prior, in the context of realistic RI simulations. Building on these developments, we propose to promote sparsity of the radio map {{via a log-sum prior, enforcing sparsity more strongly than the $\ell_1$-norm}}. 
The resulting minimisation task is addressed via a sequence of non-convex minimisation tasks composed of re-weighted $\ell_1$ image priors, which are solved approximately. We demonstrate the efficiency of the approach on RI observations of the celebrated radio galaxy Cygnus~A obtained with the Karl G. Jansky Very Large Array at X, C, and S bands. More precisely, we showcase that the approach enhances data fidelity significantly while achieving high resolution high dynamic range radio maps, confirming the suitability of the priors considered for the unknown DDEs and radio image. As a clear qualitative indication of the high fidelity achieved by the data and the proposed approach, we report the detection of three background sources in the vicinity of Cyg~A, at S band.
\end{abstract}



\section{Introduction} \label{sec:intro}
With the extreme resolution and sensitivity of modern radio telescopes, and their tremendous amount of data, radio-interferometric (RI) imaging has never been more challenging. On the one hand, RI imaging algorithms to be devised need to scale to RI data. On the other hand, efficient calibration need to be performed jointly with imaging in order to achieve the deep and highly resolved radio maps expected from modern arrays, namely The Karl G. Jansky Very Large Array ({{VLA}}), MeerKAT, The Low-Frequency Array ({{LOFAR}}), and the upcoming The Square Kilometre Array (SKA).

Antenna gains are typically classified into two categories: (i) Direction-dependent effects (DDEs) and (ii) Direction-independent effects (DIEs). On the one hand, DDEs are spatially-variable complex-valued modulations in the image domain, which can be equivalently modelled via baseline-dependent convolutions in the spatial Fourier space. DDEs can be geometric, namely the so-called $w$-effect originating from the non-coplanar RI Fourier sampling {{\citep{Cornwell1992}}}. They can also be instrumental, such as antenna pointing errors, or atmospheric. With the exception of the $w$-effect, DDEs are unknown. {On the other hand,} DIEs are spatially-constant complex modulations in the image domain. These can be equivalently modelled as complex scalar multipliers in the spatial Fourier domain. In general, antenna gains are varying in time and frequency.

Traditionally, the calibration process is limited to the estimation of the {DIEs} and the approach consists in the alternation between a calibration step and an imaging step \citep{Mitchell2008,Salvini2014}. DIEs mis-modelled during imaging and lack of DDE model can limit the fidelity of the recovered radio map. Consequently, the recovery of weak radio sources can be severely hampered in the presence of very bright emissions which can be particularly problematic in the context of low frequency imaging. 

Novel calibration approaches aiming to estimate the DDEs have been devised in the recent years. Typically, they alternate between calibration and imaging. Building on the works of \citet{Cornwell2005,Cornwell2008} addressing the $w$-effect, \citet{Bhatnagar2008} proposed the A-projection framework where DDEs can be accounted for in the spatial Fourier domain as 2-Dimensional (2D) convolutional kernels. In practice, the A-Projection algorithm is devised to correct for the antenna primary beam in the context of full polarisation and has been further extended to the wide-band wide-field imaging \citep{Bhatnagar2013}. 
A faceting-based framework was adopted in \citet{Tasse2014b, Smirnov2015, Weeren2016}, that is based on the assumption of piece-wise constant DDEs across the field of view (FoV). Within this framework, the radio sky is partitioned into facets, where the facet centre is determined by the brightest source. Estimates of the antenna gains are then obtained for each facet. The approach has been extended to wide-band wide-field imaging \citep{tasse2018}. The resulting state-of-the-art framework corrects for generic direction-dependent effects. More recently, {{Bayesian inference approaches for calibration have been developed. \citet{arras2019} has devised a joint DIE calibration and imaging framework leveraging multivariate Gaussian probability distribution.}} \citet{Albert20} propose to probabilistically infer smooth ionospheric phase screens on each facet from gain solutions of few relatively bright sources in the field of view.

Recently, \citet{Repetti2017} and \citet{Repetti20172} proposed a joint calibration and imaging approach, consisting in the estimation of the DDEs in the spatial Fourier domain jointly with the radio image by leveraging state-of-the-art non-convex optimisation. Given the severe ill-posedness of the inverse problem, suitable regularisation for both the DDEs and the radio image were considered. On the one hand, the DDEs are modelled as smooth functions of the sky imposed via spatially band-limited convolutional kernels in the spatial Fourier domain. On the other hand, {the intensity map of} the radio sky is assumed to be non-negative and sparse in a data representation space. More precisely, average sparsity-by-analysis in a collection of orthogonal bases is promoted via the $\ell_1$ norm. The resulting non-convex minimisation task is addressed via a block-coordinate forward-backward algorithm, that is shipped with {guarantees of convergence to a critical point estimate}~\citep{Chouzenoux2016}. The method has been extended to account for the time and frequency variability of the DDEs by assuming temporal smoothness \citep{Thouvenin2018} in the context of monochromatic imaging and spectral smoothness in the context of wideband imaging \citep{Dabbech2019}, imposed via band-limited kernels in their respective Fourier domains. A generalisation of the approach to the full polarisation RI problem has been provided by \citet{Birdi2019}.

Building on the works of \citet{Repetti2017} and \citet{Thouvenin2018}, we propose to promote sparsity of the unknown image in $\ell_0$ sense via a {non-convex} log-sum penalty. Inspired by the recent developments of \citet{Repetti2019b}, consisting in a generalisation of re-weighting methods, we address the minimisation task {associated with} the log-sum penalty by solving a sequence of re-weighted $\ell_1$ minimisation tasks, approximately, i.e. for a finite number of iterations. We validate the joint calibration and imaging approach for the first time on high resolution high sensitivity real data of the radio galaxy Cyg~A. The data are observations with the {{VLA}}, acquired at the frequency bands X ($8-12~$GHz), C ($6-8~$GHz) and S ($2-4~$GHz), and {utilising} its four configurations.
The remainder of the article is structured as follows. In Section~\ref{sec:pb}, we revisit the RI observation model and provide formulations of the inverse problem from two perspectives. These are (i) RI imaging problem, given estimates of the DDEs, and (ii) RI calibration problem given an estimate of the radio map. In Section~\ref{sec:JCI}, we present the joint calibration and imaging framework. We explain the considered global non-convex minimisation task and the underlying algorithmic structure to solve it, that is the block-coordinate forward-backward
algorithm \citep{Chouzenoux2016}. Cyg~A reconstruction results from the different data sets are presented in Section~\ref{sec:cyga}. We discuss the impact of temporal and spatial smoothness of the estimated DDEs. We also showcase the suitability of strong sparsity via the high reconstruction quality of the recovered radio maps at the different bands exhibited in the achieved resolution and {depth} while presenting a high fit-to-data. Interestingly, we report the detection of three compact sources north and south east of Cyg~A at S band indicating the high fidelity achieved via the adopted joint calibration and imaging framework from the highly sensitive data of the {{VLA}}. Finally, conclusions are stated in Section~\ref{sec:cc}.
\section{Problem statement} \label{sec:pb}

\subsection {RI observation model}\label{ssec:RIobs}

RI data are Fourier components of the radio sky, modulated with the antenna gains and primary beam shapes. 
Assuming a monochromatic and non-polarised radiation, each RI measurement, also termed visibility, acquired by an antenna pair $(\alpha, \beta)\in \{1, \ldots, n_a\}^2$ at a time instance $t\in \{1, \ldots,T\}$ relates the radio map to the relative position between the antenna pair. {{Let $(u_{t,\alpha,\beta},v_{t,\alpha,\beta},w_{t,\alpha,\beta})$ denote the coordinates of the baseline associated with the antenna pair  $(\alpha,\beta)$, measured in units of the wavelength, where $ w_{t,\alpha,\beta}$  is the coordinate along the line of sight and $(u_{t,\alpha,\beta},v_{t,\alpha,\beta})$ lie on its perpendicular plane, dubbed the $uv$-plane.}} 
 Let $\Omega$ denote the observed FoV and ${\lb}\in \Omega$ the coordinates of a source in the plane tangent to the celestial sphere. Theoretically, a visibility, expressed as a function of the antennas coordinates and denoted by $\ds V:\eR^3 \mapsto \eC$, relates to the sky surface brightness ${\ds I}:\Omega\mapsto \eR_+$ as follows
\begin{multline}
 \label{eq:model-cont}
 {\ds V}(\kb_{t,\alpha, \beta},w_{t,\alpha, \beta}) =\\\int_{\Omega} e^{-2i\pi \kb_{t,\alpha, \beta}\cdot \lb}{\ds C}(\lb, w_{{{t}},\alpha,\beta}) {\ds G}_{t,\alpha}(\lb) {\ds G}^{\,\ast}_{t,\beta}(\lb) {\ds I}(\lb) d^2\lb,
\end{multline}
where $(.)^{\,\ast}$ denotes complex conjugation, $ \kb_{t,\alpha, \beta}=({{u_{t,\alpha,\beta} ,v_{t,\alpha, \beta}}}) $ is the measured Fourier mode at time instance $t$. The term ${\ds C}:\Omega \times \eR \mapsto \eC$ describes the $w$-effect and is given by ${\ds C}(\lb, w_{{{t}},\alpha,\beta}) =e^{-2i\pi {w_{{{t}},\alpha,\beta}}(\sqrt{1- \vert {\lb} \vert^2}-1)}/{\sqrt{1-\vert\lb\vert^2}},~{\rm{for~ any}}~ \lb \in \Omega$. Yet, in practice, accurate incorporation of the $w$-effect in the mapping operator is computationally demanding. Different approaches have been devised to address it approximately \citep[e.g.][]{Cornwell2008,Offringa2014,Dabbech2017,Pratley2019,arras2021}.
Note that, for the data utilised herein, we consider the approximation ${\ds C}(\lb, w_{{{t}},\alpha-\beta}) \simeq 1$,{~for~any~}${\lb}\in \Omega$ due to the narrow imaged FoV (i.e. $\vert {\lb} \vert^2\ll 1$). The modulation $\ds{G}_{t,\alpha}:\Omega \mapsto \eC$ (resp. $\ds{G}_{t,\beta}:\Omega \mapsto \eC$) is the antenna gain associated with antenna $\alpha$ (resp. $\beta$), which, as opposed to the $w$-effect, is unknown. These gains can have multiple origins \citep{Smirnov2011}. They often encompass important primary beam effects, which include but are not limited to pointing inaccuracies due to mechanical issues, errors induced by wind pressure and by the non-circular beams rotating on the sky due to parallactic angle rotation. Antenna gains can also involve phase delays incurring during the propagation of the signal through the atmosphere. Some of these effects are strongly frequency dependent.
Recovering the unknown radio map and the antenna gains from the visibilities is a non-linear inverse problem.
\subsection{Imaging problem formulation}\label{ssec:impbf}

In modern RI, using the Discrete Fourier Transform in solving the problem \eqref{eq:model-cont} is impracticable. Since the collection of the measured Fourier modes $ \kb_{\alpha,\beta,t}$, 
 for $ 1 \le \alpha < \beta \le n_a$ and ${1\le t \le T}$, does not lie on a uniformly sampled grid, the common recourse is to adopt 2D Non-Uniform Fast Fourier Transforms (NUFFT) \citep[see][]{memo,Thompson2007}. The measurements are mapped from the uniformly sampled spatial Fourier components of the radio map via convolutional interpolation kernels {\citep[e.g.][]{Fessler2003}}. Assuming known and spatially band-limited antenna gains ${\db}_{\alpha,t} \in \eC^{N}$ for any $\alpha \in \{1,\dots,n_a\}$ and time instance $t\in \{1,\dots,T\}$, the discrete version of problem \eqref{eq:model-cont} reads \citep{onose2016}
\begin{equation} 
 \label{eq:model-dist}
 \yb =\Phib{\xb} +\bb, \text{~ with } \Phib = \Gb{\Fb_{\rm2D} {\Zb}},
\end{equation}
where $\yb \in \eC^{M^{\prime}}$ is the measurement vector, whose dimension is given by ${M^{\prime}}=T n_a (n_a-1)/2$, $\xb\in\eR^N$ is the unknown radio map, $\bb\in \eC^{M^{\prime}}$ is additive white Gaussian noise and $\Phib\in \eC^{M^{\prime} \times N}$ is the mapping operator. The operator ${\Fb}_{\rm 2D}\in\eC^{\oN\times \oN}$ describes the 2D Fourier transform. The matrix ${\Gb} \in \eC^{{M^{\prime}}\times\oN}$ incorporates the spatially band-limited antenna gains and the NUFFT interpolation kernels. More specifically, each row $\gb_{\alpha,\beta,t} \in \eC^{\oN}$, associated with the antenna pair $(\alpha, \beta)$ and time instance $t$, is given by $ \gb_{\alpha,\beta,t}=\widehat{\db}^{\ddagger}_{\alpha,t} \Conv \widehat{\db}_{\beta,t}^{\,\ast} \Conv {\gb^\prime}_{\kb_{\alpha,\beta,t}}$, where $(\Conv) $ denotes the linear convolution and $(^\ddagger) $ denotes the flipping operation\footnote{Considering a vector $\zb \in \eC^{M'}$, the flipped version of $\zb$ is given by $z^{\ddagger}(s)=z(M'-s+1), $ for all $s\in\{1,\dots,M'\}$.}.
$\widehat{\db}_{\alpha,t} \in \eC^{S}$ and $\widehat{\db}_{\beta,t} \in \eC^{S}$ are the respective Fourier transforms of the antenna gains ${\db}_{\alpha,t} \in \eC^{N}$ and ${\db}_{\beta,t} \in \eC^{N}$, each having a support of size $S\le N$. ${\gb^\prime}_{\kb_{{\alpha,\beta,t}}} \in \eC^{\oN}$ is the NUFFT interpolation kernel centred at the Fourier mode $\kb_{\alpha,\beta,t}$, with a support of size ${S^\prime}\ll N$, which also acts as anti-aliasing filter. Note that the spatial bandwidth of the image to be recovered is set beyond the angular resolution of the instrument (given by the maximum projected baseline), allowing to accommodate the result of the linear convolutions between the spatially band-limited antenna gains and the NUFFT interpolation kernels. Finally, the operator $\Zb \in \eC^{\oN\times N}$ allows for a fine discrete uniform sampling in the Fourier domain via zero-padding of the unknown image. It also involves a grid correction function to pre-compensate for the convolutions with the NUFFT interpolation kernels. On a further note, we consider the oversampling factor ${o}=\oN{/N}=2\times 2$ and Kaiser-Bessel interpolation kernels with a support size ${S^\prime}=7\times 7$ \citep{Fessler2003}.

\subsection{Calibration problem formulation}\label{ssec:calpbf}
 Antenna gains are modulations varying smoothly both in space and time. Indeed, given the physical origins of the antenna gains, small-scale spatial and temporal variations of the gains cannot occur
arbitrarily. It is also worth noting that, a first order gain correction obtained from scans at a known calibrator, termed calibration transfer, is typically applied to the data. By doing so, significant changes in
direction-independent antenna gains are removed. \citet{Repetti2017} and \citet{Thouvenin2018} leverage this \emph{a priori} information and propose to model the antenna gains as spatially and temporally band-limited signals. Following this model, the antenna gains are embedded into a Fourier space whose dimension is lower than that of their original temporal and image space.

\begin{figure*}
\begin{minipage}[t]{1\linewidth}
\centering
\includegraphics[width=0.7\linewidth]{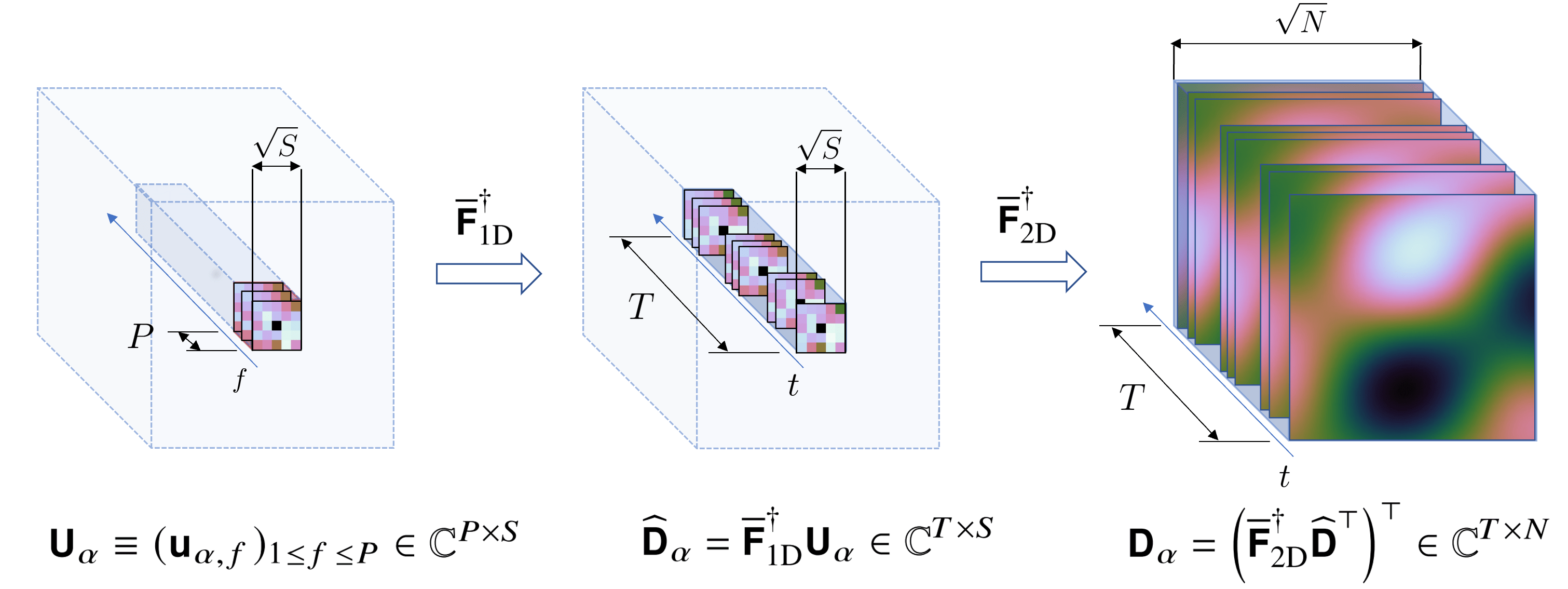}
\end{minipage}

\caption[]{{Illustration of the antenna gains Fourier model \citep[courtesy][]{Thouvenin2018}. The gains ${\Db}_{\alpha} \in \eC^{T\times N}$, associated with antenna $\alpha$, are complex-valued modulations of the sky, varying in time. Given their compact spatio-temporal Fourier representation $\Ub_{\alpha}\in \eC^{P \times S}$, the gains are obtained as $ \Db_{\alpha} =  \overline{\Fb}_{\textrm{1D}}^\dagger \Ub_{\alpha} \overline{\Fb}_{\textrm{2D}}^\ast $. For illustration purposes, in their original temporal and image space, the gains are assumed to have a dimension $\sqrt{N}$ in both spatial directions. Similarly, the compact Fourier representation of the gains is assumed to have a dimension $\sqrt{S}$ in both spatial directions.}
}
\label{fig:ddeIllustration}
\end{figure*}

Considering the matrix formulation of the gains associated with antenna $\alpha \in \{1,\dots,n_a\}$ and denoted by $\Db_{\alpha} =\left(\db_{\alpha,t}\right)_{{1\le t \le T}}\in{\eC^{T\times N}}$, smoothness in time and space is implicitly enforced by assuming that temporal and spatial Fourier transform of the gains, denoted by $\Ub_{\alpha} \in \eC^{P \times S}$, has a compact support, i.e. $P\ll T$ and $S \ll N$. Illustration of the antenna gains Fourier model is provided in Figure~\ref{fig:ddeIllustration}. From this low-dimension Fourier representation, the antenna gains in their original temporal and image space can be obtained as
$\Db_{\alpha} = \overline{\Fb}_{\textrm{1D}}^\dagger \Ub_{\alpha} \overline{\Fb}_{\textrm{2D}}^\ast$,
where $(.)^\dagger$ denotes the conjugate transpose operation of its argument, $\overline{\Fb}_{\textrm{2D}}\in \eC^{S\times N}$ is the spatial 2D Discrete Fourier transform and $\overline{\Fb}_{\textrm{1D}}$ is the temporal 1-Dimensional (1D) NUFFT \citep{Fessler2003}. The latter is adopted due to the fact that RI data are not usually acquired at equally-spaced time intervals. This can be the result of the occasional need to observe calibrator sources, typically in alternation with the source of interest, for calibration purposes or the desire of the observer to observe multiple objects of interest during the allotted time. The editing process can also contribute significantly to the non-uniform time sampling of RI data, where measurements corrupted by radio-frequency interference or other {{instrumental}} issues, need to be removed.

Assuming an estimate $\xb \in \eR^{N}$ of the radio map, in what follows, we describe the matrix formulation of the RI data model from the perspective of a given antenna $\alpha$. Let $\Yb^{\alpha}\in\eC^{ n_a\times T}$ denote the re-ordered visibilities associated with antenna $\alpha$, where $\Yb^{\alpha}_{\beta,t} = y_ {\alpha, \beta,t}$ if $\alpha < \beta$, $\Yb^{\alpha}_{\beta,t} = y_{\beta,\alpha,t}^*$ if $\alpha > \beta$ and $\Yb^{\alpha}_{\beta,t}=0$ otherwise \citep{Repetti2017}. 
For each antenna $\beta \neq \alpha$, let $\boldsymbol{\Xi}^{\alpha,\beta} \in \eC ^{ST \times S T}$ denote the diagonal per block matrix that encompasses the convolution of the oversampled Fourier transform of the radio image $\widehat{\xb}={\Fb_{\rm 2D}\Zb} \xb \in \eC^{\oN}$ with the NUFFT interpolation kernels $\left( {\gb^\prime}_{\kb_{{\alpha,\beta,t}}}\right)_{{1\le t \le T}}$. Each block $\boldsymbol{\Xi}^{\alpha,\beta}_{t}$ has on its rows truncated versions of the vectors
${{\boldsymbol{\xi}}}_{\kb_{{\alpha,\beta,t}}} =\widehat{\xb}\Conv {\gb^\prime}_{\kb_{\alpha,\beta,t}} \in \eC^{\oN}$ and its conjugate, flipped and shifted by the discrete Fourier modes $\lbrace s, s\in \{-S+1,\dots,S-1\} \rbrace $. The data model associated with antenna $\alpha$ reads

\begin{equation}
\label{eq:pb_calib_sup}
(\forall \beta \in \{1,\ldots, n_a \} \smallsetminus \{\alpha \})\quad 
\Yb^{\alpha}_{\beta} = \widehat{\Phib}_{\beta}\big( \Ub_{\alpha}\big) + \Bb^{\alpha}_{\beta}, 
\end{equation} 
\begin{equation}
{\textrm{with ~}}\widehat{\Phib}_{\beta} \left( \Ub_{\alpha} \right) ={{\textrm{Diag}}} \left( \left(\overline{\Fb}^{\dagger} _{{\rm 1D}} \Ub_{\alpha} \right)^{\ddagger} \check{\Ib}~{\boldsymbol{\Xi}^{\alpha,\beta}}~{\check\Ib}^\dagger\left( \overline{\Fb}^\dagger_{{\rm 1D}} \Ub_{\beta} \right)^\dagger \right),
\end{equation}
where $\Bb^{\alpha}\in \eC^{n_a\times T}$ is the matrix formulation associated with the realisation of a white Gaussian additive noise in \eqref{eq:model-dist}. The operator $\widehat{\Phib}_{\beta}:\eC^{P\times S} \mapsto \eC^{n_a\times T}$ maps the antenna gains $\Ub_{\alpha}$ to the data space, given estimates of the unknown image $\xb $ and gains $\Ub_{\beta}$ associated with antenna $ \beta \in \{1,\ldots, n_a \} \smallsetminus \{\alpha \}$. The matrix ${\check\Ib} \in \eC^{T \times S T}$ is diagonal per block, where each row indexed by $t$ is composed of ones at positions $\left\lbrace (t-1)S+1,\dots, tS\right\rbrace $ and zeros otherwise. Finally, the operator ${\textrm{Diag}}:\eC^{T\times T} \mapsto \eC^{T}$ returns a vector corresponding to the diagonal of its argument matrix. 
In this formulation of the problem, each RI measurement is taken into account twice (itself and its conjugate). 
 In the remainder of this article, antenna gains will be described by {their tensor representation} ${\Ub} \in \eC^{n_a \times P\times S}$.
\section{Joint Calibration and Imaging via non-convex optimisation}\label{sec:JCI}
\subsection{Global minimisation problem }\label{ssec:globalmin}
Given an estimate of the radio image, the calibration inverse problem posed in \eqref{eq:pb_calib_sup} is non-linear with respect to the antenna gains. To circumvent this non-linearity, \citet{Salvini2014} proposed a bi-linear formulation of the problem with respect to the antenna gains in the particular case of DIEs estimation, by introducing two intermediary variables $\left( {\Ub}_1,{\Ub}_2 \right)$, both representing the antenna gains, that is ${\Ub}_1 ={\Ub}_2 ={\Ub}$. Recently, \citet{Repetti2017} has generalised the bi-linear model for the estimation of the DDEs jointly with the radio image, {yielding a tri-linear inverse problem, that is addressed by} solving a non-convex minimisation task of the form
\begin{equation} 
 \label{eq:min-gen-th}
 \underset{\Ub_1,\Ub_2, \xb}{\operatorname{minimise}} \;
 {h}(\Ub_1,\Ub_{2}, \xb) + p(\Ub_1,\Ub_{2})+ r(\xb)
\end{equation}
where $\xb\in\eR^N_{+}$ is the unknown radio map, $\Ub_1\in\eC^{n_a \times P\times S}$ (resp. $\Ub_2\in\eC^{n_a \times P\times S}$) are the unknown antenna gains. 

The data fidelity function $h:\eC^{n_a \times S \times P}\times \eC^{n_a \times S \times P}\times \eR^N\mapsto \eR_+$ reads
\begin{equation}
 	{h}(\Ub_1,\Ub_{2}, \xb)= \Vert \Gc \left( \Ub_1,\Ub_2\right) {\Fb\Zb} \xb - \yb \Vert^2_2, 
\label{eq:data_fid_im}	
\end{equation}
where the operator $\Gc:\eC^{n_a \times P \times S}\times \eC^{n_a \times P \times S} $, is such that $\Gc\left( \Ub_1,\Ub_2\right)=\Gb$. The rows of $\Gb$ incorporates the convolutions between the spatial Fourier representation of the antenna gains $\overline{\Fb}_{\rm{1D}}^{\dagger}{\Ub}_{1,\alpha}$ and $ \overline{\Fb}_{\rm {ID}}^{\dagger}{\Ub}_{2,\beta}$, for every $ 1\leq \alpha< \beta \leq n_a$. 
{The $\ell_2$ criterion emanates from the negative log-likelihood associated with the probability distribution of the noise considered in our model \eqref{eq:model-dist}, that is a additive white Gaussian variable}. Note that the above definition of the data fidelity term equivalently reads 

\begin{align}
~{h}(\Ub_{1}, \Ub_{2},\xb )	&= \sum_{\alpha = 1}^{n_a}\frac{1}{2} \Vert \Hc_{\alpha,1} \left( \Ub_{\alpha,1}\right) - \Yb^{\alpha} \Vert_F^2
\label{eq:data_fid_d_C1}	\\
 &= \sum_{\alpha = 1}^{n_a}\frac{1}{2} \Vert \Hc_{\alpha,2}\left( \Ub_{\alpha,2}\right) - \Yb^{\alpha} \Vert_F^2,
 \label{eq:data_fid_d_C2}
\end{align} 
where for any antenna $\alpha \in \left\lbrace 1,\dots,n_a\right \rbrace$, the operators $\Hc_{\alpha,1}:\eC^{n_a \times P \times S} \mapsto \eC^{n_a \times T}$ (resp. $\Hc_{\alpha,2}:\eC^{n_a \times P \times S} \mapsto \eC^{n_a \times T}$) are given by $\Hc_{\alpha,1}\left( \Ub_{\alpha,1}\right)= \left[\widehat{\Phib}_{\beta}\left( \Ub_{\alpha,1}\right) \right]_{\underset{\beta\neq \alpha}{ 1 \le \beta \le n_a}} $ (resp. $\Hc_{\alpha,2}\left( \Ub_{\alpha,2}\right)=\left[\widehat{\Phib}_{\alpha}\left( \Ub_{\beta,1}\right)\right]_{\underset{\beta\neq \alpha}{ 1 \le \beta \le n_a}} $ ) and incorporates convolution operations of the NUFFT interpolation kernels, the oversampled Fourier transform of the radio image $\widehat{\xb}$ and the spatial Fourier transform of the antenna gains $ \overline{\Fb}^\dagger_{{\rm 1D}}{\Ub}_{2,\beta}$ (resp. $ \overline{\Fb}^\dagger_{{\rm 1D}}{\Ub}_{1,\alpha}$), for all antennas $\beta \in \{1,\ldots, n_a \} \smallsetminus \{\alpha \}$ (resp. $\alpha \in \{1,\ldots, n_a \} \smallsetminus \{\beta \}$). 

\noindent
The function ${p}:\eC^{n_a \times P \times S} \times \eC^{n_a \times P \times S}\mapsto \left]-\infty,+\infty\right]$ denotes the regularisation function associated with the antenna gains. In addition to the implicit temporal and spatial smoothness prior imposed via the tensor representation of the antenna gains, the regularisation function reads
\begin{equation}
\label{eq:dde-prior}
 p(\Ub_1,\Ub_{2}) = \frac{1}{2} \nu \| \Ub_{1} -\Ub_{2} \|_2^2 + \iota_{\eD^{n_a}} (\Ub_{1})+ \iota_{\eD^{n_a}} (\Ub_{2})
\end{equation}
where $\nu $ is a regularisation parameter. Assuming that calibration transfer has been done, the set $\eD=\mathcal{B}_\infty(\boldsymbol{\theta},\mu)$ is the $\ell_\infty$ complex ball, of radius $\mu \ll 1$, and centred at ${\boldsymbol \theta}\in \eC^{P\times S}$, whose zero spatial and temporal frequency component is set to 1 and the remaining coefficients are fixed to 0. 

\noindent
The function $r:\eR^{N}\mapsto \left]-\infty,+\infty\right]$ denotes the image regularisation which consists of a sparsity prior and the non-negativity constraint. Supported by compressed sensing theory, sparsity-promoting regularisation functions have been widely adopted in astronomy in general, and radio astronomy in particular, both explicitly \citep[e.g.][]{Wiaux09,li11,Dabbech2012,Carrillo2012,Garsden2015,Dabbech2015} and implicitly \citep[the celebrated \alg{clean} framework, e.g.][]{hogbom74,cornwell08b}. In this work, we adopt {{the}} {SARA} prior \citep{Carrillo2012}, that is a state-of-the-art image model in RI, consisting in the average sparsity-by-analysis in a redundant dictionary $\Psib\in \eR^{ N\times B}$ {in $\ell_0$ sense}, promoted via a log-sum regularisation function, and non-negativity constraint. The sparsity basis is composed of nine orthogonal bases. These are the Dirac basis and the first eight Daubechies wavelet bases. SARA regularisation, composed of the log-sum function and denoted by $r$, reads

\begin{equation}
\label{eq:model-log-prior}
 r(\xb) = \eta \sum_{n=1}^{B} {{\rho \log\left( 1 +\rho^{-1}{\left | \left(\Psib^\dagger \xb \right)_n\right|} \right)}} + \iota_{\eR^{N}_{+}}(\xb),
\end{equation} 
where $\left(.\right)_n$ denotes the $n^{th}$ coefficient of its argument vector {and} $(\eta,\rho) \in \eR^2_+$ are regularisation parameters. The function $\iota_{\eR^{N}_{+}}:\eR^{N}\mapsto \left[0,+\infty\right]$ is the indicator function imposing the non-negativity constraint of the radio image, which, for a given $\xb \in \eR^N$, reads
\begin{equation}
	 ~\iota_{{\eR^N_{+}}} (\xb) \overset{\Delta}{=} \left\{ \begin{aligned}
					0 & \qquad \xb \in {\eR^N_{+}} \\
					+\infty & \qquad \xb \notin {\eR^N_{+}}
				 \end{aligned} \right. 
				 \end{equation}
Note that the prior is {additively} separable with respect to the orthogonal bases. 

\citet{Repetti2017} propose to model the radio map as a sum of two components. These are (i) a known component consisting of estimates of the strong emissions in the imaged FoV and (ii) an error component to be estimated, encompassing the faint emissions and correcting for the inaccuracies of the known fixed component. In this framework, additional priors enforcing pixel-wise bounds on the amplitude of the error component at the pixel positions of the known component are introduced. The resulting joint calibration and imaging problem is reformulated {{with}} the aim {{of estimating}} the error component and the DDEs. In this work, we {consider simplified image priors, where the full image is estimated,} as it yields high quality reconstructions of Cyg~A.
\subsection{Re-weighted $\ell_1$-minimisation}\label{ssec:rw_algo}
{In the context of imaging with given estimates of the antenna gains, the non-convex SARA prior, proposed by \citet{Carrillo2012}, has been addressed by leveraging an iterative majorise-minimise algorithm consisting in solving a sequence of convex re-weighted $\ell_1$-minimisation tasks \citep{Candes2008}. The iterative procedure is summarised as follows. At each iteration $k\in \eN$, a local point estimate $\tilde{\xb}^{(k)}\in \eR^N$ is utilised to define the function $\tilde{r}^{(k)}:\eR^{N}\times \eR^{N}\mapsto \left]-\infty,+\infty\right]$, that is a convex majorant function of the non-convex image regularisation function $r$, given by
\begin{equation}
\label{eq:model-prior}
 \tilde{r}^{(k)}(\xb,\tilde{\xb}^{(k)}) = \eta \sum_{n=1}^B {\omega_n(\tilde{\xb}^{(k)})} \left | \left(\Psib^\dagger \xb \right)_n\right| + \iota_{\eR^{N}_{+}}(\xb).
\end{equation} 
The weights $\boldsymbol{\omega}^{(k)}=\big({\omega_n(\tilde{\xb}^{(k)})} \big)_{1 \leq n \leq B}$ are determined from the local point estimate $\tilde{\xb}^{(k)}$ as follows
\begin{equation}
{\omega_n(\tilde{\xb}^{(k)})} = \frac{{{\rho}}}{\rho+ \left | \left(\Psib^\dagger \tilde{\xb}^{(k)} \right)_n\right| },~\textrm{for every~} n \in \{1,\dots,B\}.
\label{eq:rw}
\end{equation}
{{Under this scheme, the weights are in the interval $]0,1]$. The regularisation parameter $\rho$ is set as the standard deviation of the noise in the sparsity basis $\Psib$. By doing so, the significant analysis coefficients are severely down-weighted (i.e. their associated weights are close to zero), thus non-penalised, whereas, the analysis coefficients which are at the noise level or below, remain highly penalised (i.e. their associated weights are close to one). In practice, the estimate of the noise level is obtained as follows. Let $\tilde{\Phib}$ denote the measurement operator in the absence of antenna gain models, thus consisting in the 2D-NUFFT transform. Given a realisation of a zero-mean white Gaussian noise $\tilde{\bb} \in \eC^{M^{\prime}}$ in the visibility domain, with a standard deviation equal to one (the RI data are whitened via natural weighting), a noise map is obtained as $\tilde{\nb}=\tilde{\beta}\tilde{\Phib}^\dagger \tilde{\bb} $. The normalisation factor $\tilde{\beta}$ is such that the Point Spread Function, associated with the measurement operator $\tilde{\Phib}$, and defined at the phase centre, has a peak value equal to 1. From this noise map, the regularisation parameter ${\rho}$ is set to the standard deviation of the vector $\Psib ^{\dagger}\tilde{\nb}$, which corresponds to the analysis coefficients of the modelled noise map.}}

The resulting weighted $\ell_1$-minimisation task at iteration $k$ is convex and reads
\begin{equation}
\label{eq:im-min} 
\underset{\xb \in \eR^{N}}{\rm{minimise}}{~\overline{h}(\xb)+ \tilde{r}^{(k)}(\xb,\tilde \xb^{(k)})},
\end{equation}
where the function $\overline{h}:\eR^N\mapsto \eR_+$, enforces fidelity to data. The solution to the convex minimisation task \eqref{eq:im-min} constitutes the local point estimate $\tilde{\xb}^{(k+1)}$ that defines the majorant function $\tilde{r}^{(k+1)}$ of the form \eqref{eq:model-prior} and consequently the associated weighted $\ell_1$-minimisation task of the form \eqref{eq:im-min}, that is to be solved at the iteration $k+1$. The iterative procedure can be initialised by fixing the initial local point estimate to $\tilde \xb^{(0)} = \mathbf{0}$.

Recently, \citet{Repetti2019b,Repetti2020} have proposed a generalisation of the above re-weighting procedure, shipped with convergence guarantees, where the convex re-weighted $\ell_1$-minimisation tasks are tackled via inexact forward-backward iterations.
Inspired by these results, we extend this approach to address the non-convex log-sum penalty within the adopted non-convex joint calibration and imaging framework. More precisely, we propose to address the minimisation task of the form \eqref{eq:min-gen-th} by solving, approximately, a sequence of non-convex minimisation tasks of the form 
\begin{equation} 
 \label{eq:min-gen}
 \underset{\Ub_1,\Ub_2, \xb}{\operatorname{minimise}} \;
 {h}(\Ub_1,\Ub_{2}, \xb)+ p(\Ub_1,\Ub_{2})+ \tilde{r}^{(k)}(\xb,\tilde{\xb}^{(k)}) ,
\end{equation}
where $\tilde{r}^{(k)}$, defined in \eqref{eq:model-prior}, is a local majorant function of $r$ at the local point estimate $\tilde{\xb}^{(k)}$. Each of these minimisation tasks are solved via the block-coordinate forward-backward (BCFB) algorithm \citep{Chouzenoux2016}, approximately, i.e. for a finite number of iterations.
Although the convergence of the overall procedure to a critical point is not proven in the context of the joint calibration and imaging, in practice, we have noticed that such inexact estimation could be crucial to avoid local minima.}

The global algorithmic structure is summarised in Algorithm~\ref{algo:rwl1}. At a given iteration $k\in \eN$, a minimisation task of the form \eqref{eq:min-gen}, associated with the weights {$\boldsymbol{\omega}^{(k-1)}$}, is solved via the BCFB algorithm within an inner loop for a finite number of iterations $I$. Details of the BCFB algorithmic structure are provided in Section~\ref{ssec:BCFWA}. 
The resulting inexact image estimate $\xb^{(k)}$ is deployed in the definition of the next minimisation task through the update of {$\boldsymbol{\omega}^{(k)}$} which is to be solved at iteration $k+1$. Together with the inexact DDE estimates $\left( \Ub_1^{(k)},\Ub_2^{(k)}\right)$, the image estimate $\xb^{(k)}$ is fed to BCFB algorithm as initial points. To indicate convergence, let ${\varphi}^{(k)}$ denote the value of the objective function{{ minimised in \eqref{eq:min-gen-th}}}  evaluated at the current estimate point $\left(\xb^{(k)},\Ub_1^{(k)},\Ub_2^{(k)}\right)$. Given ${\varrho}^\star\in \eR$, such that $ 0< \varrho^\star \ll 1$, the iterative procedure stops, once the maximum number of iterations $K$ is reached or the relative variation  {{of the objective function at two consecutive iterations}} satisfies the following condition
\begin{equation}
 {|{\varphi}^{(k-1)} -{\varphi}^{(k)}|}/ {{\varphi}^{(k-1)}} < {\varrho}^\star.
 \end{equation} 

Given the non-convexity of the global minimisation task of the form \eqref{eq:min-gen-th}, {in addition to the non-convexity of the image regularisation}, initialisation is highly important to avoid local minima. {We consider the following procedure, proposed by \citet{Repetti20172}}. The first minimisation task defined for ${\omega}_n^{(0)}=1$, for every $n\in\{1,\dots,B\}$, is initialised via the image estimate obtained with the Adaptive Preconditioned Primal-dual ({Adaptive~PPD}) \citep{Dabbech2018}. In imaging, no antenna gain solutions have been considered, and the mapping operator consists in the NUFFT transform. A hard threshold is then applied to the image estimate obtained with {Adaptive~PPD}, discarding artefacts and keeping significant signal only. As for DDEs, these are initialised randomly, while satisfying the constraints defined in \eqref{eq:dde-prior} i.e. their associated zero spatial and temporal Fourier components are set within the $\ell_\infty$ complex ball centred in 1 and the remaining coefficients are set within a $\ell_\infty$ complex ball centred in 0. 

Our experiments, presented in Section~\ref{sec:cyga} shows the efficiency of this methodology when it comes to the recovery of high quality RI images with a strong fidelity to data.

\begin{algorithm}[t]
\caption{Re-weighting iterative procedure.}
\label{algo:rwl1}
\begin{algorithmic}[1]
\small

\Given{$ {\xb^{(0)}\in \eR^N_{+}},  { \boldsymbol{\omega}^{(0)}},\left(\Ub_{1}^{(0)}, \Ub_{2}^{(0)}\right) \in \eD^{n_a} \times \eD^{n_a},~K\in \eN$}
\RepeatFor{$k=1,\ldots,K$}
\State {$
\left[{\xb}^{(k)},{\Ub_{1}^{(k)}}, {\Ub_{2}^{(k)}} \right ] = \mathrm{BCFB}
\label{algo:rw-step}
~\big({\xb}^{(k-1)},{\Ub_{1}^{(k-1)}}, {\Ub_{2}^{(k-1)}},{  \boldsymbol{\omega}^{(k-1)}},\cdots \big) 
$}
\State { {\bf update} { $\boldsymbol{\omega}^{(k)}=\big(\omega_n({\xb}^{(k)}) \big)_{1 \leq n \leq B}$}}
\State compute $\varphi^{(k)}$

\Until \textbf{convergence}: {{{ $\frac{|{\varphi}^{(k-1)} -{\varphi}^{(k)}|} {{\varphi}^{(k-1)}} < {\varrho}^\star$}}}
\end{algorithmic}
\end{algorithm}
\subsection{Block-coordinate forward-backward algorithm}\label{ssec:BCFWA}

In solving the non-convex minimisation problem of the form \eqref{eq:min-gen}, we adopt the block-coordinate forward-backward algorithm proposed in \citet{Repetti2017}. The algorithmic structure alternates between the estimation of the unknown
image $\xb$ and the antenna gains represented by $\Ub_{1}$ and $\Ub_{2}$ for a fixed number of iterations $I$. Since the minimisation problem \eqref{eq:min-gen} consists in the sum of smooth and non-smooth functions, the different variables of interest are updated sequentially via a gradient step (forward step) on the differentiable functions followed by a proximity step (backward step) on the non-smooth functions. In this section, we provide a general overview of the algorithm, that is summarised in Algorithm~\ref{algo:VMBCFB}. For full details, we advise the reader to refer to \citet{Repetti2017}.

\begin{algorithm}[]
\caption{Block-coordinate forward-backward (BCFB) algorithm.}\label{algo:VMBCFB}
\begin{algorithmic}[1] 
\footnotesize
\vspace*{0.1cm}
\State 
\textbf{Initialisation:}
Let $\xb^{(0)} \in \eR^N_{+} $, $(\Ub_{1}^{(0)}, \Ub_{2}^{(0)}) \in \eD^{n_a} \times \eD^{n_a}$, $\Gb^{(0)}$, for every $1\leq\alpha \leq n_a$, $\Hc_{\alpha,2}^{(0)}$, $\Hc_{\alpha,1}^{(0)}$, $\left(I,J,L,M \right) \in \eN^4$. 
\vspace*{0.2cm}
\State 
\textbf{repeat for} $i = 0,\dots, I-1 $ 
\vspace*{0.1cm}
\State 
\quad
$\left( \Ub_{1}^{(i,0)} , \Ub_{2}^{(i,0)}\right) = \left( \Ub_{1}^{(i)}, \Ub_{2}^{(i)}\right)$
\vspace*{0.15cm}
\State
\quad
\label{algo:step:startU}
{{{\textbf{repeat for} $l = 0, \ldots, L-1$} }}{~\algc{$//$ DDEs update: outer cycle}}
\vspace*{0.1cm}
\State
\quad\quad
$ \Ub_{1}^{(i,l,0)}= \Ub_{1}^{(i,l)}$
\vspace*{0.1cm}
\State
\quad\quad
Compute the operators $\Hc_{\alpha,1}^{(i,l)}$\label{algo:step:h1}
\vspace*{0.1cm}

\State
\quad\quad
\textbf{repeat for} $m = 0, \ldots, M-1${~\algc{$//~\Ub_1$ update: inner cycle  }}\label{algo:step:startu1}
\vspace*{0.1cm}
\State
\quad\quad\quad
\label{algo:gu1}
$\check{\Ub}_1^{(i,l,m)} = \Ub_1^{(i,l,m)}- \Upsilon^{(i,l)} \cdot \Big(  \nu \big( \Ub_1^{(i,l,m)}-\Ub_2^{(i,l)}\big) +{\hspace{1.8cm} \nabla_{\Ub_1} {h}^{(i,l)} \big( \xb^{(i)}, \Ub_1^{(i,l,m)}, \Ub_2^{(i,l)}\big) }\Big) $ \\
\quad\quad\quad
$\Ub_1^{(i,l,m+1)} = \Pc_{\eD^{n_a}} \left( \check{\Ub}_1^{(i,l,m)} \right)$    
\State
\quad\quad
\textbf{end for}\label{algo:step:endu1}
\vspace*{0.1cm}

\State
\quad\quad
$\left( \Ub_{1}^{(i,l+1)} , \Ub_{2}^{(i,l,0)}\right) = \left( \Ub_{1}^{(i,l,M)}, \Ub_{2}^{(i,l)}\right)$
\State
\quad\quad
Compute the operators $\Hc_{\alpha,2}^{(i,l)}$\label{algo:step:h2}
\vspace*{0.1cm}
\State
\quad\quad
\textbf{repeat for} $m = 0, \ldots, M-1	${~\algc{$//~\Ub_2$ update: inner cycle  }}\label{algo:step:startu2}
\vspace*{0.1cm}
\State\label{algo:gu2}
\quad\quad\quad
$\check{\Ub}_2^{(i,l,m)} = \Ub_2^{(i,l,m)} -\Upsilon_2^{(i,l)} \cdot  \Big(  \nu \big( \Ub_2^{(i,l,m)} - \Ub_1^{(i,l+1)}\big) + {\hspace{1.8cm} \nabla_{\Ub_2} {h}^{(i,l)} \big( \xb^{(i)}, \Ub_1^{(i,l+1)}, \Ub_2^{(i,l,m)} \big)  } \Big)$ \\
\quad\quad\quad
$\Ub_2^{(i,l,m)} = \Pc_{\eD^{n_a}} \left( \check{\Ub}_2^{(i,l,m)} \right) $ 
\vspace*{0.1cm}

\State
\quad\quad
\textbf{end for}\label{algo:step:endu2}
\vspace*{0.1cm}

\State
\quad\quad
$\Ub_2^{(i,l+1)} = \Ub_2^{(i,l,M)}$
\vspace*{0.1cm}
\State
\quad
{{\textbf{until} $ \max \big(\frac{|| \Ub_1^{(i,l+1)} -\Ub_1^{(i,l)}||_2}{|| \Ub_1^{(i,l)}||_2}, \frac{|| \Ub_2^{(i,l+1)} -\Ub_2^{(i,l)}||_2} { || \Ub_2^{(i,l)}||_2} \big) < \varsigma$} }\label{algo:step:stopDDEs} 
\vspace*{0.15cm}
\State 
\quad
$\left(\Ub_1^{(i+1)}, \Ub_2^{(i+1)} \right) = \left( \Ub_1^{(i,L)}, \Ub_2^{(i,L)} \right)$
\State
\label{algo:step:updateG}
\quad $\Gb^{(i)}=\Gc\left( \Ub^{(i+1)}_1,\Ub^{(i+1)}_2\right)$
\State
\quad
$\xb^{(i,0)} = \xb^{(i)}$
\vspace*{0.1cm}
\State
\quad
{{\textbf{repeat for} $j = 0, \ldots, J-1$}} {~\algc{$//$ Update  image $\xb$ }}\label{algo:step:startim}

\State\label{algo:gx} 
\quad\quad\quad

$\check{\xb}^{(i,j)} =\xb^{(i,j)}- \lambda^{(i)} {\nabla_{\xb} h^{(i)}\left( \xb^{(i,j)}, \Ub_1^{(i)}, \Ub_2^{(i)} \right)}$ ~	\\
\quad\quad
\label{algo:step:proxx} 
$\xb^{(i,j+1)} = \prox_{\eta\lambda^{(i)} \tilde{r}} \left( \check{\xb}^{(i,j)} \right)$ ~

\State
\quad
{{\textbf{until} $\frac{|| \xb^{(i,j+1)} -\xb^{(i,j)}||_2}{ || \xb^{(i,j)}||_2 }< \vartheta$\label{algo:step:endim}}}
\vspace*{0.1cm}

\State
\quad
$\xb^{(i+1)}= \xb^{(i, J)}$

\vspace*{0.1cm}

\State 
\quad
compute $\tilde{\varphi}^{(i)}$
\vspace*{0.1cm}

\State
{{\textbf{until}   $\frac{{|\tilde{\varphi}^{(i+1)} -\tilde{\varphi}^{(i)}|}}{ {\tilde{\varphi}^{(i)}}} < \check{\varrho}$} }

\end{algorithmic}
\end{algorithm}
At each global iteration $i < I$, the updates of the three variables of interest are performed within two main blocks, dedicated to the estimation of the DDEs and the radio image, respectively. Within each block, a finite number of iterations is performed yielding inexact estimates of the associated variables, that is crucial to reach convergence {to a critical point estimate} \citep{Repetti2017}. {{Note that, in both blocks, an additional stopping criterion is considered to avoid unnecessary computations near convergence. The criterion is based on the relative variation of the associated estimated variable between two consecutive iterations. Details of the parameters involved in the stopping criteria of the different iterative blocks are provided in Table~\ref{tab:auxvars}.}}

Concerning DDEs, these are iteratively updated within an outer cycle {(Steps~\ref{algo:step:startU}-\ref{algo:step:stopDDEs})} composed of a finite number of iterations $L\in \eN$ alternating between the updates of $\Ub_1$ and $\Ub_2$. {These updates are performed within the inner cycles (Steps~\ref{algo:step:startu1}-\ref{algo:step:endu1} and Steps~\ref{algo:step:startu2}-\ref{algo:step:endu2}, for $\Ub_1$ and $\Ub_2$, respectively), each} composed of a finite number of sub-iterations $M^{\prime}\in \eN$. More precisely, let $l\in \left \lbrace 0,\dots,L-1 \right \rbrace$ denote an iteration of the DDEs outer cycle. Given the estimates $\xb^{(i)}$ and $\Ub_2^{(i,l)}$, the update of antenna gains $\Ub_{1}$ is performed within the inner cycle described in Steps~\ref{algo:step:startu1}-\ref{algo:step:endu1} of Algorithm~\ref{algo:VMBCFB}, and consists of a gradient step involving the data fidelity term given in \eqref{eq:data_fid_d_C1} through the partial gradient of ${h}^{(i,l)}$ with respect to $\Ub_{1}$ and the smooth prior minimising the distance between $\Ub_{1}$ and $\Ub_{2}$, followed by a backward step consisting in the projection onto the set $\eD^{n_a}$. Note that these steps are performed in parallel with respect to each antenna $\alpha \in \{1,\dots,n_a\}$. The update of the antenna gains $\Ub_2$, is then performed in a symmetrical fashion using the current estimate $\xb^{(i)}$ and the latest estimate $\Ub_1^{(i,l+1)}$ via Steps~\ref{algo:step:startu2}-\ref{algo:step:endu2}. 

Once the associated mapping operator is updated from DDE estimates $\left(\Ub_1^{(i,L)},\Ub_2^{(i,L)}\right)$, the unknown radio image is updated within a cycle composed of a finite number of iterations $J\in \eN$, explained in Steps~\ref{algo:step:startim}-\ref{algo:step:endim} of Algorithm~\ref{algo:VMBCFB}. More specifically, a gradient step enforcing the fit-to-data given in \eqref{eq:data_fid_im} through the partial gradient of $h^{(i)}$ with respect to $\xb$ is performed and followed by a proximity step\footnote{Considering a lower semi-continuous and proper convex function $g$, its proximal operator is defined as 
\begin{equation}
	 (\forall\zb), ~\prox_g (\zb) \overset{\Delta}{=} \textrm{argmin}_{\bar{\zb}} g(\bar{\zb}) + \frac{1}{2} \| \zb - \bar{\zb}\|_2^2.\end{equation}}
enforcing the image regularisation. The latter step is solved via a Primal-dual forward-backward algorithm \citep{Condat2013,Vu2013}. 
 
The convergence of the iterative procedure described above to a critical point $\big( \xb^\star, \Ub_{1}^\star,\Ub_{2}^\star\big)$ is guaranteed under certain conditions {{\citep{Chouzenoux2016,Repetti2017}}}. Let $\lambda^{(i)}$ denote the {step-size} associated with the image estimate involved in Step~\ref{algo:gx} and $\Upsilon_1^{(i,l)} = \left[ \upsilon_{1,1}^{(i,l)} \unb_S \ldots | \upsilon_{1,n_a}^{(i,l)} \unb_S\right]^\top $ (resp. $
\Upsilon_2^{(i,l)} = \left[ \upsilon_{2,1}^{(i,l)} \unb_S |	\ldots | \upsilon_{2,n_a}^{(i,l)} \unb_S\right]^\top $) denote the {step-size} associated with the estimate of $\Ub_1$ (resp. $\Ub_2$), involved in Step~\ref{algo:gu1} (resp. Step~\ref{algo:gu2}) of Algorithm~\ref{algo:VMBCFB}. {{Convergence is ensured}} if at each global iteration $i\in \eN$, the following conditions are met
\begin{equation}
0 < \lambda^{(i)} < 1/ \| \Gc\left( \Ub^{(i+1)}_1,\Ub^{(i+1)}_2\right) {\Fb\Zb} \|_s^2,
\end{equation}
where $\|.\|_s$ denotes the spectral norm, and for each iteration of DDEs outer cycle, denoted by $l \in \left \lbrace 0,\dots,L-1 \right \rbrace$, 
\begin{equation}
0 < \upsilon_{\ell,\alpha}^{(i,l)} < 1 / ( \nu + \zeta_{\ell,\alpha}^{(i,l)} ), ~\forall (\alpha,\ell) \in \{1, \ldots, n_a\}\times \{1,2\}
\end{equation}
where $\left(\zeta_{1,\alpha}^{(i,l)},\zeta_{2,\alpha}^{(i,l)}\right) $ are the respective spectral norms {{squared}} of operators $\left(\Hc_{\alpha,1}^{(i,l)},\Hc_{\alpha,2}^{(i,l)}\right) $, {associated with the current estimate of the image}. {{Given these conditions, the step-sizes are set to 0.98 times the associated upper bounds.}}
In the context of the re-weighting procedure described in Algorithm~\ref{algo:rwl1}, the BCFB algorithm is stopped before convergence, that is typically once the fixed number of iterations $I$ is reached. {{Yet, to avoid unnecessary computations near convergence,  Algorithm~\ref{algo:VMBCFB} may be stopped earlier, when the relative variation of the minimised objective function in \eqref{eq:min-gen}, and denoted by $\tilde{\varphi}$ is small.}}

{{On a final note, any given DDEs estimates, such as $w$-modulations,  can be incorporated in the measurement operator as described in Section~\ref{ssec:impbf}. In this context, the algorithmic structure can be adopted to solve the imaging problem, by deactivating the DDEs updating cycle (Steps~\ref{algo:step:startU}-\ref{algo:step:updateG}).}}
{{

\begin{table}
 {{	\caption{The parameters involved in the stopping of criteria of the different iterative cycles of Algorithm~\ref{algo:VMBCFB}. In essence, each iterative cycle stops when the fixed number of iteration is reached. However, near convergence, an iterative cycle can be stopped earlier if the relative variation of the associated estimate reaches a configurable lower bound.}
 	
 	\centering
	\small
 	\begin{tabular}{p{2.1cm}p{5.5cm}}
	\hline
   \textbf{Global cycle}	 & ~\\
    \hline
    ${I} \in \eN $ & configurable; the maximum number of global iterations to be performed in Algorithm~\ref{algo:VMBCFB}, set to ${I}=10$.\\
	$0 < \check{\varrho} \ll 1 $ & configurable; the lower bound on the relative variation between two consecutive values of the objective function minimised via Algorithm~\ref{algo:VMBCFB}, indicating near convergence,  set to $\check{\varrho}=5\times10^{-3}$.\\
	\hline
	\textbf{Calibration cycle} & ~\\
	\hline
	${M} \in \eN $ & configurable; the number of iterations to be performed within the inner cycle of the DDEs updates, set to $M=5$.\\
	${L} \in \eN $ & configurable; the maximum number of iterations to be performed within the outer cycle of the DDEs updates, set to ${L}=5$.\\
	$0 < \varsigma \ll 1 $ & configurable; the lower bound on the relative variation between two consecutive DDE estimates in the outer cycle of DDEs updates, indicating near convergence, set to $\varsigma=10^{-4}$.\\
	\hline
    \textbf{Imaging cycle} & ~ \\
	\hline
	${J} \in \eN$ & configurable; the maximum number of iterations to be performed,  set to ${J}=150$.\\
	$0 < \vartheta \ll 1 $ & configurable; the lower bound on the relative variation between two consecutive image estimates, indicating near convergence, set to $\vartheta=5\times10^{-5}$.\\
	\hline 
 	\end{tabular}
\label{tab:auxvars}
}}
\end{table}
}}

\section{Application to Cyg~A observations}\label{sec:cyga}
To showcase the performance of our approach on real RI data, we consider observations of the radio galaxy Cyg~A at multiple frequency bands acquired with the {{VLA}}. Note that observations at bands X and C have been recently investigated in the context of RI imaging via the Adaptive Preconditioned Primal-dual ({Adaptive} PPD) approach \citep{Dabbech2018}. In this section, we study the performance of the joint DDE calibration and imaging with respect to joint DIE calibration and imaging and {Adaptive~PPD} imaging. In this work, we do not provide a full comparison with DDFacet \citep{tasse2018} that is a state-of-the-art framework for DDEs calibration, as it has not been validated in regimes where extended radio emissions, such as Cyg~A, span multiple facets. On a further note, reconstructed images using our joint DDE calibration and imaging approach are made available online [dataset]~\citet{Cyg21}.
\subsection{Observations details}\label{ssec:obs}

The RI data utilised herein consist of highly sensitive observations of Cyg~A with the {{VLA}}. {Data acquisition with its four configurations (A, B, C, D) was conducted over two years (2014 -- 2015),} spanning the frequency range $2-18~\rm{GHz}$. All observations utilise a single pointing centred at the core of the radio galaxy, that is given by the coordinates $\rm{RA}=19\rm{h}59\rm{mn}28.356\rm{s}$ ($J2000$) and $\rm{DEC}=+40^{\circ}44\arcmin2.07\arcsec$ at the frequencies 8.422~GHz (X band), 6.678~GHz (C band) and 2.052~GHz (S band). The original integration duration and channel-width were 2 seconds and 2~MHz, respectively. Careful self-calibration has been performed on the data in \alg{AIPS} by alternating between a DIE calibration step and an imaging step with Cotton-Schwab \alg{clean} \citep{csclean1983}. Details of the data editing process and self-calibration are provided in \citet{Sebokolodi2020}. Following this pre-processing step, both X and C band data have been averaged over a duration of about 8~seconds and channel-width of 8~MHz while S band data have been averaged over a time lapse of about 10~seconds. Additional band-specific details of the data are shown in Table~\ref{tab:timeObs}.
	\begin{table}
	
	\caption{Details of the data acquired with the configurations of the VLA at the observed frequencies.}
	
	 \hspace{-0.22cm}
	 \resizebox{0.5\textwidth}{!}{    
\centering
	\begin{tabular}{lcccr} 
		\hline
		\textbf{Num. of data points} & A & B& C & D\\
		\hline
		8.42GHz &1432161 &360998  &185648 &  101565\\
		6.67GHz &   773706   &   247050  &197641  &73482
      \\
        2.05GHz &762775  & 268376 & 202776 &47750     \\
		\hline
		\textbf{Num. of active antennas} & A & B& C & D\\
		\hline
		8.42GHz & 25& 25 & 25 & 25\\
		6.67GHz & 27 & 26 & 26& 26      \\
        2.05GHz & 26 & 27 & 26 &26     \\
		\hline
	\textbf{Total observation time }& A & B& C & D\\
		\hline
		8.42GHz~ & $14.33\rm{h}$&$6.96\rm{h}$& $9.80\rm{h}$ & $5.34\rm{h}$\\
		6.67GHz & $7.18\rm{h}$ &$6.76\rm{h}$  & $9.8\rm{h}$ &$5.28\rm{h}$       \\
			2.05GHz & $7.17\rm{h}$ &$7.07\rm{h}$  & $11.75\rm{h}$ &$5.3\rm{h}$       \\
		\hline
		\textbf{Num. of time slots }& A & B& C & D\\
		\hline
		8.42GHz~ & 2015+2229 & 1113 & 625 & 348\\
		6.67GHz &  2248 & 765 & 627 &230     \\
			2.05GHz & 2347 &927 &676 &198       \\
	
		\hline
	\end{tabular}
    	}
	 	\label{tab:timeObs}

\end{table}
%
\subsection{Joint calibration and imaging general settings}\label{ssec:settings}
For the three bands considered herein, joint calibration and imaging is applied to the aggregated observations 
utilising the four configurations of the {{VLA}}. In this context, we consider additively separate data fidelity terms and DDE regularisation terms, associated with the different
configurations. Note that, the data have been whitened using estimates of the standard deviation of the noise in order to preserve the statistics of the noise. This type of weighting is the so-called natural weighting. The following settings have been adopted for all data sets.

Let the subscript $ c \in\{A,B,C,D\}$ denote a configuration of the {{VLA}}. The impact of the temporal and spatial smoothness of the DDEs is studied by varying their spatial and temporal bandwidths. The spatial bandwidth of the DDEs is varied such that $S_c\in \{ 3\times3, 5\times5, 7\times7\}$. The DIE calibration case ($S_c=1\times1$) is also considered. Let $\mathbf{t}_c \in \eR^{T_c}$ denote the observation time domain associated with a configuration $c$, that is defined as the total observation time discretised with a sampling rate corresponding to the mean value of the integration time $\delta t$. 
The temporal variation of the DDEs is characterised via the reduction ratio $\tau_c$\footnote{The dimension in the temporal Fourier domain is the closest odd integer to $ T_c/\tau_c $}, such that the temporal bandwidth $P_c$ is set to $P_c\approx T_c/\tau_c $. In other words, $\tau_c$ represents the ratio between DDEs full temporal bandwidth and their effective temporal bandwidth. The investigated temporal scales are such that $\tau_c \in\{ 2^p\}_{0\leq p \leq7}$, where the case $\tau_c =1$ corresponds to absence of the temporal smoothness prior. Note that some of the data sets analysed herein present large gaps in the time domain, where the number of time slots (i.e. the number of
unique timestamps after time-averaging), reported in Table~\ref{tab:timeObs}, may be significantly smaller than the temporal dimension resulting from a chosen temporal reduction ratio $\tau_c$. When this is the case, no temporal smoothness is imposed.
In all experiments conducted herein, the spatial bandwidth of the DDEs and their associated temporal reduction ratio are fixed for the four configurations. Thus, for simplicity, the subscript $c$ will be omitted from the considered spatial support sizes and temporal reduction ratios of the DDEs, in the remainder of the article.

{{Considering the operator $\tilde{\Phib} =[ \tilde{\Phib}_c ]_{c\in\{A,B,C,D\}}$, that is the concatenation of the measurements operators associated with the configurations of the VLA in absence of antenna gain models, the image regularisation parameter is fixed to {{$\eta = \check{\eta} \Vert \tilde{\Phib} \Vert^2_S$}} with $\check{\eta}$ chosen in the interval $[10^{-7}, 10^{-6}]$ and $\Vert \tilde{\Phib} \Vert_S$ being the spectral norm of the operator $\tilde{\Phib}$. The estimated DDEs are enforced to be in the set $\eD=\mathcal{B}_\infty(\boldsymbol{\theta},\mu)$, with radius $\mu = 0.01$. For any VLA configuration $c$, the regularisation parameter involved in the DDE prior term is set to $\nu_c ={{\Vert \tilde{\Phib}_c \Vert^2_S}}/{n_a}_c$, where $\Vert \tilde{\Phib}_c \Vert_S$ is the spectral norm of the associated measurement operator $ \tilde{\Phib}_c$ and ${n_a}_c$ is the number of the active antennas.

In Algorithm \ref{algo:rwl1}, the image estimate is initialised from the model image obtained in the absence of antenna gain models via {Adaptive~PPD} \citep{Dabbech2018}, to which a hard-thresholding is performed, keeping only pixels that are within four orders of magnitude from the peak value. 
Note that, alternatively, imaging can be performed using the algorithmic structure described in Section~\ref{sec:JCI} by deactivating the DDE estimation block. 
To indicate convergence, the maximum number of iterations, which corresponds to the number of the weighted $\ell_1$-minimisation tasks of the form \eqref{eq:min-gen} to be solved approximately, is fixed to $K=40$ and the lower bound on the relative variation of the objective function described in \eqref{eq:min-gen-th} is set to $\varrho^\star =10^{-3}$.
With regards to Algorithm~\ref{algo:VMBCFB}, details of the parameters involved in the stopping criteria of its iterative cycles are provided in Table~\ref{tab:auxvars}. }}

To assess the reconstruction quality of the joint calibration and imaging approach, we provide a visual inspection of the recovered model images and DDE estimates, in addition to the residual images. These are obtained as ${\rb} ={\beta}^{\star}{\Phib}^\dagger(\yb-{\Phib}{\xb})$, where ${\beta}^\star$ is a normalisation factor such that, the Point Spread Function, associated with the estimated measurement operator ${\Phib}$, and defined at the phase centre, has a peak value equal to one. Numerical analysis of the results is conducted via the statistics of the residual data and images. Data fidelity is also assessed via the Signal to Noise Ratios (SNR)
$\rm{vSNR} = 20\log_{10}\left({\Vert{\yb}\Vert_2}/{\Vert\yb-{\Phib}{\xb}\Vert_2}\right)$ and $\rm{iSNR} = 20\log_{10}\left({\Vert \beta^{\star}{\Phib}{\yb}\Vert_2}/\Vert{{\rb}\Vert_2}\right)$, evaluated in the visibility and image spaces, respectively.
\vspace{-0.1cm}
\subsection{C band: the impact of DDEs spatial and temporal smoothness}
C band data are imaged at a spatial resolution that is about 1.56 times the instrumental resolution of the observations, which corresponds to a pixel size $\delta x=0.08\arcsec$. The mapped sky of interest is of size $N=2048\times 2048$, that is a FoV $\Omega= 0.0455\degr \times 0.0455\degr $. Note that the extent of Cyg~A is about $0.036\degr$. Leveraging these data, two sets of experiments have been conducted to investigate the impact of the temporal and spatial smoothness of the estimated DDEs, by probing different sizes of the DDEs spatial dimensions, at a first instance, and investigating multiple temporal reduction ratios at a second instance. 
\vspace*{-0.5cm}
\subsubsection*{DDEs spatial smoothness}
\label{sec:cyga-c}
\begin{table}
    \caption{C band: Statistics of the residual data and image for different spatial bandwidths of the antenna gains ($\tau =8$).}
   \resizebox{0.5\textwidth}{!}{         \centering
    \begin{tabular}{lllll}
	\hline
       \textbf{Residual data} 
       & STD & Skewness  & Kurtosis & vSNR{~(dB)} \\
       \hline
       Adaptive PPD     & 2.43 & -0.720  & 34.45 & 34.46 \\
       $S=1\times1$     & 1.86 &  -1.266 & 63.51 &  37.08\\
       $S=3\times3$     & 0.94 & -0.020  & 1.18  & 42.96 \\
       $S=5\times5 $    & 0.90 & 0.003   & 1.13  & 43.40\\
       $S=7\times7$     & 0.87 & 0.006   & 1.20  &  43.66\\
	\hline
      \textbf{Residual Image}& STD~{($\times10^{-4}$)} & Skewness & Kurtosis  & iSNR{~(dB)}\\
	\hline
	Adaptive PPD  &  3.39  & -1.800 & 4.65  & 45.22\\
	$S=1\times1$  & 2.13  & -0.721 & 0.91  & 50.35\\
    $S=3\times3$  & 1.25  & -0.047 & 0.28  & 55.13 \\
	$S=5\times5$  &  1.08 & 0.002  & 0.27  & 56.44 \\
	$S=7\times7$  & 0.97  & 0.031  & 0.26  & 57.38 \\

         \hline
    \end{tabular}
   } 
\label{tab:C_vis} 
     
\end{table}

In this experiment, the effective temporal bandwidth of the DDEs associated with the different configurations of the {{VLA}} have been set such that $\tau=8$. The resulting dimensions of the estimated compact DDE kernels in the temporal Fourier domain are $321$, $303$ and $439$ for configurations A, B and C, respectively. No temporal smoothness is imposed for configuration D given the sparse temporal sampling of the observations (see Table~\ref{tab:timeObs}). {{With the aim of investigating}} the spatial smoothness of the antenna gains, we vary the spatial bandwidth of the DDEs, and include the DIE case, such that $ S \in\{1\times 1,3\times 3,~5\times 5$, $7\times 7\}$.
In all joint calibration and imaging tests, the sparsity regularisation parameter is set such that $\check{\eta}=5\times10^{-7}$. The remaining parameters involved in Algorithm~\ref{algo:VMBCFB} are set as described in Section~\ref{ssec:settings}.

Reconstruction results of DDE calibration with the different spatial bandwidths are shown in Figure~\ref{fig:c_maps}. One can clearly notice the high quality of the images provided by DDE calibration, exhibited through highly sensitive model images, as opposed to DIE calibration, where the fidelity of the associated model image is severely limited by the ringing artefacts. Moreover, DDE calibration allows for the recovery of faint features of the jets whose surface brightness are about five order of magnitude lower than the peak value (that is about $1.22$ Jy/pixel), with background artefacts being very limited, in particular for DDEs spatial support sizes $S\in\{5\times 5, 7\times 7\}$ (see the zooms on regions of the west and east jets displayed in Figure~\ref{fig:c_maps}, second and third columns). Note that we do not show the recovered maps via Adaptive PPD in absence of calibration as these can be found in \citet{Dabbech2018}, nevertheless we emphasise herein their comparable quality to joint DIE calibration and imaging.

A serendipitous discovery of a transient at the inner core of the radio galaxy has been recently made at X band, with a flux of about 4~mJy \citep{perley2017}. The source, dubbed Cyg~A-2, is interpreted as a secondary super-massive black hole. It has also been imaged at C band via {Adaptive~PPD} \citep{Dabbech2018}, where \alg{CLEAN}-based approaches failed to resolve it. Inspection of the inner core of Cyg~A displayed in Figure~\ref{fig:c_maps}, fourth column, indicates that the transient, located south east from the main black hole, is well recovered with a flux\footnote{The flux is computed over a circular region centred at $\rm{RA}=19\rm{h}59\rm{mn}28.320\rm{s}~ (J2000)$ and $\rm{DEC}=+40\degr 44\arcmin 1.86\arcsec$ with a radius of $0.1\arcsec$} of about 4.7~mJy, 4.5~mJy and 4.6~mJy achieved by DDE calibration with $S$ fixed to $3\times 3$, $5\times5$ and $7\times 7$, respectively, against 4~mJy with DIE calibration and 4.5~mJy obtained with {Adaptive~PPD} imaging \citep{Dabbech2018}. These values are close to the one reported in \citet{perley2017} at X band. In addition to Cyg~A-2, one can notice structures emerging via DDE calibration around the main black hole, more precisely on the diagonal of the jets' axis. However, the hypothesis of calibration-induced artefacts can not be ruled out at this stage. 

The visual examination of the residual maps shows the higher fidelity to data reached via DDE calibration in comparison with DIE calibration and {Adaptive~PPD} imaging. Quantitative evaluation of the residual images, as well as the residual data, is provided in Table~\ref{tab:C_vis}, through the examination of their statistics\footnote{For a given complex vector $\zb\in \eC^{M^{\prime}}$, all metrics are applied on its real representation $\tilde{\zb}\in \eR^{2 {M^{\prime}}} $ resulting from the concatenation of its respective real and imaginary parts.} including measures of Non-Gaussianity (excess kurtosis and skewness). The numbers confirm the Gaussian nature of the residual images obtained with DDE calibration, where both skewness and excess kurtosis values are close to zero. Knowing that the data have been naturally-weighted, statistics of the residual data, more precisely, their standard deviation (STD) values being close to one, suggest that DDE calibration enables to reach the theoretical noise level.
This study shows the efficiency of DDE calibration, where strong spatial smoothness of the DDEs imposed via a spatial support size as small as $S=3\times3$ can achieve high quality radio maps with important fidelity to data and limited artefacts in the model images. 

\begin{figure*}
\begin{minipage}[t]{1\linewidth}
\centering
\includegraphics[width=0.33\linewidth]{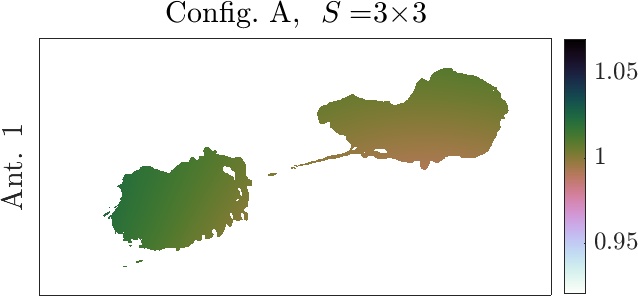}
\includegraphics[width=0.31\linewidth]{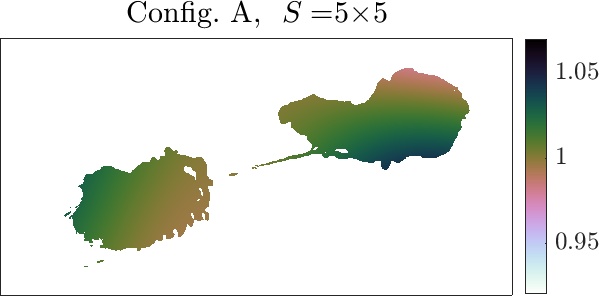}
\includegraphics[width=0.31\linewidth]{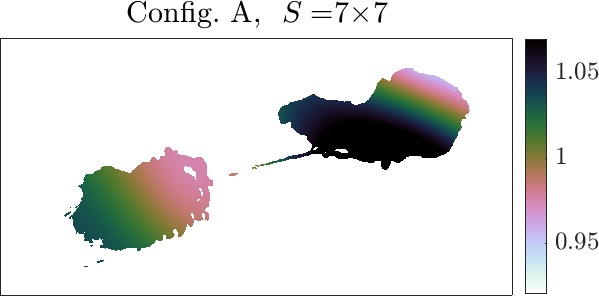}
\includegraphics[width=0.33\linewidth]{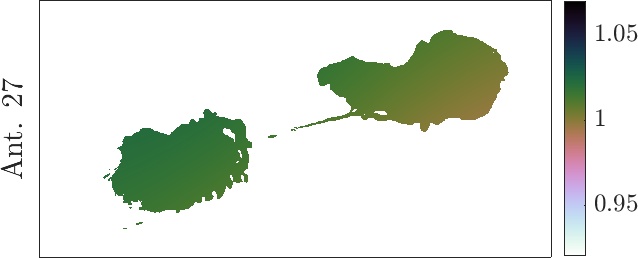}
\includegraphics[width=0.31\linewidth]{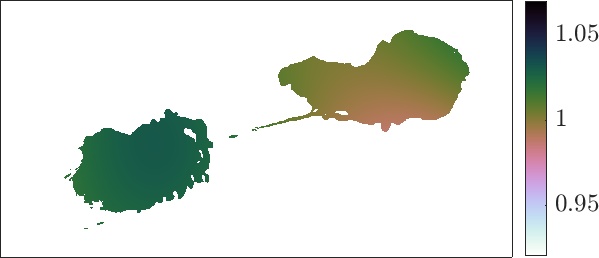}
\includegraphics[width=0.31\linewidth]{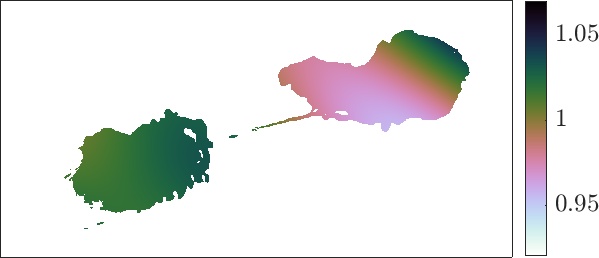}
\end{minipage}
~\\
\begin{minipage}[t]{1\linewidth}\centering
\includegraphics[width=0.33\linewidth]{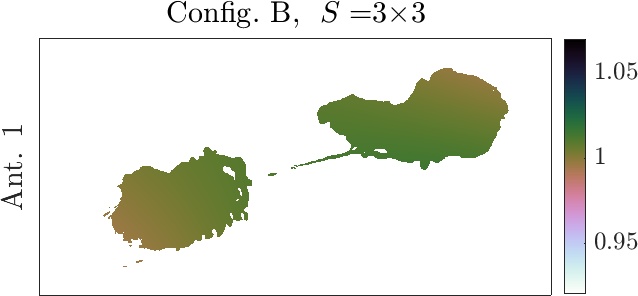}
\includegraphics[width=0.31\linewidth]{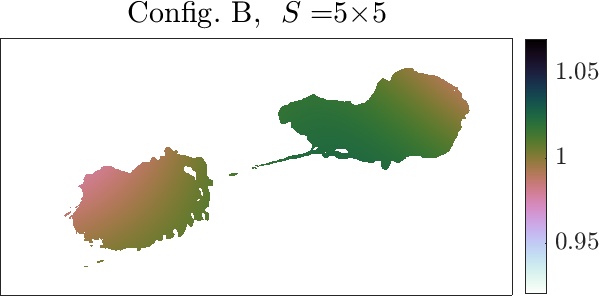}
\includegraphics[width=0.31\linewidth]{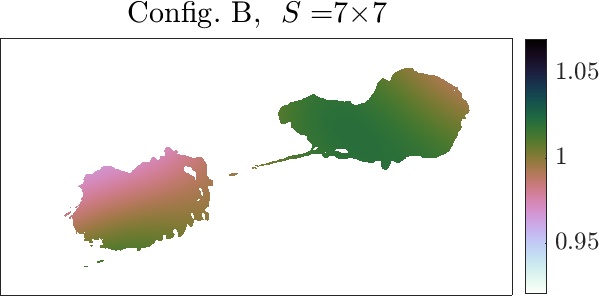}
\includegraphics[width=0.33\linewidth]{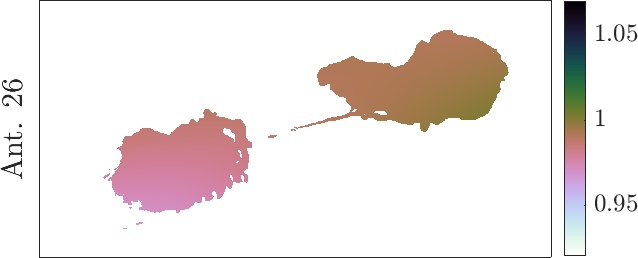}
\includegraphics[width=0.31\linewidth]{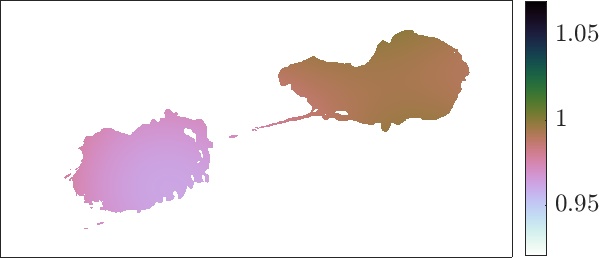}
\includegraphics[width=0.31\linewidth]{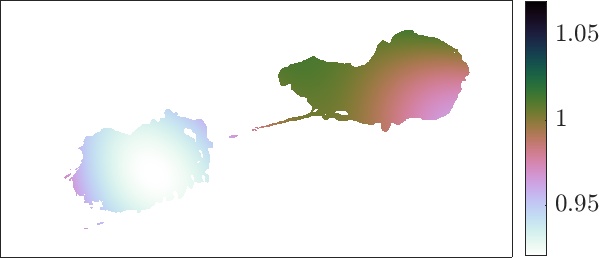}
\end{minipage}
~\\
\begin{minipage}[t]{1\linewidth}\centering
\includegraphics[width=0.33\linewidth]{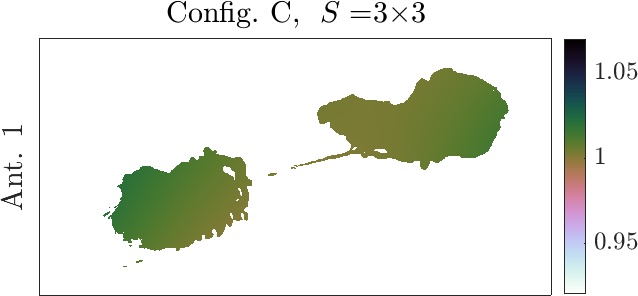}
\includegraphics[width=0.31\linewidth]{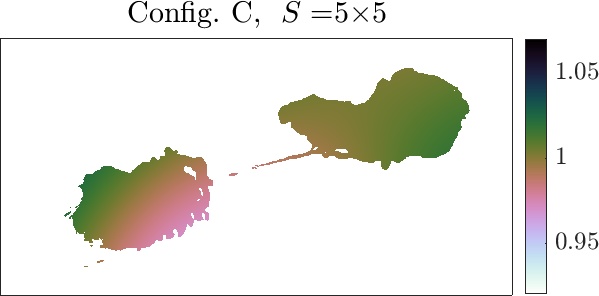}
\includegraphics[width=0.31\linewidth]{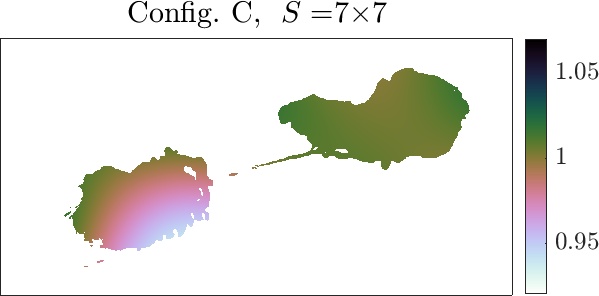}
\includegraphics[width=0.33\linewidth]{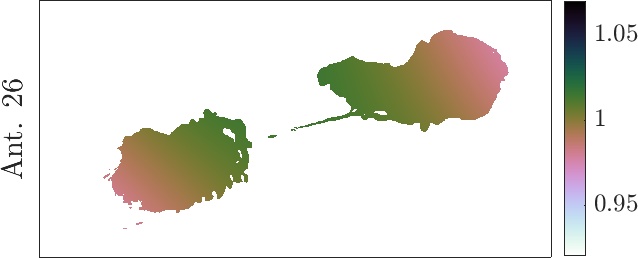}
\includegraphics[width=0.31\linewidth]{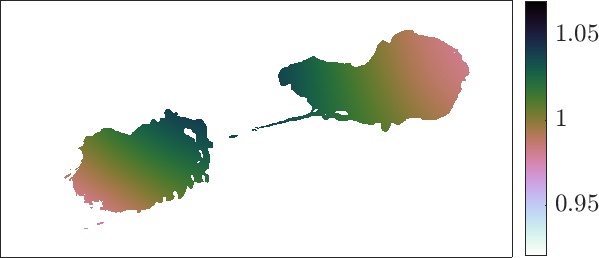}
\includegraphics[width=0.31\linewidth]{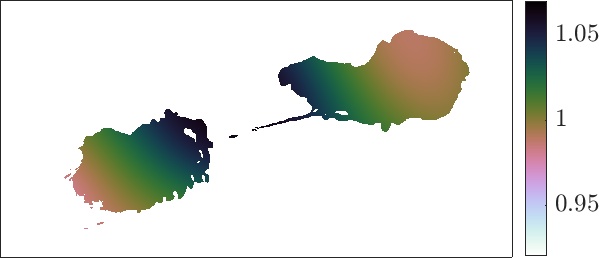}
\end{minipage}
~\\
\begin{minipage}[t]{1\linewidth}\centering
\includegraphics[width=0.33\linewidth]{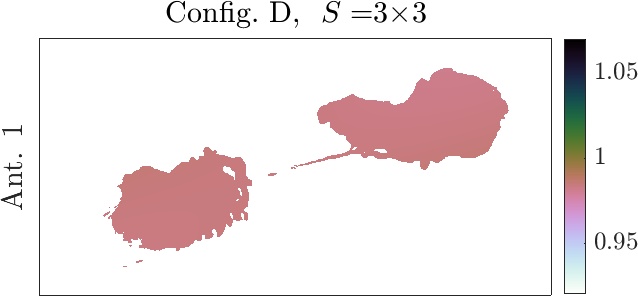}
\includegraphics[width=0.31\linewidth]{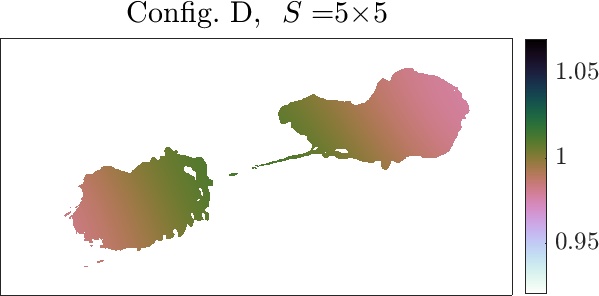}
\includegraphics[width=0.31\linewidth]{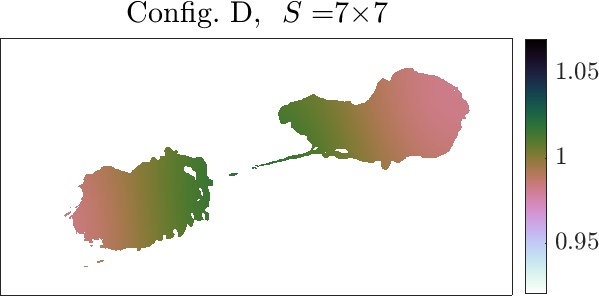}
\includegraphics[width=0.33\linewidth]{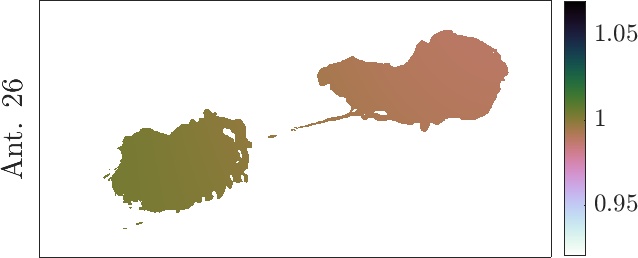}
\includegraphics[width=0.31\linewidth]{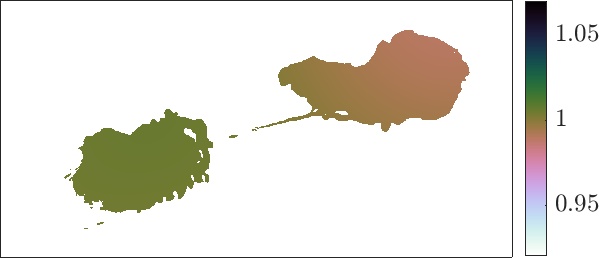}
\includegraphics[width=0.31\linewidth]{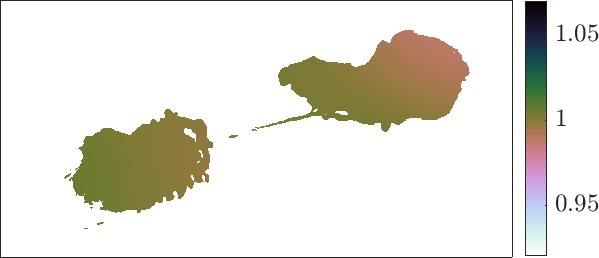}
\end{minipage}
\caption{C band: amplitudes of DDE solutions in the image domain obtained with the joint calibration and imaging approach ($\tau =8$). Estimated DDEs are displayed over a FoV of about $0.02275\degr \times 0.0455\degr$ and at pixel positions spanning six orders of magnitude of the recovered dynamic range in the image estimate. From left to right, results obtained for DDEs spatial Fourier dimension $S$ set to $3\times 3,~5\times 5,~ 7\times 7$, respectively. From top to bottom, DDE estimates of two selected antennas for each of the VLA configurations at {our chosen} time slot. Rows 1-2, configuration A, antennas 1 and 27.  Rows 3-4, configuration B, antennas 1 and 26. Rows 5-6, configuration C, antennas 1 and 26. Rows 7-8, configuration D, antennas 1 and 26. }
\label{fig:c_dde_solutions}
\end{figure*}

DDE estimates represented in the image space, obtained with the different spatial support sizes $S\in\{3\times 3,5\times 5,7\times7\}$ are displayed in Figure~\ref{fig:c_dde_solutions}. In this and subsequent figures, we choose to show the amplitudes of DDE realisations associated with two antennas at a particular time instance, in this case corresponding to the $10^{\textrm{th}}$ integration of each data set, that is about 90~seconds into the observation.
 Given the fact that DDEs contribution is modelled as multiplications with the radio map in the image domain (see the data model described in \eqref{eq:model-cont}), there is a high ambiguity in the obtained DDE solutions at pixel positions with zero amplitude. Therefore, DDE estimates are shown at the significant non-zero pixel positions of the associated estimated radio map. {{ DDEs estimates spanning the imaged FoV are displayed in  Figures~\ref{fig:c_dde_solutions_full1},~\ref{fig:c_dde_solutions_full2}.}} From a global perspective, one can notice the smoothness of the DDEs given their limited bandwidth in the spatial Fourier domain. One also observes the consistency of the amplitudes gradients with the different support sizes and for the different observations acquired with the four antenna configurations of the {{VLA}}, with subtle differences between support sizes $S=7\times 7$ and $S=5\times 5$, and more noticeable differences in comparison with the support size $S=3\times 3$. In fact, in the case of estimated DDEs with the spatial support size $S=3\times 3$, lower amplitudes of the recovered DDEs are noticed. Given the presence of some artefacts in the associated recovered map, this observation might suggest a some extent of over-fitting in the image domain in the case of calibration with DDEs of spatial support size $S=3\times 3$. Similar tendencies have been observed on different antennas at different time slots. 
\vspace{-0.5cm}
\subsubsection*{DDEs temporal smoothness}
In this experiment, spatial smoothness of the DDEs is imposed via the spatial support size $S=5\times 5$, as this setting has shown to yield high reconstruction quality. Temporal smoothness of the DDEs is investigated by varying the recovered temporal bandwidths of the DDEs associated with the configurations of the {{VLA}}, where the probed temporal reduction ratio $\tau$ is varied in the set $\{ 2^p\}_{0\leq p \leq7}$. The resulting temporal dimensions of the estimated compact DDE kernels are provided in Table~\ref{tab:dde_temporal_dim}. The remaining parameters involved in the joint DDE calibration and imaging approach are fixed similarly to the previous experiment.
\begin{table}
	\caption{C band: Temporal dimensions of the DDE solutions, associated with the configurations of the VLA, and obtained for the investigated temporal reduction ratio values. {{Note that, when the value of temporal reduction ratio results in a temporal dimension of the antenna gains larger than the number of the time slots, temporal smoothness is not considered. }} }	
\centering
	\begin{tabular}{lcccr} 
		\hline
		\textbf{Temporal dimension $P_c$} & A & B& C & D\\
		\hline
		 $\tau=1$ & 2248 & 765 & 627 & 230 \\ 
		$\tau=2$ & 1291 & 765 & 627 & 230\\
 $\tau=4$ & 645 & 607 & 627 & 230\\
 $\tau=8$ & 321 & 303 & 439 & 230\\
 $\tau=16$ & 159 & 151 & 219 & 117\\
 $\tau=32$ & 79 & 75 & 109 & 57\\
 $\tau=64$ & 39 & 37 & 53 & 27\\
  $\tau=128$ & 19 & 17 & 25 & 13\\

		\hline
	\end{tabular}	
	\label{tab:dde_temporal_dim}	
\end{table}

\begin{landscape}
\begin{figure}
\centering
\includegraphics[scale=0.17]{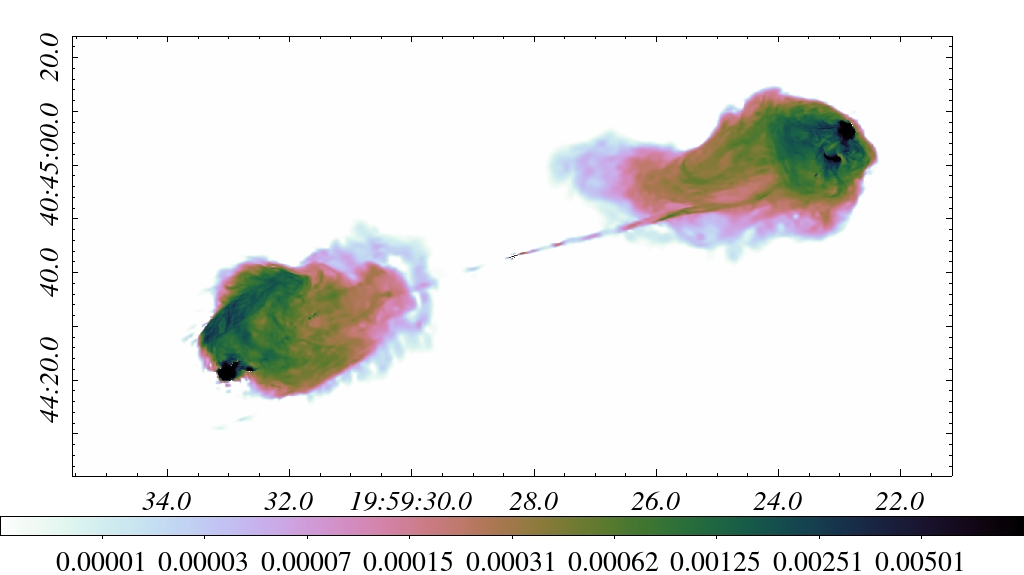}
\includegraphics[scale=0.17]{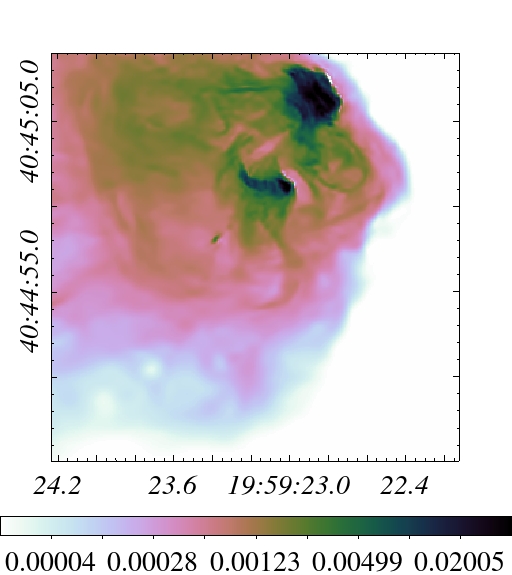}
\includegraphics[scale=0.17]{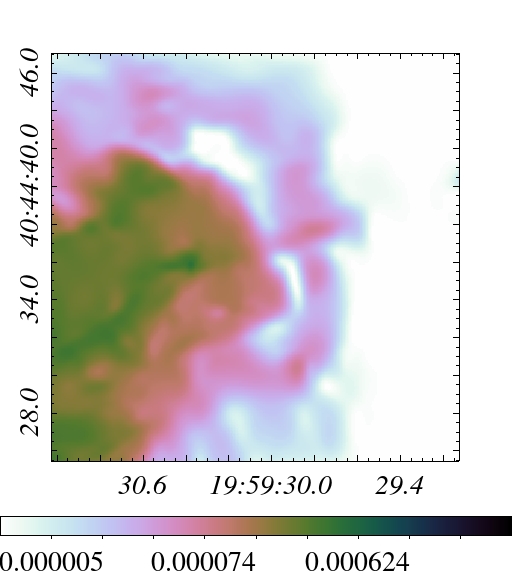}
\includegraphics[scale=0.17]{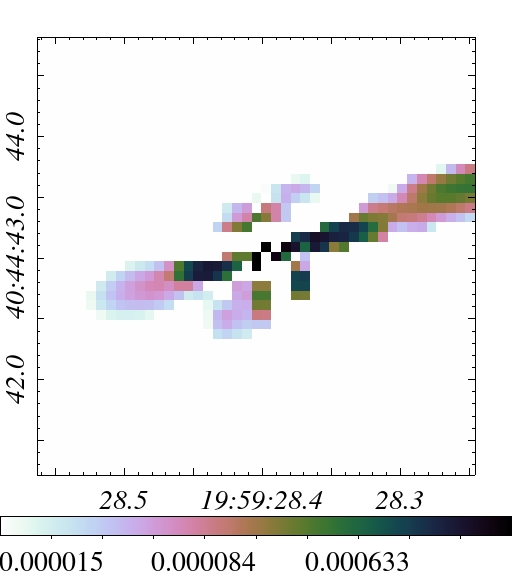}
\includegraphics[scale=0.17]{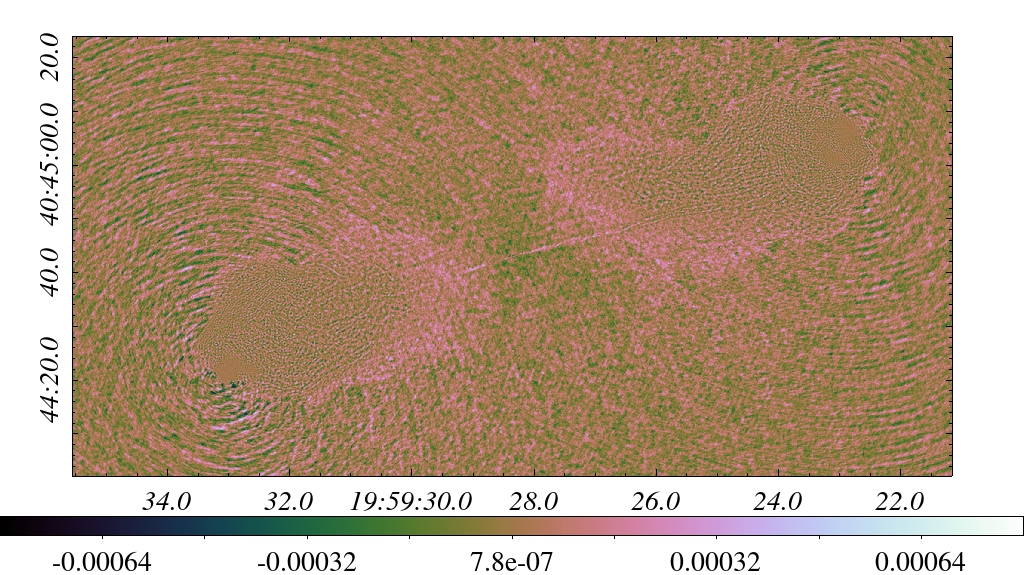}
\includegraphics[scale=0.17]{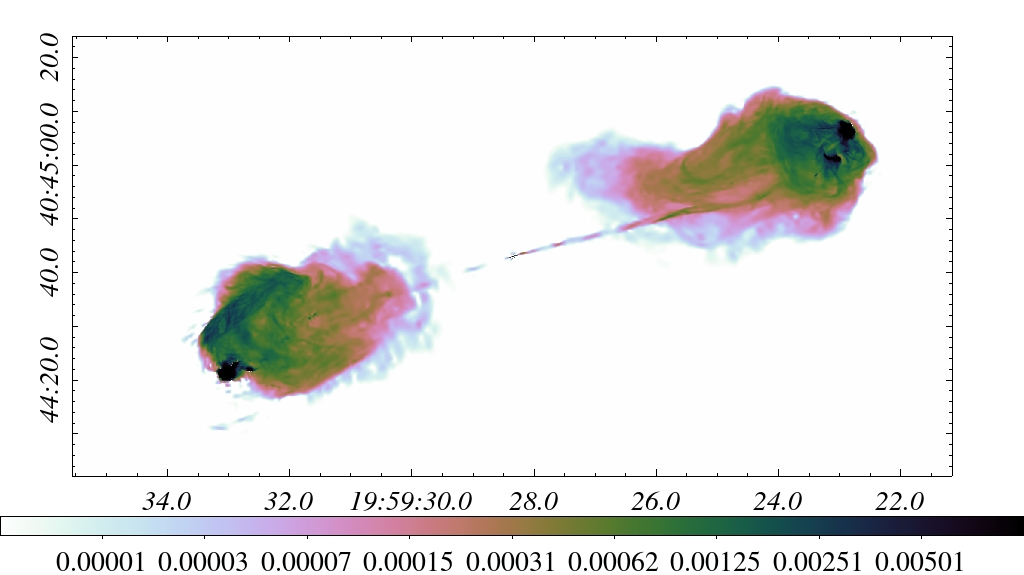}
\includegraphics[scale=0.17]{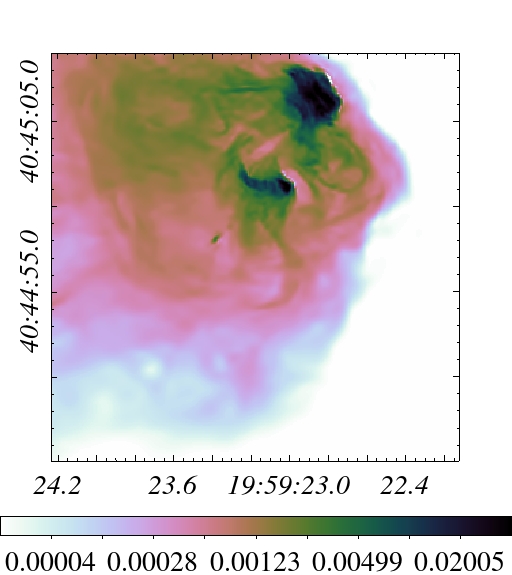}
\includegraphics[scale=0.17]{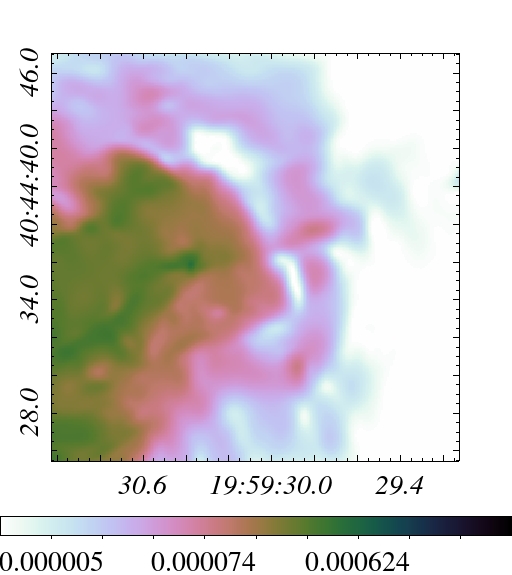}
\includegraphics[scale=0.17]{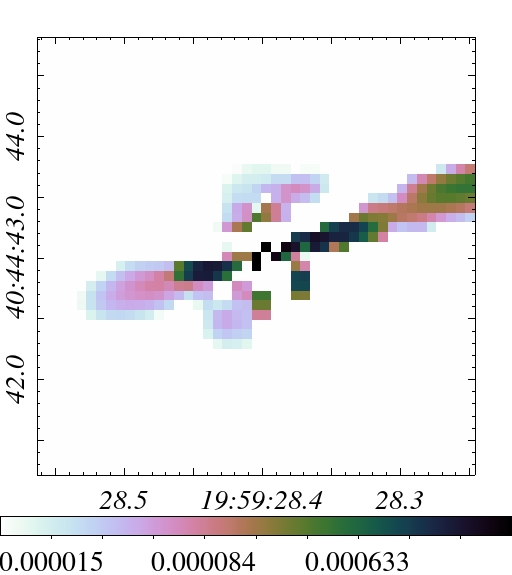}
\includegraphics[scale=0.17]{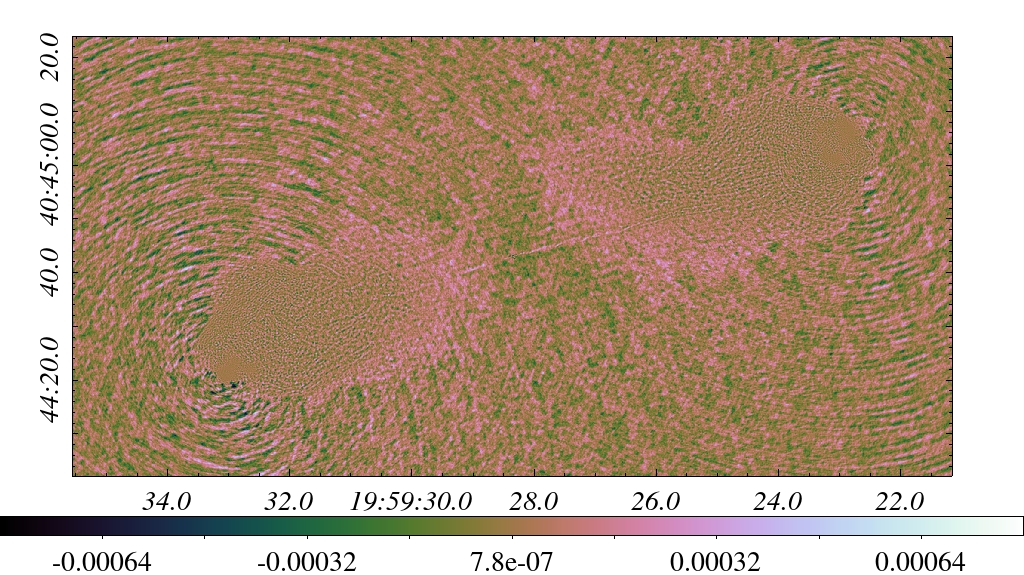}
\includegraphics[scale=0.17]{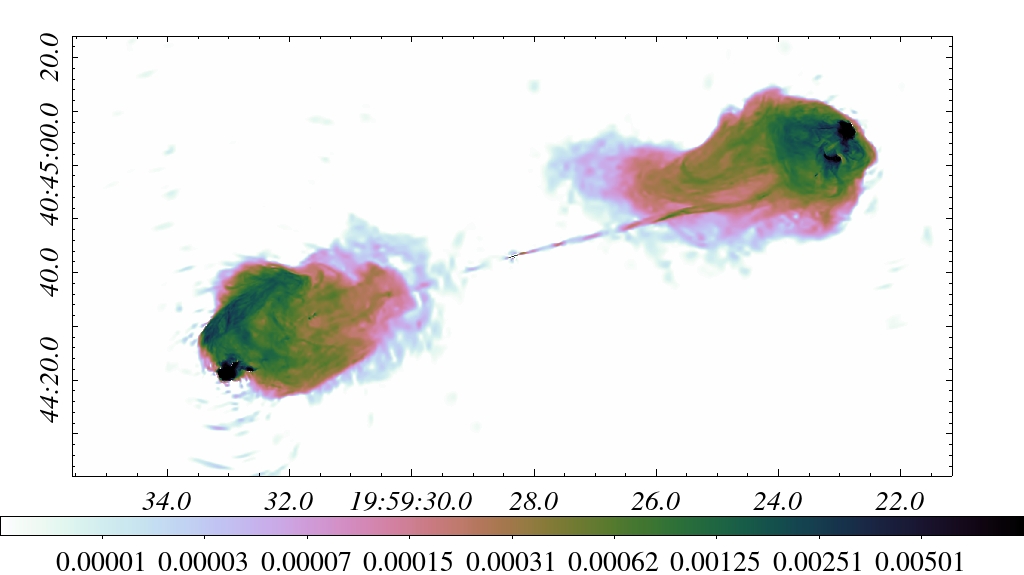}
\includegraphics[scale=0.17]{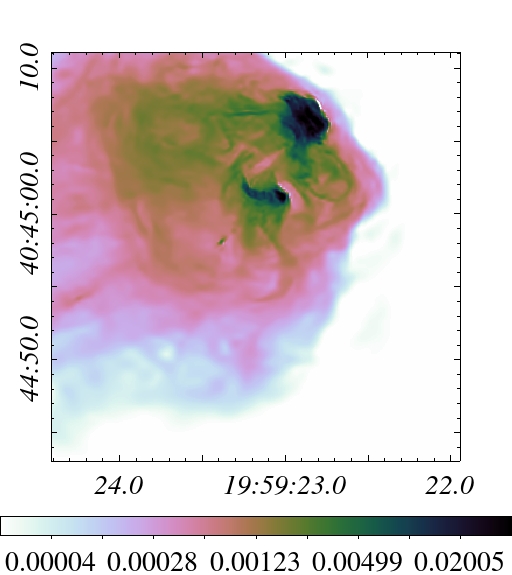}
\includegraphics[scale=0.17]{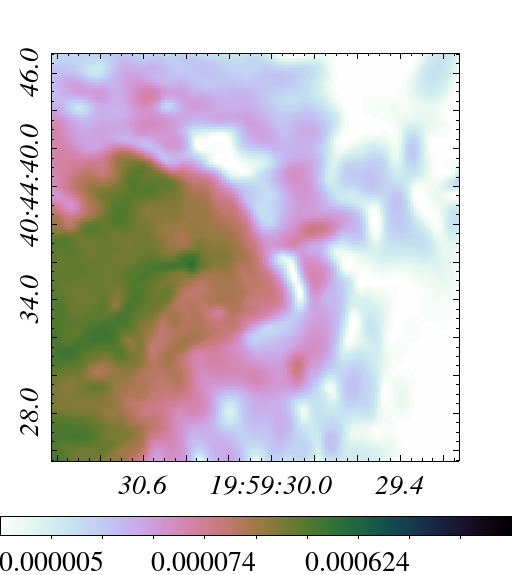}
\includegraphics[scale=0.17]{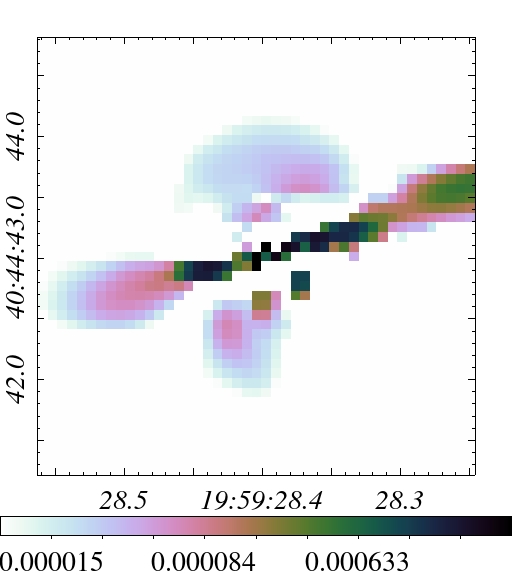}
\includegraphics[scale=0.17]{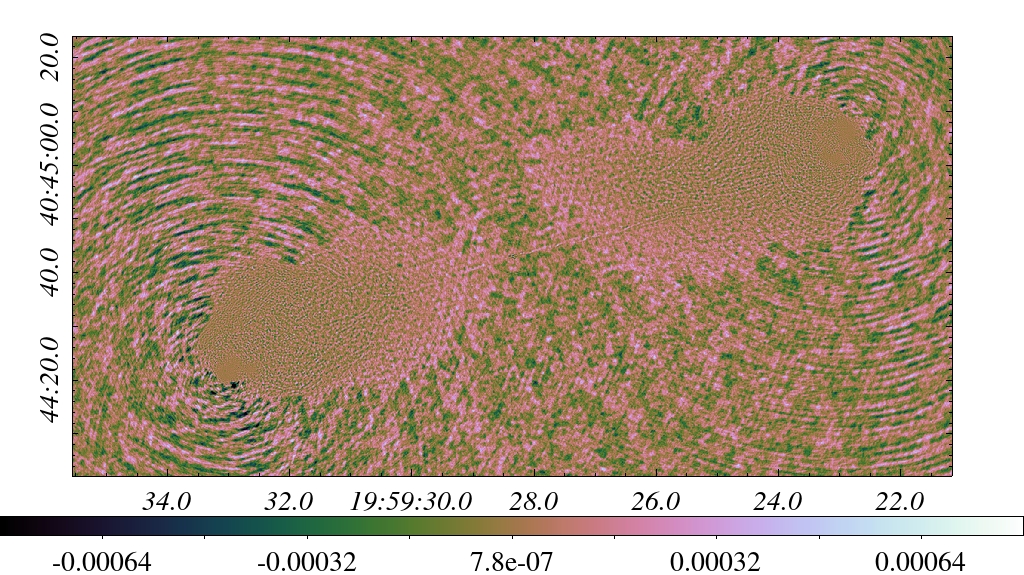}
\includegraphics[scale=0.17]{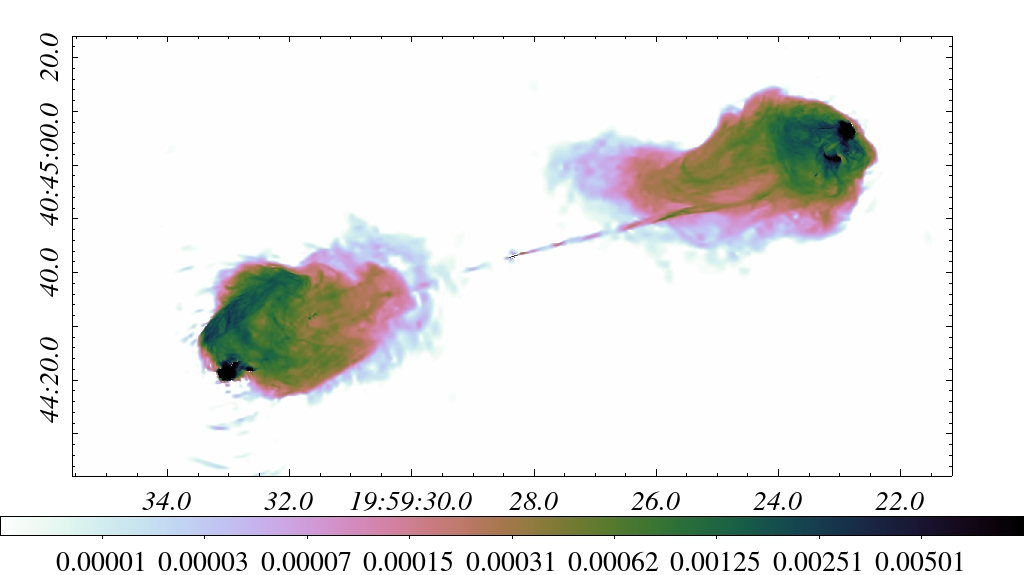}
\includegraphics[scale=0.17]{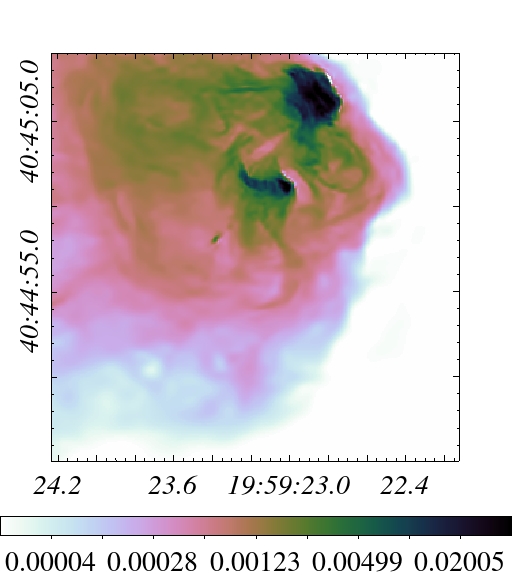}
\includegraphics[scale=0.17]{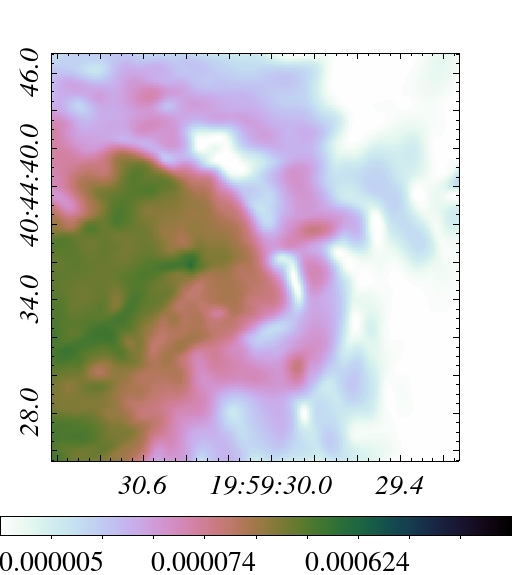}
\includegraphics[scale=0.17]{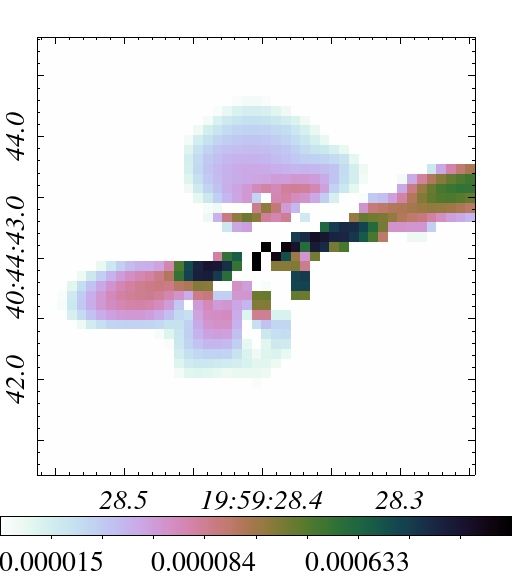}
\includegraphics[scale=0.17]{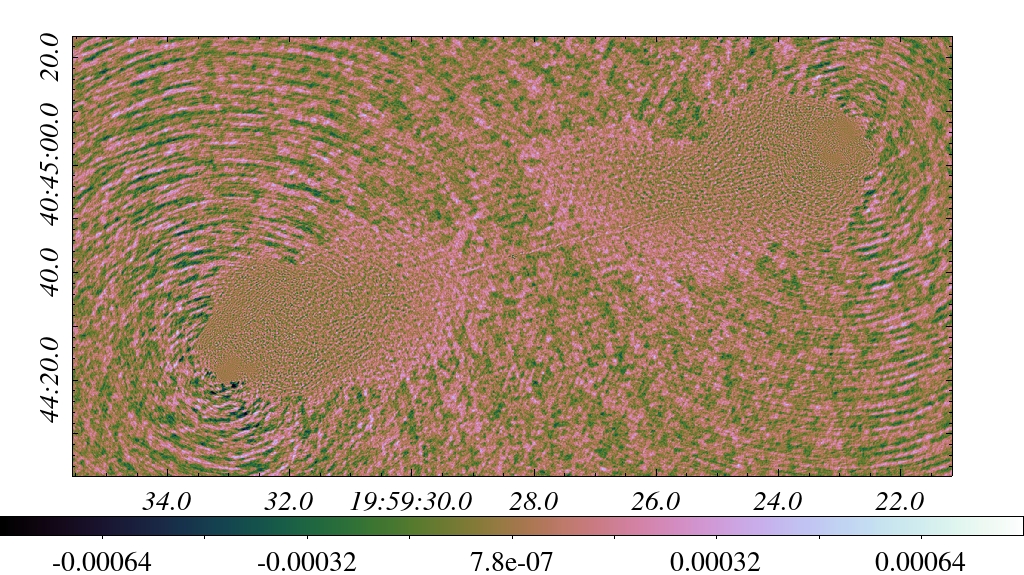}
\includegraphics[scale=0.17]{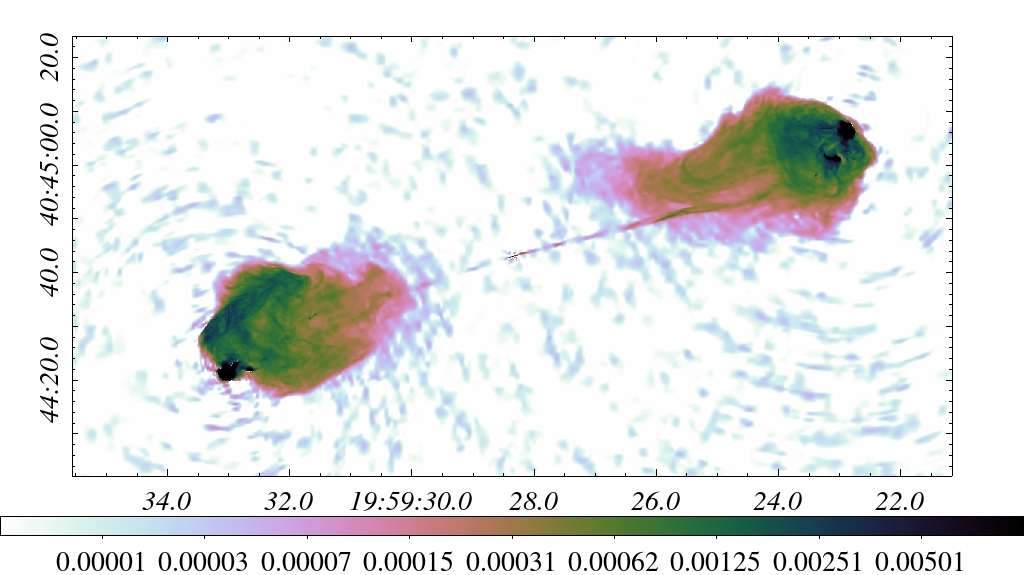}
\includegraphics[scale=0.17]{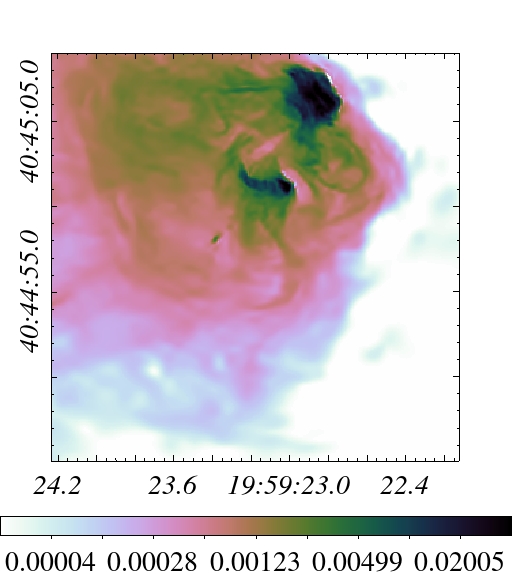}
\includegraphics[scale=0.17]{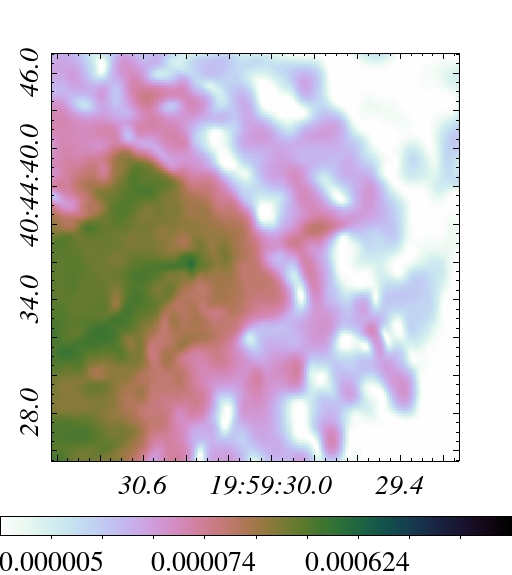}
\includegraphics[scale=0.17]{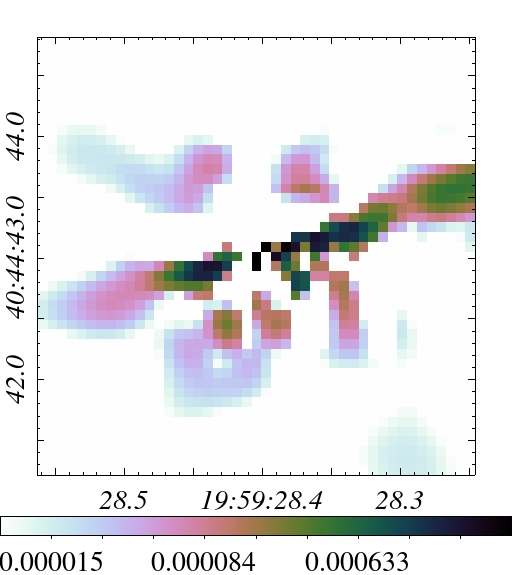}
\includegraphics[scale=0.17]{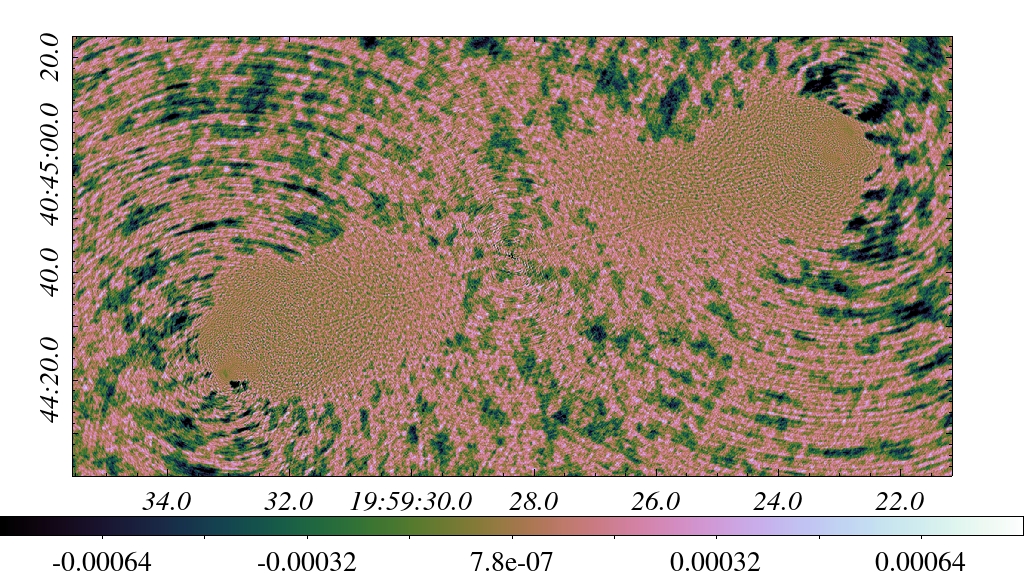}
%

\caption{C band: From top to bottom, reconstructed maps via joint DDE calibration and imaging, associated with the respective spatial dimensions and temporal reduction ratios; $(S=7\times 7,\tau=8)$, $(S=5\times 5,\tau =8)$ ,~$(S= 5\times 5,\tau =1)$, $(S=3\times 3,\tau =8)$ and ($S=1\times1,~\tau =8$). Recall that  cases $S=1\times1$ and $\tau =1$ correspond to DIE calibration and absence of temporal smoothness, respectively. From left to right, estimated model images ($\log_{10}$ scale), {{displayed over a FoV of about $0.02275\degr \times 0.0455\degr$}}, zooms on selected regions of the east and west jets and the inner core of the galaxy ($\log_{10}$ scale), and the residual images (linear scale). The surface brightness of the estimated model images is in Jy/pixel with a pixel size set to $0.08\arcsec$.}
\label{fig:c_maps}

\end{figure}
\end{landscape}

To evaluate the impact of DDEs' temporal variation, statistics of the residual data and residual images associated with the different temporal reduction ratios are provided in Table~\ref{tab:C_vis_time} as well as the SNR values in both data and image domains. 
On the one hand, one notices the sub-optimal fidelity obtained in the image domain in the absence of temporal smoothness ($\tau=1$), despite the relatively high fit in the visibility domain. On the other hand, temporal smoothness resulting from reduction ratios $4\leq \tau \leq 8$ yields important fit-to-data in both domains. The numbers are confirmed via the visual inspection of the reconstructed residual images obtained by enforcing temporal smoothness with reduction ratio $\tau=8$ when compared to DDE estimation with no temporal prior (see Figure~\ref{fig:c_maps}, second and third rows, fifth column). Furthermore, the reconstructed model image of the latter exhibits more ringing artefacts around the hotspots (see first column, second and third rows of the same figure). These findings suggest the suitability of the chosen settings for the temporal variation of the DDEs. Strong temporal smoothness obtained via a temporal reduction ratio $\tau\geq 16$ yields limited fidelity in both visibility and image domains, reflected in the obtained SNR values. In this regime, one notices that the higher the temporal reduction ratio is, the higher is the excess kurtosis value, suggesting that the associated residual visibilities depart from a Gaussian distribution. Such behaviour can be explained by (i) the long total observation times, spanning several hours (see Table~\ref{tab:timeObs}) and (ii) the large gaps in the temporal sampling of the data, in particular during acquisitions with {{VLA}} configurations C and D.
\begin{table}
    \caption{C band: Statistics of the residual data and image for the different temporal bandwidths of the DDEs ($S =5 \times 5$).}
   \resizebox{0.5\textwidth}{!}{         \centering
    \begin{tabular}{lllll}
	\hline
       \textbf{Residual data} 
       & STD & Skewness  & Kurtosis & vSNR{~(dB)} \\
       \hline
     $\tau =1$ & 0.88   & -0.011  & 4.14 & 43.57 \\
     $\tau =2$ & 0.85   & -0.018  & 3.90 & 43.81\\
     $\tau =4$ & 0.87   & -0.008  & 3.92  & 43.70\\
     $\tau =8$ & 0.90   & 0.003 & 4.13  & 43.40\\
     $\tau =16$ & 0.94  & 0.012  & 4.60 & 42.97\\
     $\tau =32$ & 0.99  & 0.022  & 5.37 & 42.51\\
     $\tau =64$ & 1.05  & 0.035  & 6.31 & 42.00\\
     $\tau =128$ & 1.14 & 0.055  & 8.14 & 41.28\\
	\hline
      \textbf{Residual Image}& STD~{($\times10^{-4}$)} & Skewness & Kurtosis  & iSNR{~(dB)}\\
	\hline
$\tau =1$   &  1.38 &  -0.088 & 3.31 & 54.36  \\
$\tau =2$   &  1.13 &  -0.008 & 3.21 & 55.98   \\
$\tau =4$   &  1.08 &   0.006 & 3.24 & 56.48  \\
$\tau =8$   &  1.08 &   0.002 & 3.27 & 56.44  \\
$\tau =16$  &  1.13 &  -0.019 & 3.32 & 56.05  \\
$\tau =32$  &  1.21 &  -0.041 & 3.37 & 55.45  \\
$\tau =64$  &  1.32 &  -0.083 & 3.34 & 54.71  \\
$\tau =128$ &  1.49 &  -0.132 & 3.16 &  53.67  \\
         \hline
    \end{tabular}
   } 
     \label{tab:C_vis_time} 
\end{table}

\begin{figure*}
\centering
\includegraphics[width=0.98\linewidth]{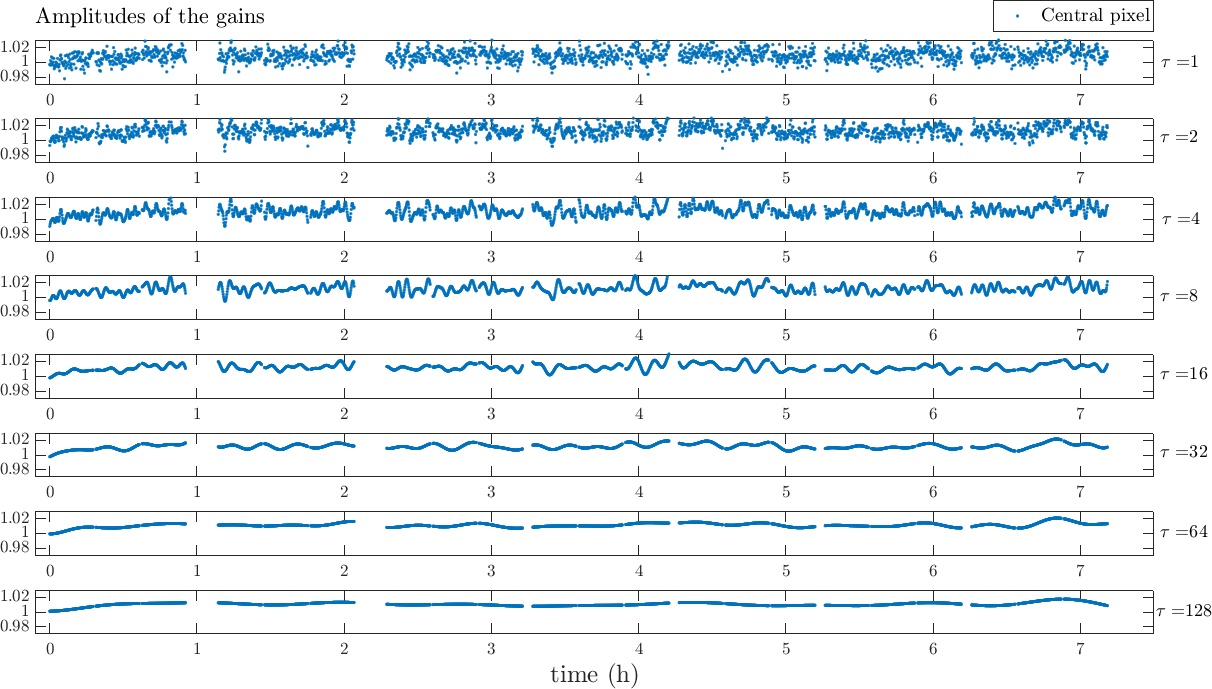}
\includegraphics[width=0.98\linewidth]{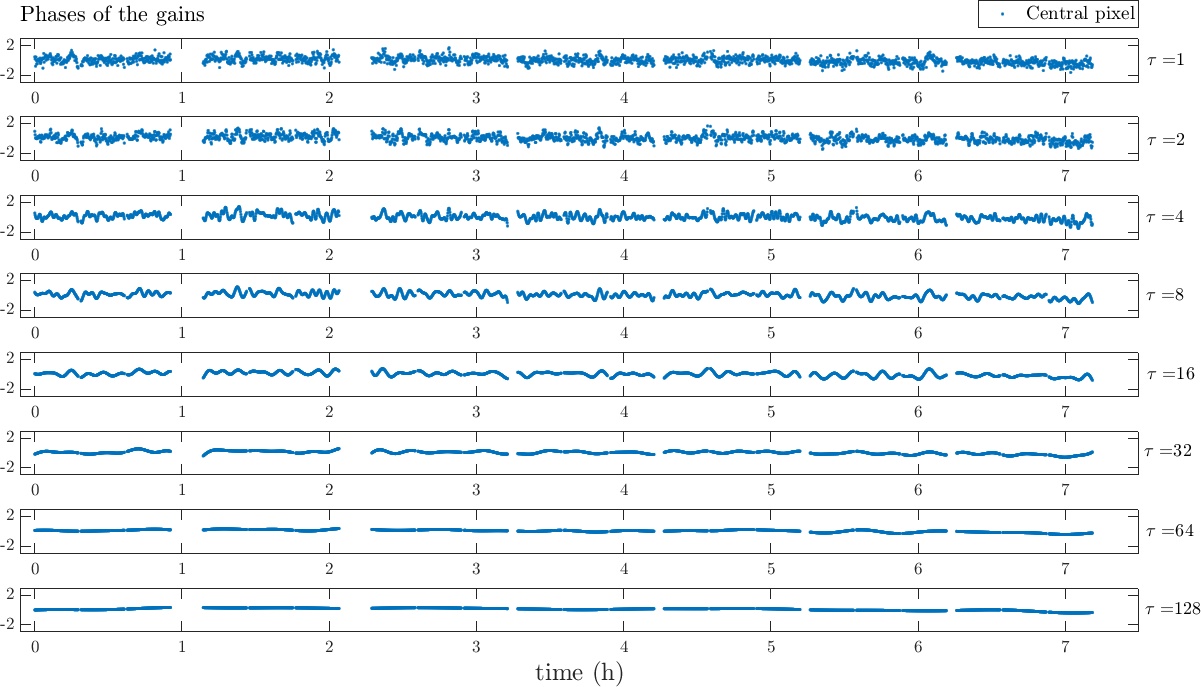}
\caption{{C band: variation in time of the DDE solutions associated with Antenna 1 of VLA configuration A at the central pixel position, which corresponds to the main black hole. Various temporal bandwidths of the DDEs are considered, such that $\tau\in\{2^p\}_{0\leq p \leq 7}$. From top to bottom, amplitudes and phases (in degrees) of DDE solutions in their original temporal and image space are displayed.}}
\label{fig:c_ddes_time_1}
\end{figure*}

\begin{figure*}
\centering
\includegraphics[width=0.98\linewidth]{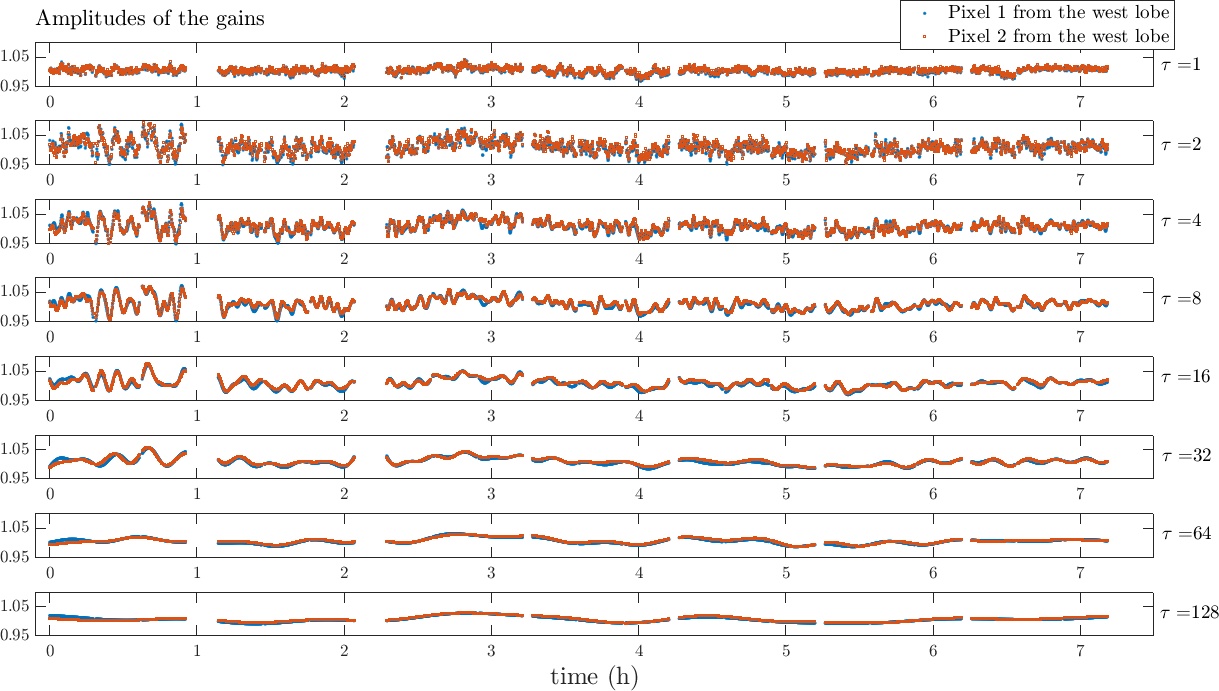}
\includegraphics[width=0.98\linewidth]{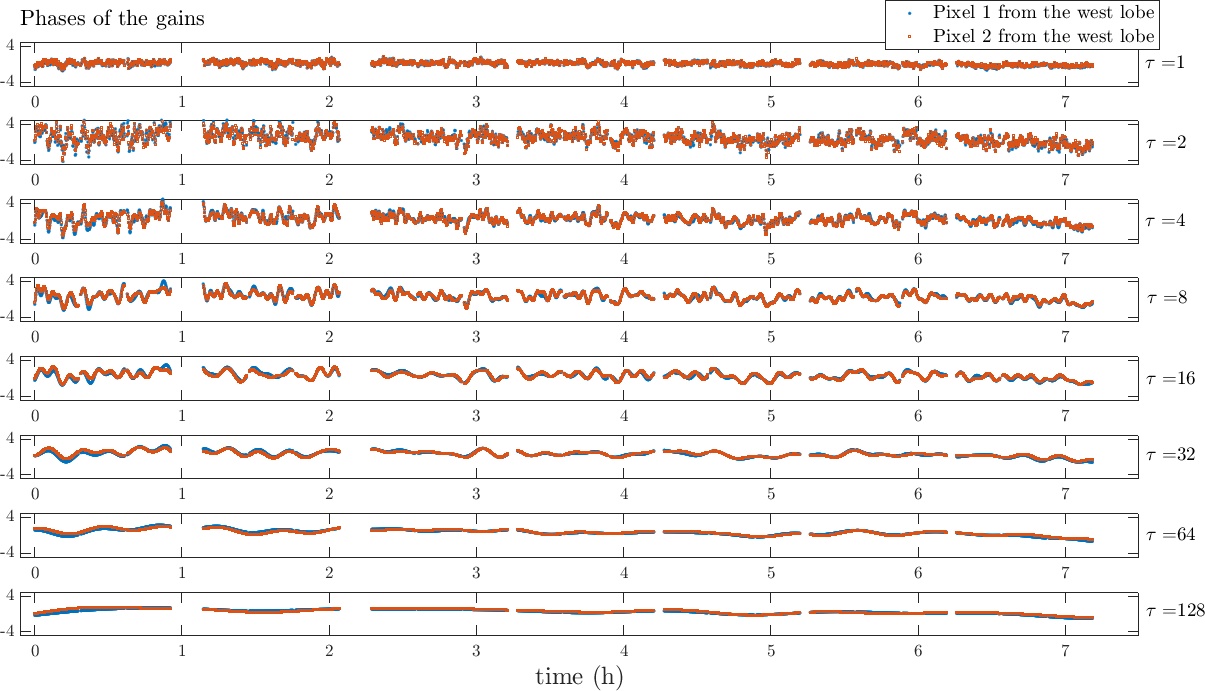}
\caption{{C band: variation in time of the DDE solutions associated with Antenna 1 of VLA configuration A at two pixel positions from the west lobe ($14\arcsec$ apart). Various temporal bandwidths of the DDEs are considered, such that $\tau\in\{2^p\}_{0\leq p \leq 7}$. From top to bottom, amplitudes and phases (in degrees) of DDE solutions in their original temporal and image space are displayed.}}
\label{fig:c_ddes_time_2}
\end{figure*}

\begin{figure*}
\centering
\includegraphics[width=0.98\linewidth]{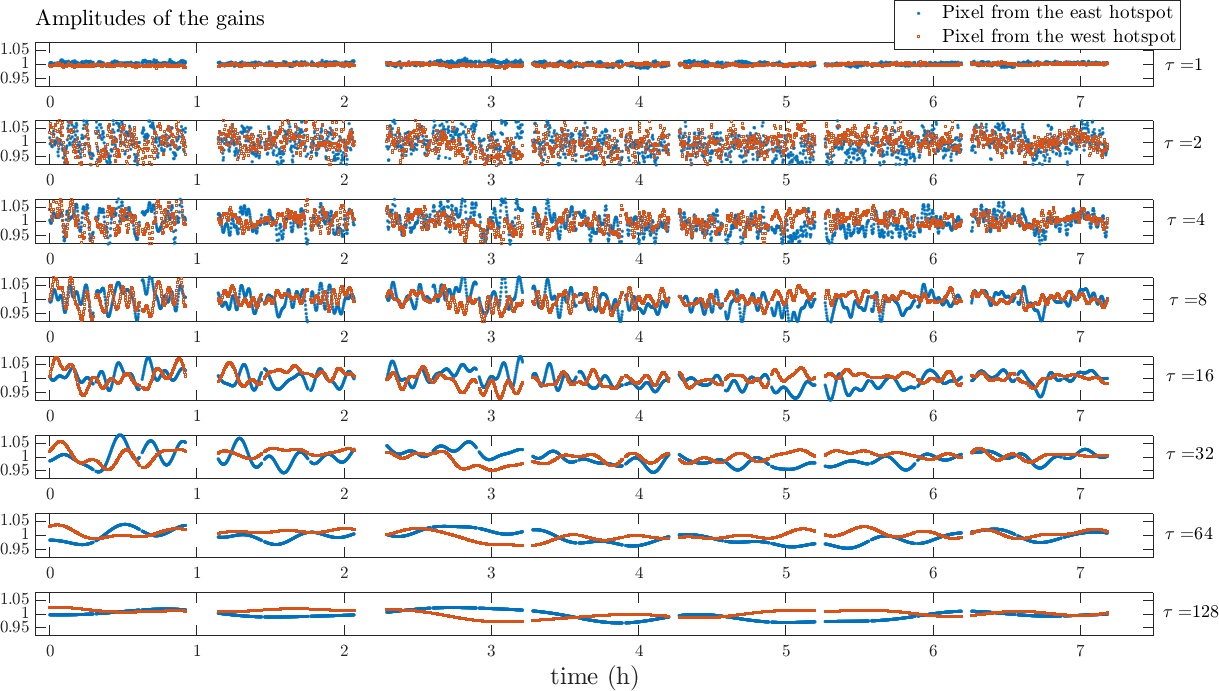}
\includegraphics[width=0.98\linewidth]{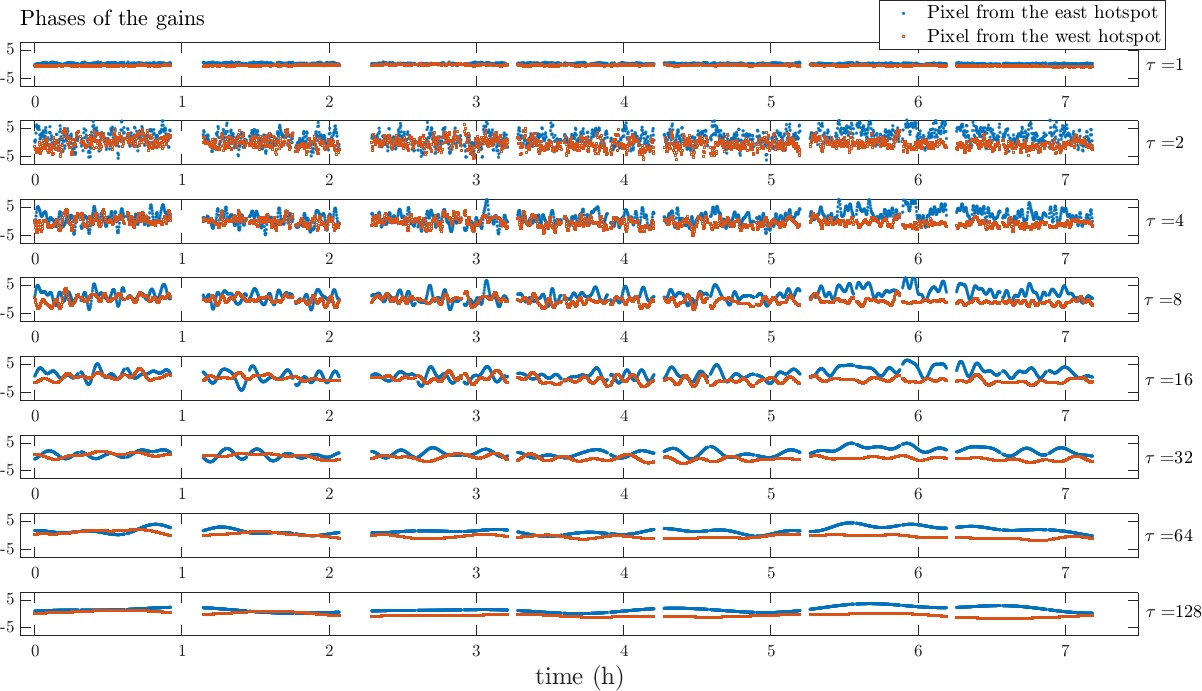}
\caption{{C band: variation in time of the DDE solutions associated with Antenna 1 of VLA configuration A at two pixel positions from the hotspots. Various temporal bandwidths of the DDEs are considered, such that $\tau\in\{2^p\}_{0\leq p \leq 7}$. From top to bottom, amplitudes and phases (in degrees) of DDE solutions in their original temporal and image space are displayed.}}
\label{fig:c_ddes_time_3}
\end{figure*}

To analyse the temporal variation of the estimated DDEs with respect to the investigated temporal bandwidths, we display the evolution in time of the DDEs associated with Antenna 1 at {{VLA}} configuration A in Figures~\ref{fig:c_ddes_time_1}-\ref{fig:c_ddes_time_3}. The visualised antenna gains correspond to selected directions from Cyg~A. These are (i) the central pixel of the estimated radio map, corresponding to the main black hole, (ii) two adjacent pixels from the west lobe that are $14\arcsec$ apart, and (iii) two pixels from the west and east hotspots, that are 130$\arcsec$ apart.
Generally, one can clearly notice that the lower the effective temporal bandwidth is, the smoother the gains are. In fact, gains whose temporal bandwidth is given by a reduction ratio $\tau \geq 32$ are nearly flat.

From the gain solutions at the core of the galaxy (Figure~\ref{fig:c_ddes_time_1}), one can see that both amplitude and phase are relatively small. This is indeed expected as careful self-calibration has been performed as a pre-processing step, where large errors have been already corrected for. Note that when the temporal prior is not considered, the temporal variation of the gains is highly irregular as opposed to the case of temporal bandwidth associated with the reduction ratio $\tau =8$, that is chosen as the best compromise between temporal smoothness and fidelity to data (see Table~\ref{tab:C_vis_time}). When inspecting the gains at adjacent directions (Figure~\ref{fig:c_ddes_time_3}), one can observe their high correlation. Indeed, atmospheric and instrumental direction-dependent effects are not expected to vary at such small spatial scales. In comparison, antenna gains at pixel positions widely separate such as the ones located at the two hotspots (Figure~\ref{fig:c_ddes_time_2}) vary in time more significantly. The differences at these two directions are likely compatible with effects expected from the atmosphere and antenna pointing errors.


\subsection{X band: the impact of SARA prior }
X band data are imaged at a spatial resolution that is about 1.25 times the nominal resolution of the observations, which corresponds to a pixel size $\delta x=0.08\arcsec$. The mapped sky of interest is of size $N=2048\times 2048$, that is a FoV $\Omega= 0.0455\degr \times 0.0455\degr $. Temporal specifications of the observations at X band are similar to those at C band. In light of the results of the previous section, the temporal dimensions of the antenna gains, associated with the different {{VLA}} configurations, are set such {{that}} the temporal reduction ratio is $\tau=8$. The resulting dimensions of the estimated kernels in the temporal Fourier domain are $321$, for both sets of configuration A and $311$ and $439$ for configurations B and C, respectively. No temporal smoothness is imposed for configuration D. Furthermore, smoothness of the antenna gains is enforced via the dimension $S =5 \times 5$ in the spatial Fourier domain. For both DIE and DDE calibration experiments, sparsity regularisation parameter of the problem \eqref{eq:model-prior} is set to $\check{\eta} =5 \times 10^{-7}$. The remaining parameters involved in Algorithm~\ref{algo:VMBCFB} are fixed as described in Section ~\ref{ssec:settings}. 

\begin{landscape}
\begin{figure}
\centering
\includegraphics[scale=0.1855]{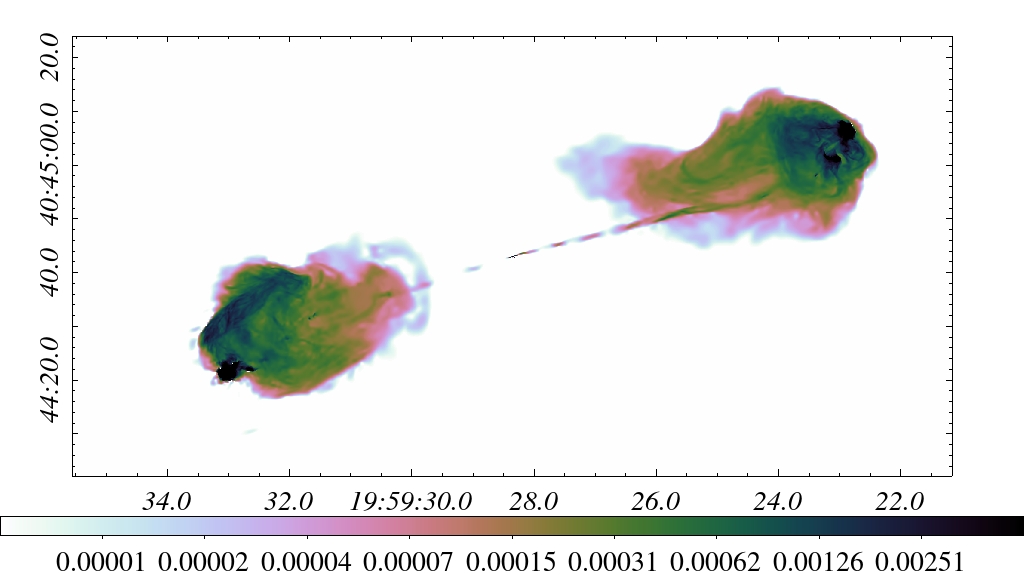}
\includegraphics[scale=0.185]{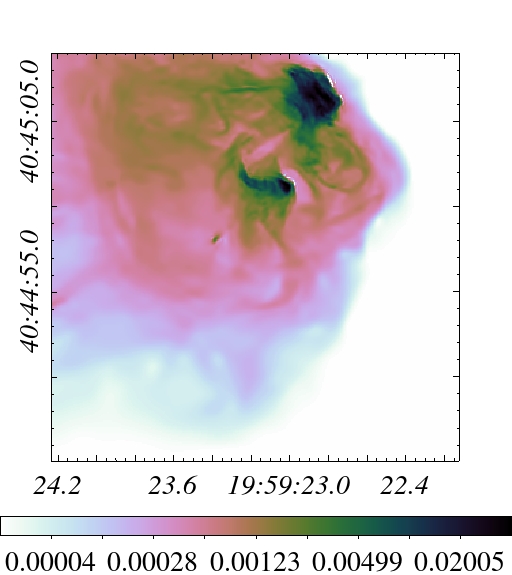}
\includegraphics[scale=0.185]{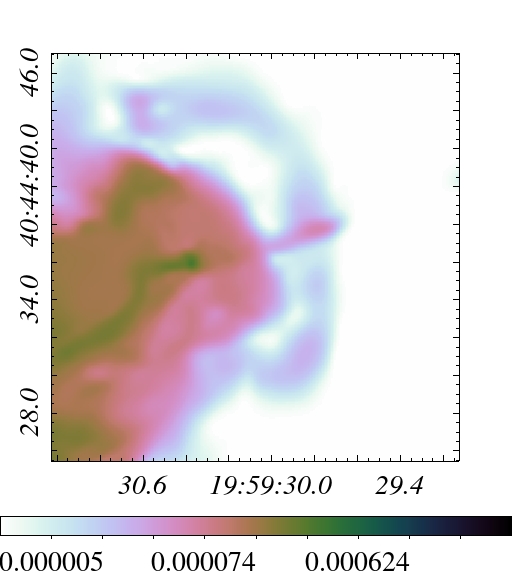}
\includegraphics[scale=0.185]{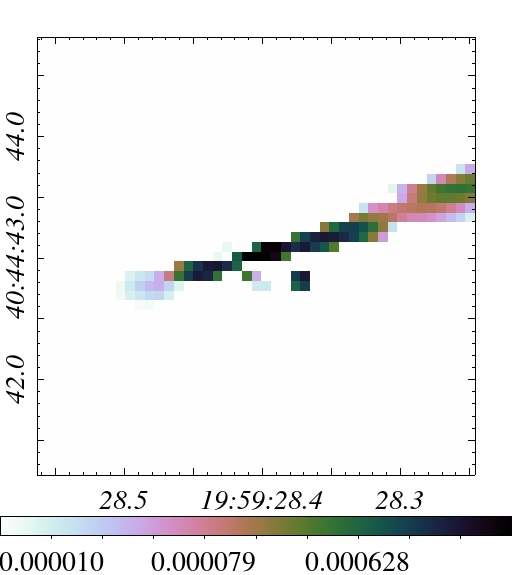}
\includegraphics[scale=0.185]{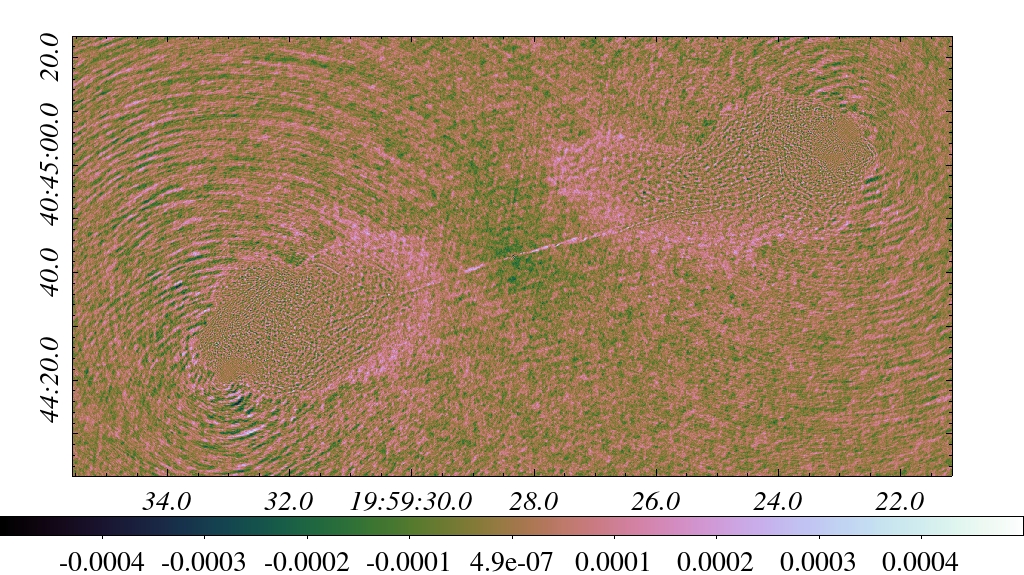}
\includegraphics[scale=0.185]{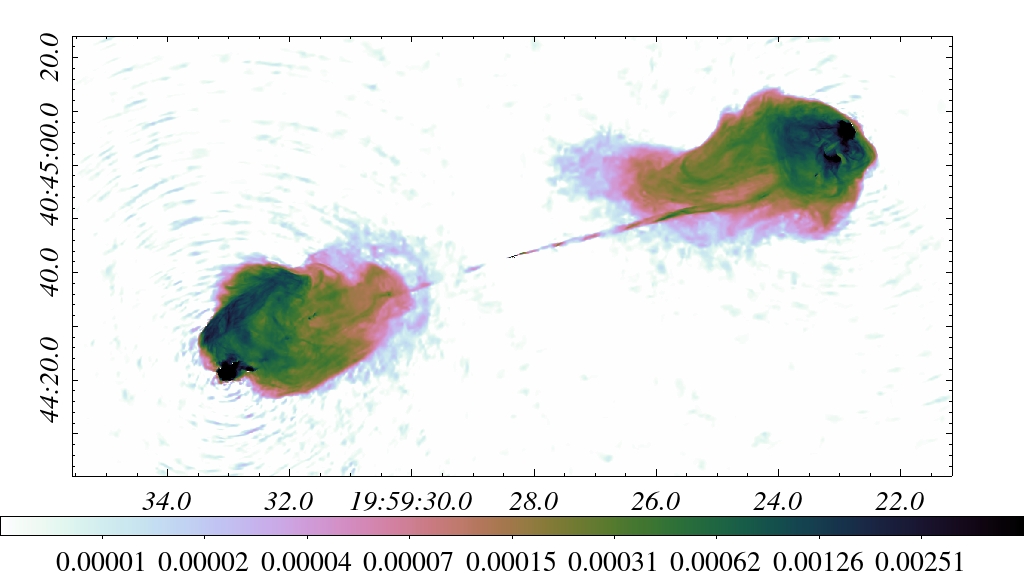}
\includegraphics[scale=0.185]{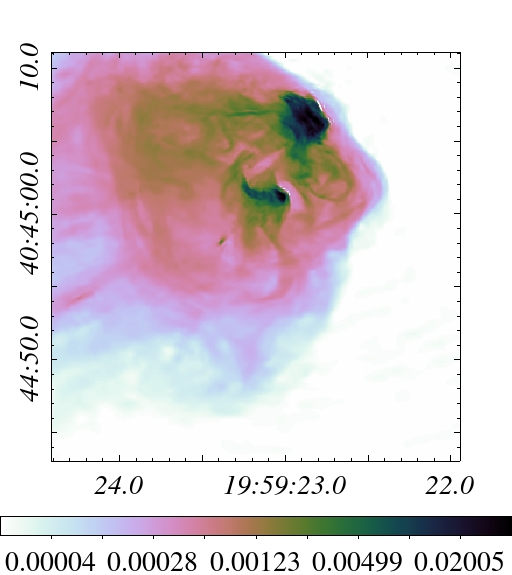}
\includegraphics[scale=0.185]{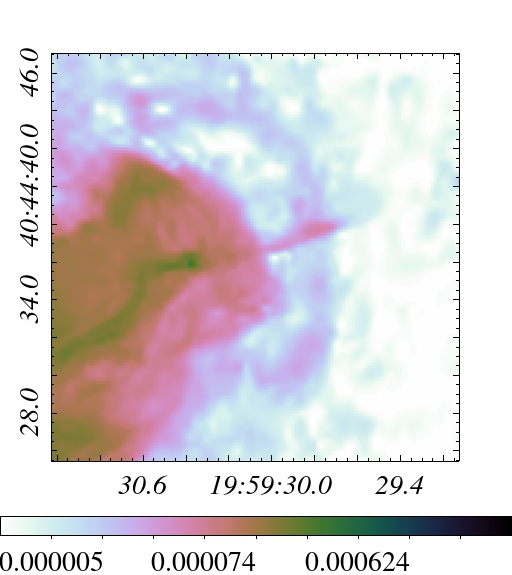}
\includegraphics[scale=0.185]{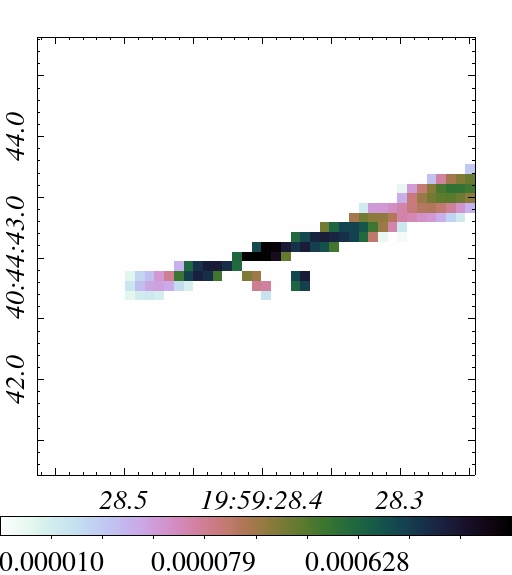}
\includegraphics[scale=0.185]{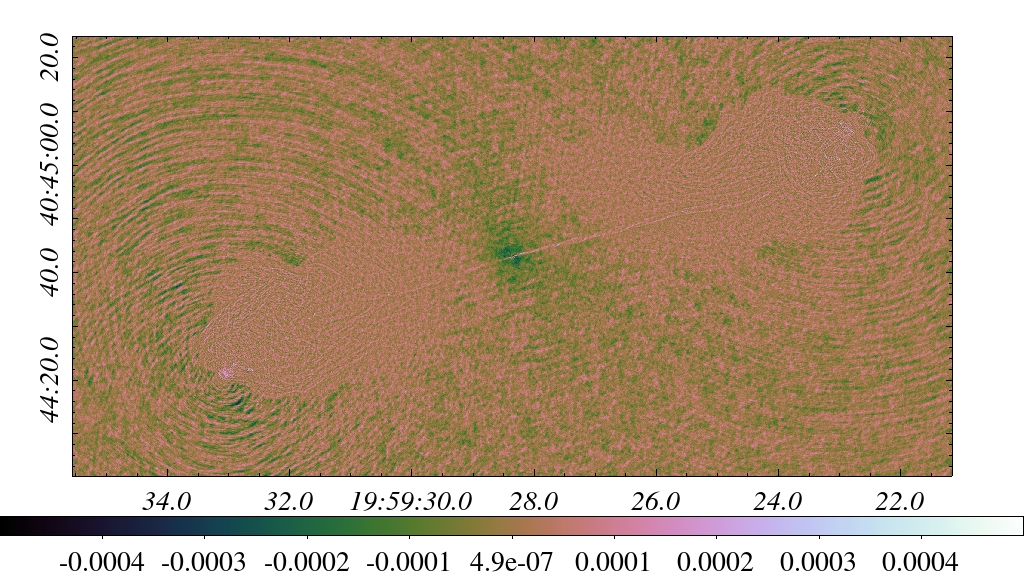}
\includegraphics[scale=0.185]{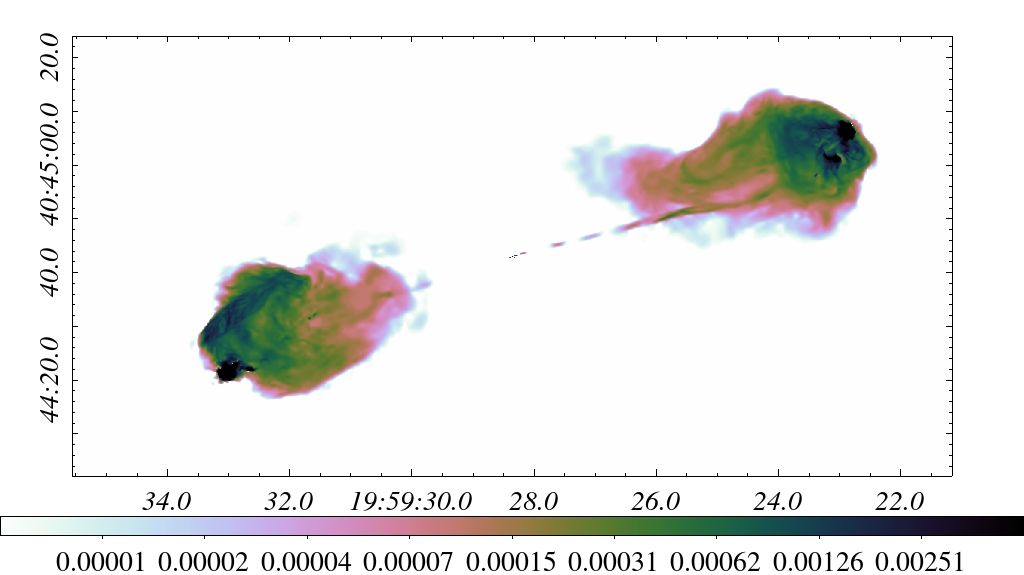}
\includegraphics[scale=0.185]{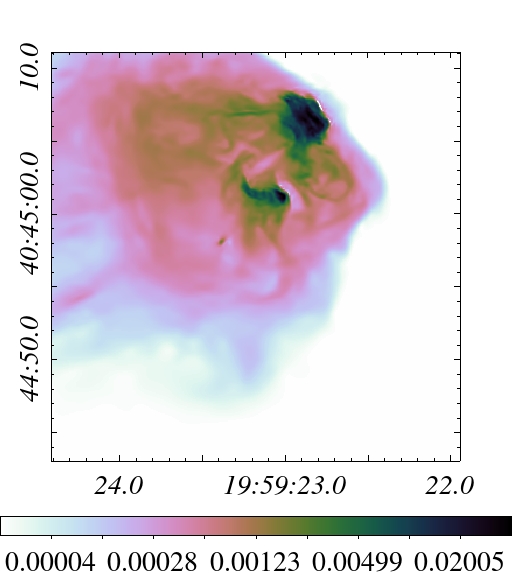}
\includegraphics[scale=0.185]{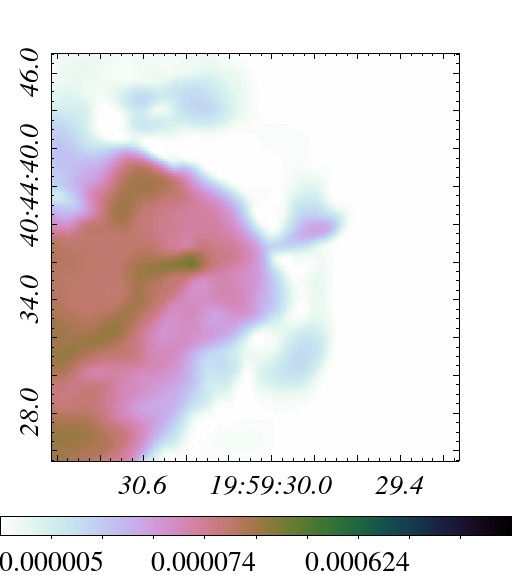}
\includegraphics[scale=0.185]{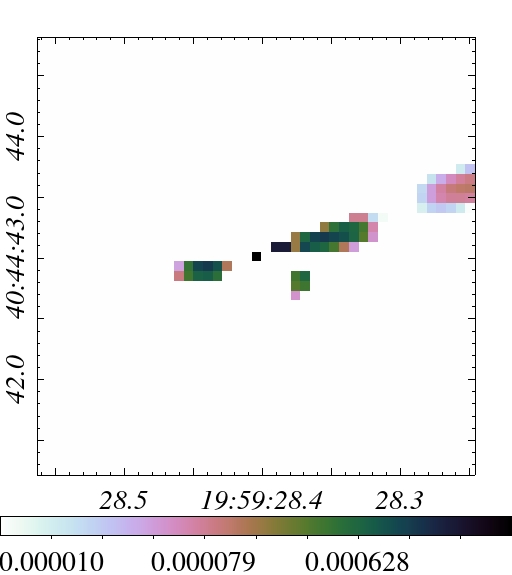}
\includegraphics[scale=0.185]{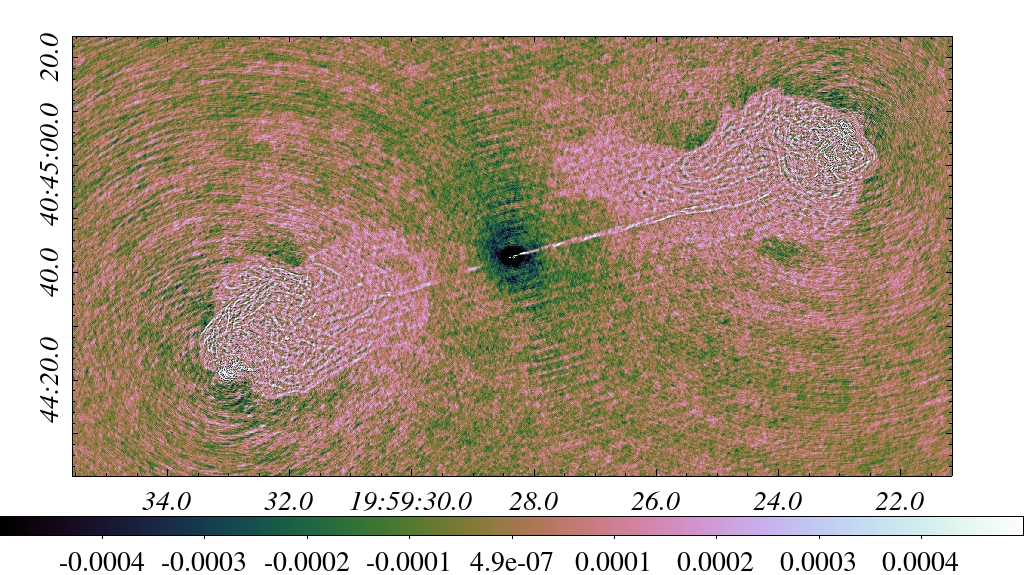}
\includegraphics[scale=0.185]{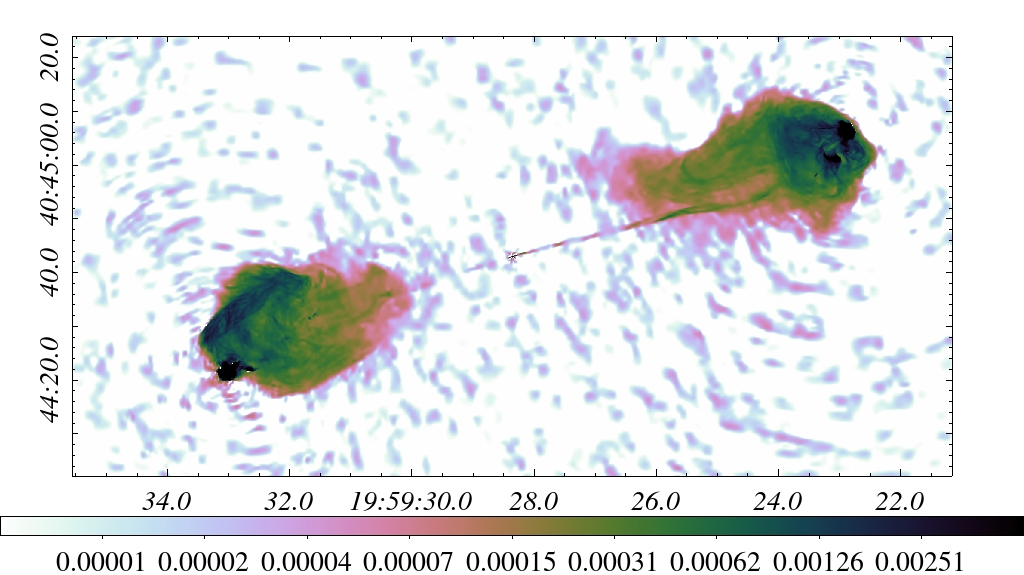}
\includegraphics[scale=0.185]{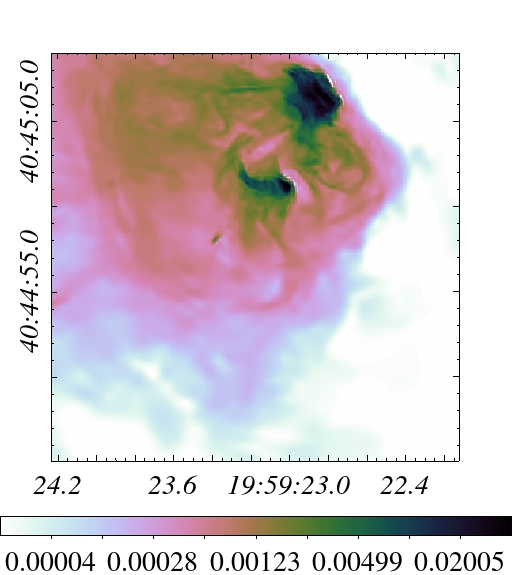}
\includegraphics[scale=0.185]{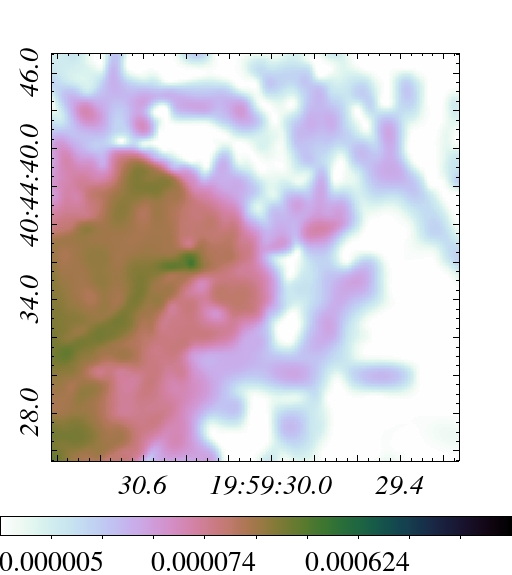}
\includegraphics[scale=0.185]{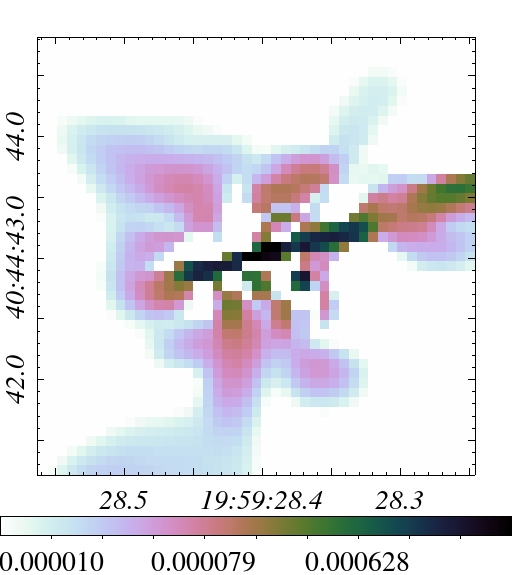}
\includegraphics[scale=0.185]{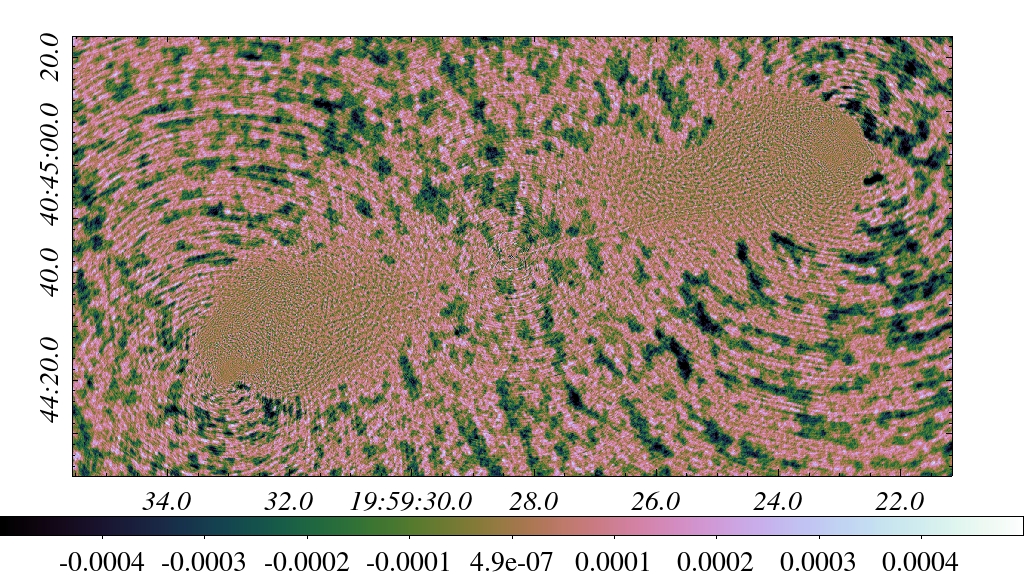}

\caption{X band: Rows 1-3, joint calibration and imaging results for DDE calibration ($S=5\times 5,~\tau=8$), adopting SARA image prior,  $\ell_1$-sparsity with $\check{\eta} =10^{-7}$, $\ell_1$-sparsity with $\check{\eta} =5\times 10^{-7}$, respectively. Bottom row, DIE calibration  ($S=1\times 1,~\tau=8$) adopting SARA image prior. From left to right,  estimated model images ($\log_{10}$ scale), {{displayed over a FoV of about $0.02275\degr \times 0.0455\degr$}}, zooms on selected regions of the east and west jets and the inner core of the galaxy ($\log_{10}$ scale), respectively, and the residual images (linear scale). The surface brightness of the estimated model images is in Jy/pixel with a pixel size is set to $0.08\arcsec$.}
\label{fig:x_maps_dde}

\end{figure}
\end{landscape}

We study the performance of our approach in the context of DDE and DIE calibration in comparison with {Adaptive~PPD} imaging~\citep{Dabbech2018}. Reconstructed images are displayed in Figure~\ref{fig:x_maps_dde}.
The superior performance of DDE calibration exhibited through the high sensitivity and high resolution recovered map (top row) in comparison with DIE calibration (bottom row). Note that, once again, the recovered maps via Adaptive PPD are not displayed as these can be found in \citet{Dabbech2018}. Yet, we emphasise on Adaptive PPD comparable reconstruction quality to joint DIE calibration and imaging experiment. Inspection of the inner core of the radio galaxy (Figure~\ref{fig:x_maps_dde}, fourth column) shows the radio transient, well resolved, south east from the main black hole. DDE calibration {{enables the recovery of}} the source with a flux of about 5.6~mJy against 5.3~mJy and 4.9~mJy via DIE calibration and {Adaptive~PPD} imaging, respectively. These findings are in agreement with the reconstruction results at C band. 

Residual maps are displayed in Figure~\ref{fig:x_maps_dde}, right column. The highest fidelity to data in the image space is achieved via DDE calibration, showcased thorough its low and homogeneous residual image, as opposed to DIE calibration. Quantitative evaluation of the residual images, as well as the residual data, is provided in Table~\ref{tab:X_vis}, through the examination of their statistics. 
Statistics of DDE calibration residual image reveal their Gaussian nature as opposed to residual images obtained with DIE calibration and {Adaptive~PPD}. The latter are heavily tailed and highly skewed. Here again, STD value of the residual data obtained with DDE calibration is close to 1, suggesting that the residual data have reached the instrumental noise level. 
\begin{table}
    \caption{ X band: Statistics  of the residual data and image. }
  \resizebox{0.5\textwidth}{!}{    
    \begin{tabular}{lllll} 
	\hline
     \textbf{Residual data}   & STD & Skewness  & Kurtosis & vSNR{~(dB)} \\
	\hline
	Adaptive PPD      & 3.07 &  -0.013 &  51.94 &  32.64 \\
	$S=1\times1$  &  2.37 &  -0.101 &  45.98 & 34.89\\
	$S=5\times5$   &  1.03 & 0.008  &  3.97  & 42.11\\
	\hline
     \textbf{Residual image}   & STD~{($\times10^{-4})$} & Skewness & Kurtosis  & iSNR{~(dB)}\\
	\hline
	Adaptive PPD    & 2.41 & -1.880 & 5.98 &  46.66\\
	$S=1\times1$   & 1.43 & -0.680 & 0.74 &  49.60\\  
	$S=5\times5$   &  0.61 &  0.064  & 0.40 &  57.16\\       
     \hline
    \end{tabular}
   }  
     \label{tab:X_vis}

\end{table}

\subsubsection*{Impact of SARA image prior}
\begin{figure}
\centering
\includegraphics[width=1.0\linewidth]{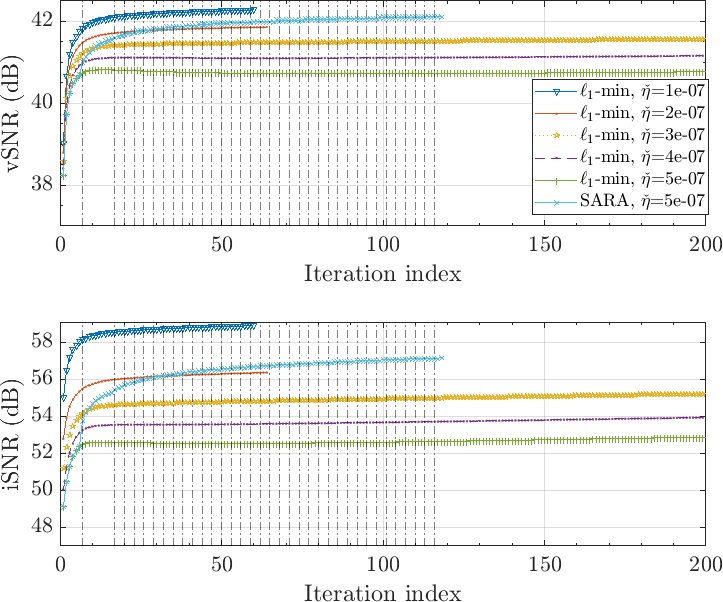}
\caption{X band: vSNR (top) and iSNR (bottom) evolution along the iterations for joint DDE calibration and imaging ($S=5\times 5,~\tau=8$) experiments, corresponding to SARA prior with $\check{\eta} =5\times 10^{-7}$ and $\ell_1$-minimisation, with different image regularisation parameter values fixed such that $\check{\eta} \in\{ 10^{-7},2\times 10^{-7},3\times 10^{-7},4\times10^{-7},5\times10^{-7}\}$. In the context of SARA experiment, the displayed vertical lines demarcate the last iteration of Algorithm~\ref{algo:VMBCFB}, solving a re-weighted minimisation task. }
\label{fig:xsnr}

\end{figure}

To illustrate the efficiency of SARA prior, we compare its performance to $\ell_1$-minimisation {which corresponds to ${\boldsymbol\omega}= \mathbf{1}_{B}$, where $\mathbf{1}_{B}\in \eR^{B}$ denotes the vector whose elements are set to one.} for every $n \in \{1,\dots,B\}$. In this context, the image regularisation parameter is set such that $\check{\eta} =5\times 10^{-7}$ for SARA and is varied within the range $[ 10^{-7}, 5\times10^{-7}]$ for the $\ell_1$-sparsity experiments. The latter experiments are stopped at convergence\footnote{Convergence is achieved if one of the following two criteria is satisfied; (i) the maximum number of iterations $I=200$ is reached and (ii) the relative variation between two consecutive values of the
objective function minimised via Algorithm \ref{algo:VMBCFB} is smaller than the lower bound $\check{\varrho}=3\times 10^{-4}$. }. The performance of the regularisers is evaluated numerically via the inspection of the evolution of data fidelity metrics iSNR and vSNR throughout the iterations, shown in Figure~\ref{fig:xsnr}. Note that for SARA prior, the evolution of the two metrics is studied by aggregating the values obtained at the inner iterations of Algorithm~\ref{algo:VMBCFB}, with vertical lines demarcating the last iteration of each re-weighted $\ell_1$-minimisation task.

In the context of the $\ell_1$-sparsity, one can observe that the higher the regularisation parameter $\check{\eta} $, the lower the fidelity to data in both spaces. Given the unconstrained formulation of the minimisation task, such behaviour is expected, as the regularisation parameter plays the role of a trade-off between model prior and data fidelity. In fact, for the highest value of the regularisation parameter set such that $\check{\eta} =5\times10^{-7}$, the estimate of the radio map is sparse and smooth, with much of the details left in the residual image as shown in Figure~\ref{fig:x_maps_dde}, third row. As for the lowest value corresponding to $\check{\eta} =10^{-7}$, while the fit-to-data is the highest numerically and visually (see Figure~\ref{fig:x_maps_dde}, second row, right panel), the model image (Figure~\ref{fig:x_maps_dde}, second row, left) present large ringing artefacts contaminating its background and limiting its dynamic range. Interestingly, SARA prior with regularisation parameter fixed such that $\check{\eta}=5\times 10^{-7}$ (Figure~\ref{fig:x_maps_dde}, top row) ensures both high fit-to-data and image reconstruction quality. When compared to $\ell_1$-sparsity with a regularisation parameter $\check{\eta} =5 \times 10^{-7}$, the evolution of the SNRs in both domains showcases that re-weighting procedure boosts the fit-to-data. These results confirm the suitability of SARA prior for image recovery within the adopted joint calibration and imaging framework. 

\begin{figure*}
\begin{minipage}[t]{1\linewidth}
\centering
\includegraphics[width=0.27\linewidth]{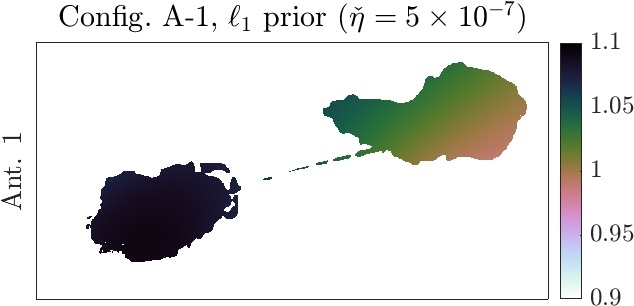}
\includegraphics[width=0.26\linewidth]{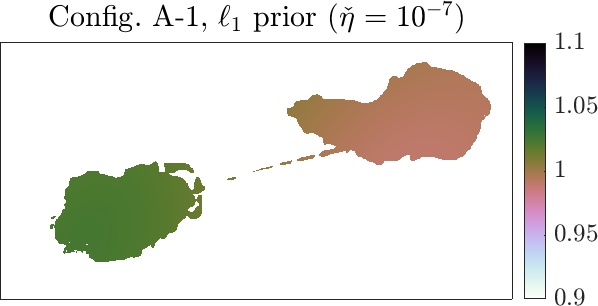}
\includegraphics[width=0.26\linewidth]{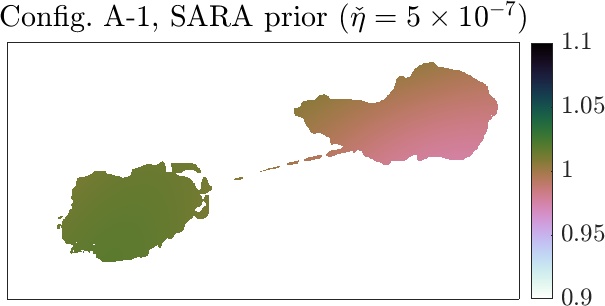}
\includegraphics[width=0.27\linewidth]{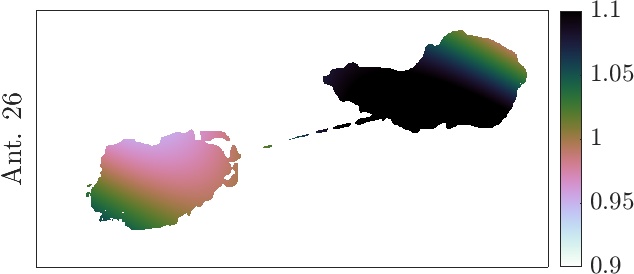}
\includegraphics[width=0.26\linewidth]{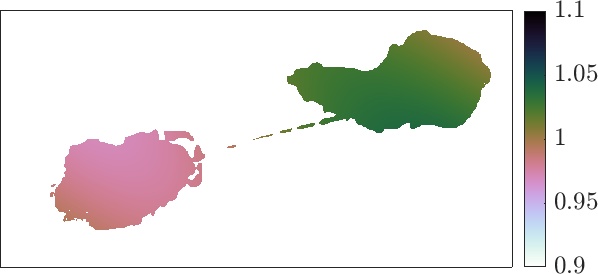}
\includegraphics[width=0.26\linewidth]{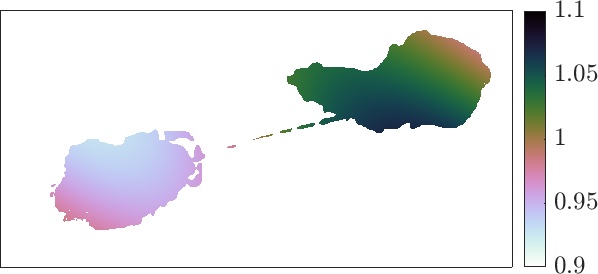}
\end{minipage}
~\\
\begin{minipage}[t]{1\linewidth}
\centering
\includegraphics[width=0.27\linewidth]{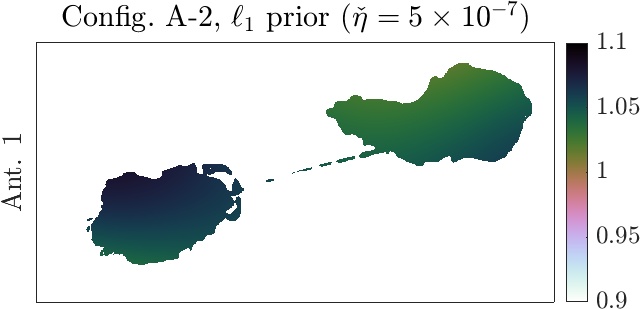}
\includegraphics[width=0.26\linewidth]{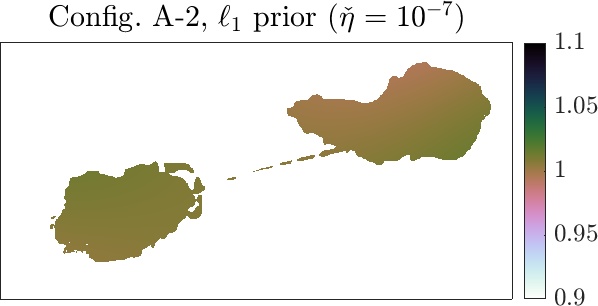}
\includegraphics[width=0.26\linewidth]{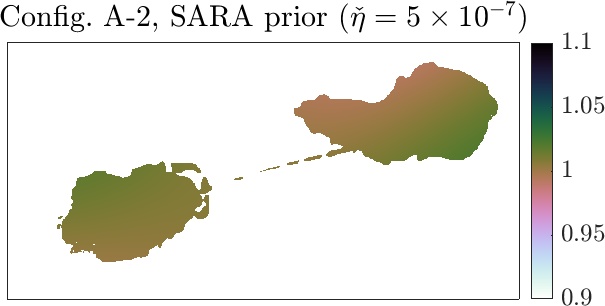}
\includegraphics[width=0.27\linewidth]{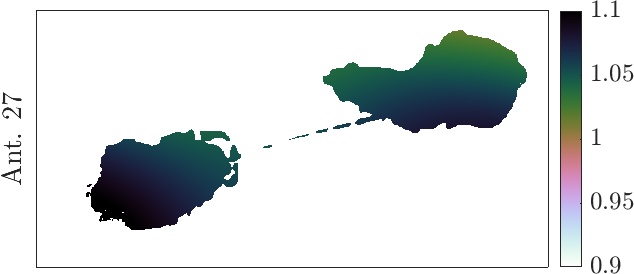}
\includegraphics[width=0.26\linewidth]{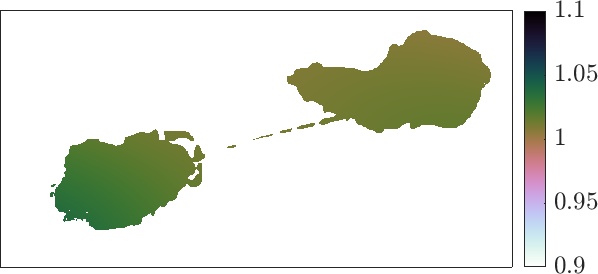}
\includegraphics[width=0.26\linewidth]{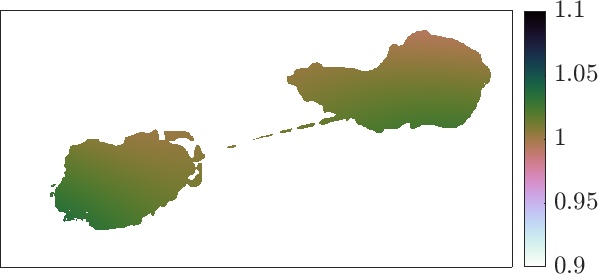}
\end{minipage}
~\\
\begin{minipage}[t]{1\linewidth}\centering
\includegraphics[width=0.27\linewidth]{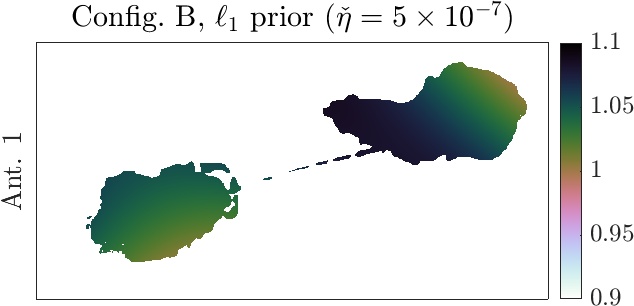}
\includegraphics[width=0.26\linewidth]{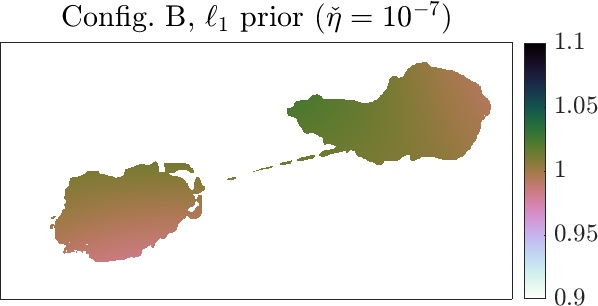}
\includegraphics[width=0.26\linewidth]{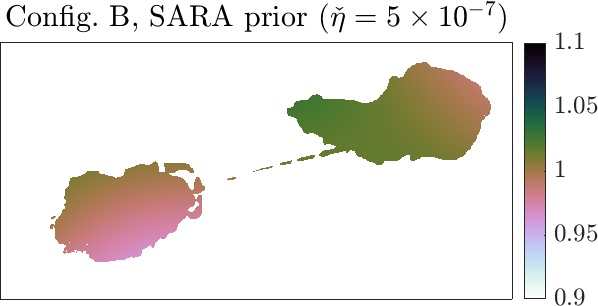}
\includegraphics[width=0.27\linewidth]{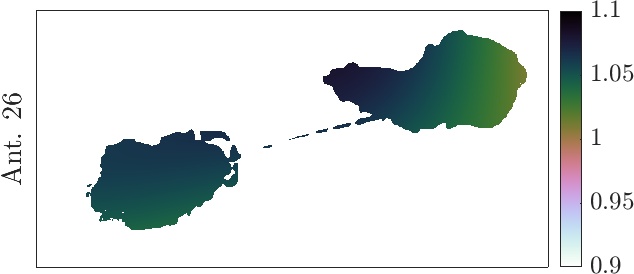}
\includegraphics[width=0.26\linewidth]{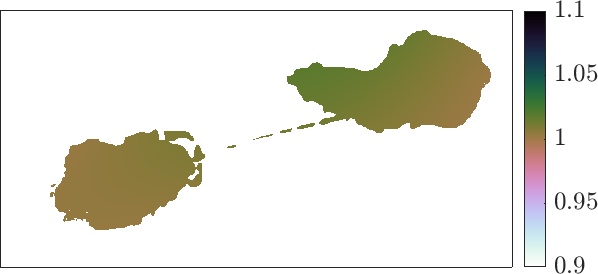}
\includegraphics[width=0.26\linewidth]{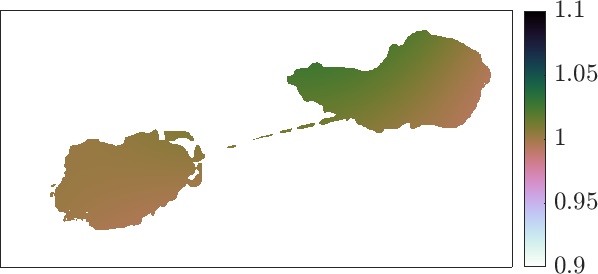}
\end{minipage}
~\\
\begin{minipage}[t]{1\linewidth}\centering
\includegraphics[width=0.27\linewidth]{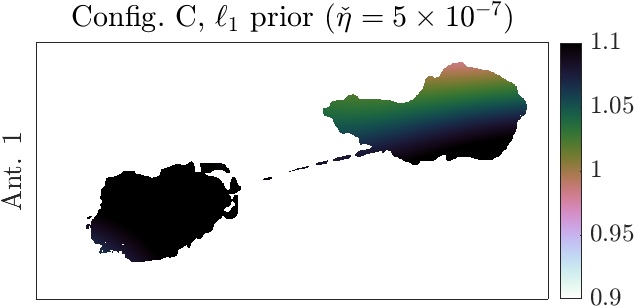}
\includegraphics[width=0.26\linewidth]{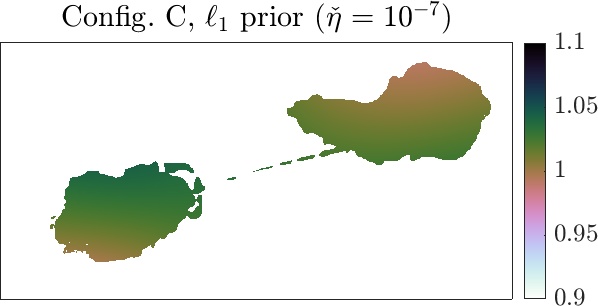}
\includegraphics[width=0.26\linewidth]{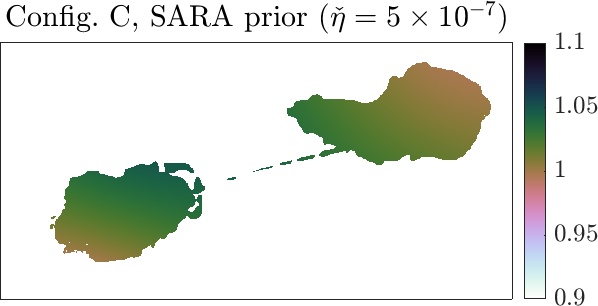}
\includegraphics[width=0.27\linewidth]{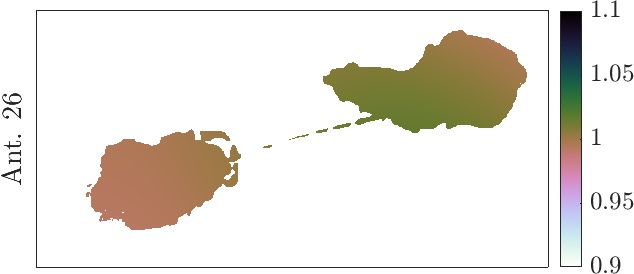}
\includegraphics[width=0.26\linewidth]{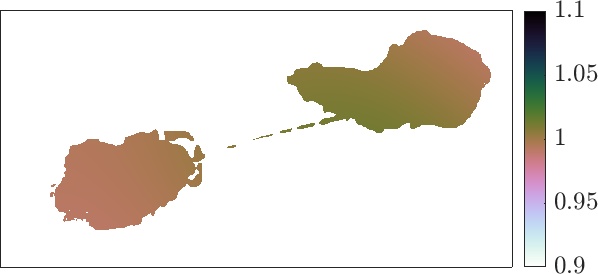}
\includegraphics[width=0.26\linewidth]{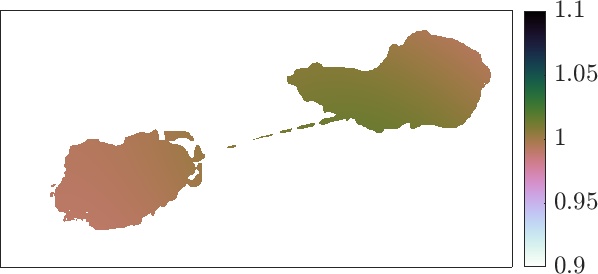}
\end{minipage}
~\\
\begin{minipage}[t]{1\linewidth}\centering
\includegraphics[width=0.27\linewidth]{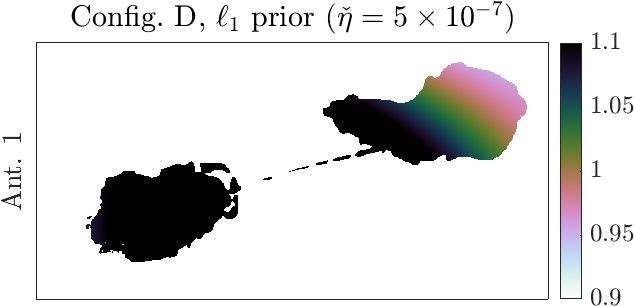}
\includegraphics[width=0.26\linewidth]{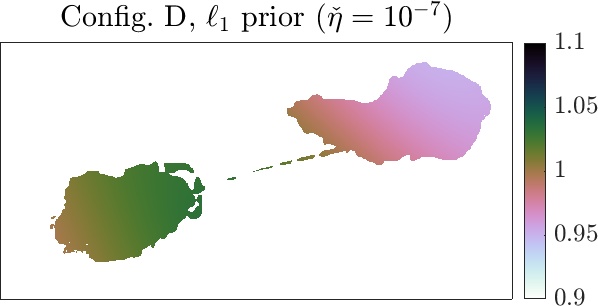}
\includegraphics[width=0.26\linewidth]{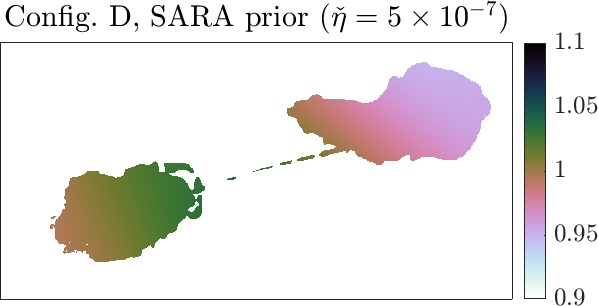}
\includegraphics[width=0.27\linewidth]{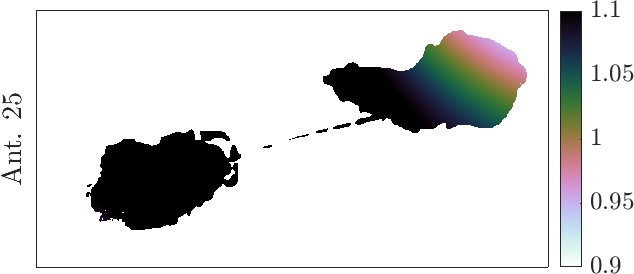}
\includegraphics[width=0.26\linewidth]{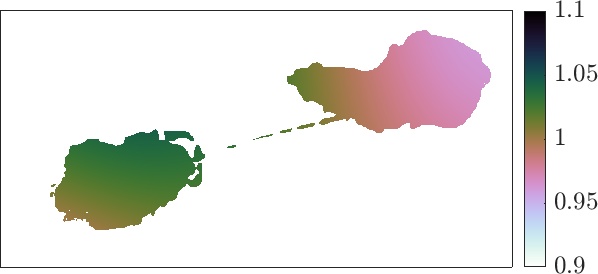}
\includegraphics[width=0.26\linewidth]{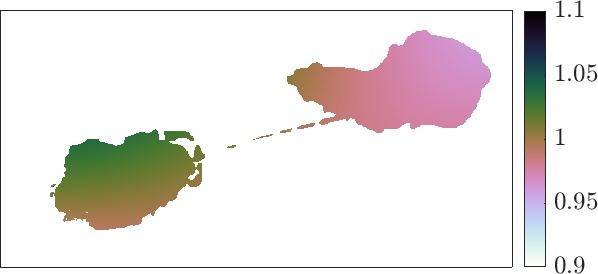}
\end{minipage}
\caption{X band: amplitudes of DDE solutions in the image domain obtained with our approach ($S=5\times 5,~\tau =8$). Estimated DDEs are displayed over a FoV of about $0.02275\degr \times 0.0455\degr$ at pixel positions spanning six orders of magnitude of the recovered dynamic range in the image estimate. From left to right, results obtained with {the $\ell_1$ prior associated with regularisation} parameters $\check{\eta}=5\times 10^{-7}$ and $\check{\eta}= 10^{-7}$, respectively, and SARA prior with regularisation parameter $\check{\eta}=5 \times10^{-7}$. From top to bottom, DDE estimates of two chosen antennas for each of the VLA configurations at our {{chosen time slot}}. Rows 1-2, configuration A, first data set, antennas 1 and 26.  Rows 3-4, configuration A, second data set, antennas 1 and 27. Rows 5-6, configuration B, antennas 1 and 26. Rows 7-8, configuration C, antennas 1 and 26. Rows 9-10, configuration D, antennas 1 and 26. }
\label{fig:x_dde_solutions}
\end{figure*}

Selected DDE estimates represented in the image space, obtained by $\ell_1$-sparsity experiments with regularisation parameters fixed such that $\check{\eta} \in\{10^{-7}, 5\times 10^{-7}\}$, and SARA image prior with regularisation parameter $5\times 10^{-7}$, are displayed in Figure~\ref{fig:x_dde_solutions}. More precisely, amplitudes of DDE solutions associated with our chosen time slot and antenna pair for the different configurations of the {{VLA}}, are displayed at the pixel positions of the radio galaxy. On the one hand, one observes that DDE solutions obtained with $\ell_1$-sparsity with $\check{\eta} =5\times 10^{-7}$ present significantly larger amplitude values in comparison with SARA prior and $\ell_1$-sparsity with $\check{\eta} =10^{-7}$ experiments. Given the low fit-to-data obtained by $\ell_1$-sparsity with $\check{\eta} =5\times 10^{-7}$ experiment, this suggest that the amplitudes of the obtained DDE solutions could be over-estimated. On the other hand, one observes the high similitude between DDE solutions of SARA prior and $\ell_1$-sparsity with $\check{\eta} = 10^{-7}$ at the pixel positions of the radio galaxy, with subtle differences noticed for some DDE solutions where $\ell_1$-sparsity with $\check{\eta} =10^{-7}$ has reached relatively smaller amplitude values. This observation, combined with the large amount of artefacts present in the associated recovered image, suggests the under-estimation of DDEs amplitudes to a small degree via $\ell_1$-sparsity with $\check{\eta} =10^{-7}$. 
Such tendencies have been observed on DDE solutions of different antennas at different time slots. These findings suggest that SARA prior achieves the best of the two $\ell_1$-sparsity priors, with DDE solutions and image estimates presenting high fidelity to data while ensuring the high quality of the recovered radio map in terms of resolution and dynamic range.

\subsection{S band reconstructions}

%
S band data are imaged at a spatial resolution that is about 2.18 times the nominal resolution of the observations, corresponding to a pixel size $\delta x=0.2\arcsec$. The mapped FoV is $\Omega= 0.0569\degr \times 0.0569\degr $ such that the reconstructed image of the sky is of size $N=1024\times 1024$. 
In this experiment, the effective spatial and temporal bandwidths of the DDEs are set such that $S=5\times5$ and $\tau=8$, as these values have shown to achieve high imaging quality at the previous bands. Parameters of the joint calibration (both DDE and DIE cases) are fixed as described in Section~\ref{ssec:settings}, with the sparsity regularisation parameter set to $\check{\eta} =8\times 10^{-7}$.

\begin{table}
    \caption{S band: Statistics  of the residual data and image.}
      \centering
           \hspace{-0.3cm}
    \resizebox{0.5\textwidth}{!}{    
    \begin{tabular}{lllll} 
	\hline
       \textbf{Residual data} 
       & STD & Skewness  & Kurtosis~ & vSNR{~(dB)}  \\
     	\hline
     	Adaptive PPD   & 8.79  & 0.30   &  3.64  & 43.85 \\
     	$S=1\times 1$    &6.07 &   -0.10  &  6.59 &  46.38\\
     	$S=5\times 5$    &3.08   &-0.60  &  3.78 & 53.80\\  
     	
	\hline
    \textbf{Residual Image}& STD~{($\times10^{-3}$)} & Skewness & Kurtosis & iSNR{~(dB)}\\
	\hline
	Adaptive PPD      & 5.4  & -0.07  &  5.05 & 66.11\\
	$S=1\times 1$    & 3.04 & -1.36  & 2.71   & 68.48\\
	$S=5\times 5$    & 2.52 & 0.80  &  3.92  & 75.76 \\

	\hline
    \end{tabular}
   } 
   \label{tab:S_vis} 
\end{table}

\begin{figure*}

\begin{minipage}[t]{0.9\linewidth}
\centering
\includegraphics[width=0.3\linewidth]{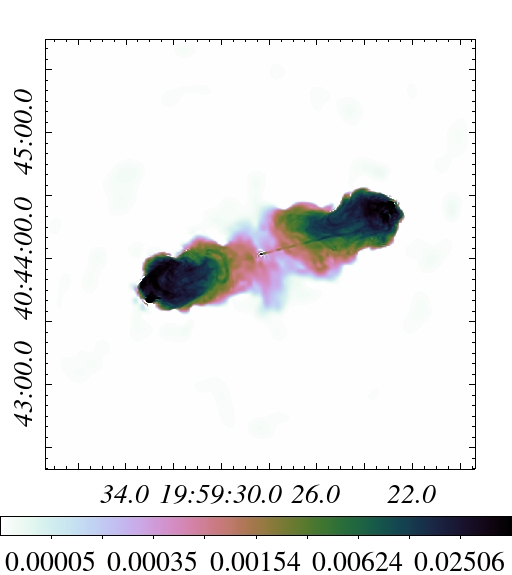}
\includegraphics[width=0.3\linewidth]{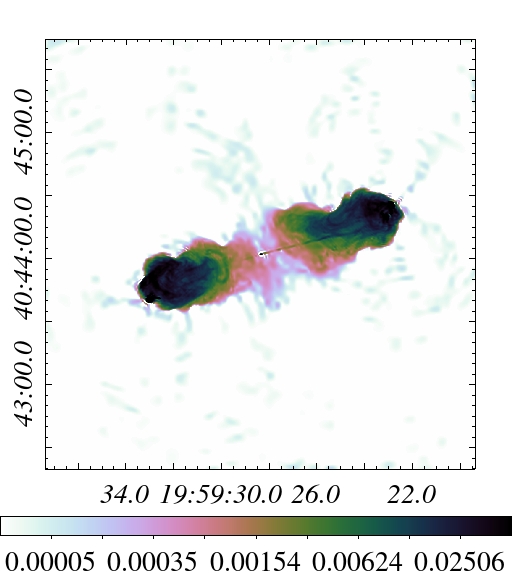}
\includegraphics[width=0.3\linewidth]{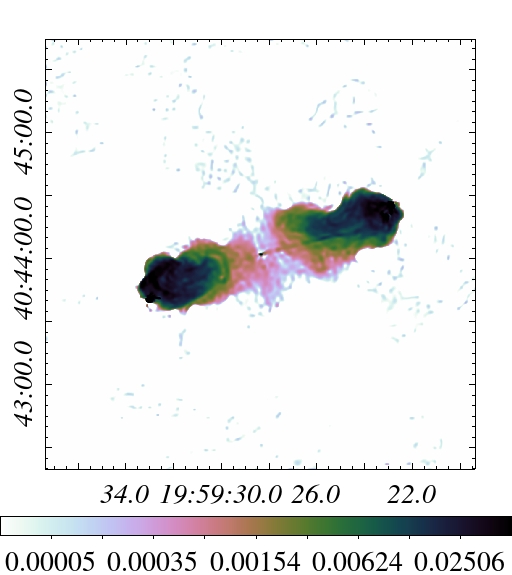}

\includegraphics[width=0.3\linewidth]{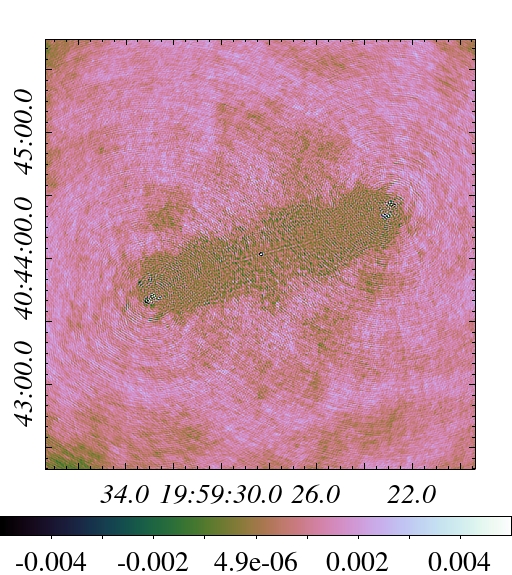}
\includegraphics[width=0.3\linewidth]{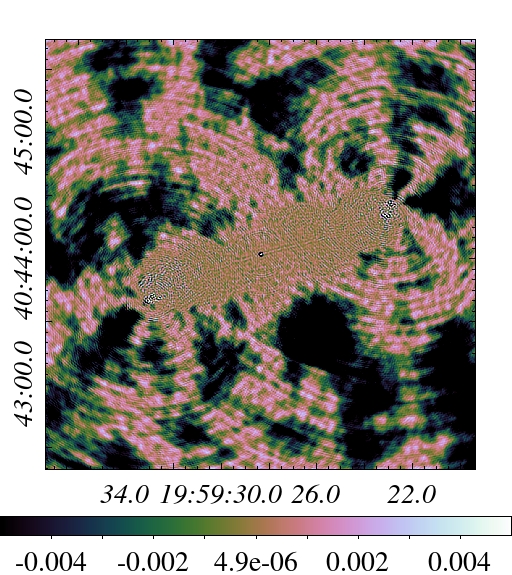}
\includegraphics[width=0.3\linewidth]{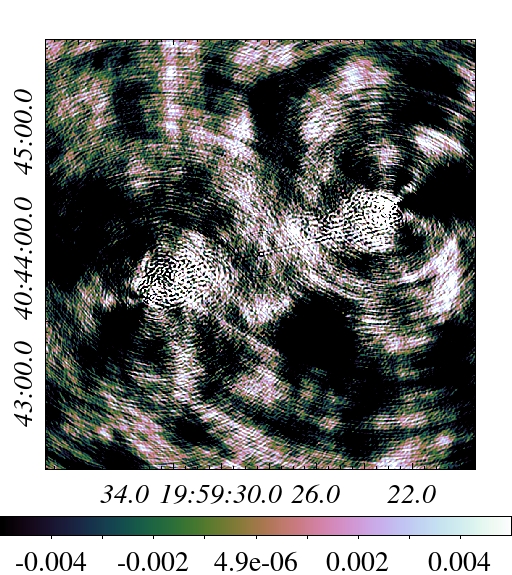}
\end{minipage}
\caption{S band: From left to right, joint calibration and imaging results for DDE calibration ($S =5\times 5,~\tau =8$) and DIE calibration ($S =1\times 1,~\tau =8$) and {Adaptive~PPD} imaging.
From top to bottom, estimated model images ($\log_{10}$ scale), and residual images (linear scale). {{The imaged and displayed FoV is $\Omega= 0.0569\degr \times 0.0569\degr$}}. The surface brightness of the estimated model images is in Jy/pixel with the pixel size set to $0.2\arcsec$.}
\label{fig:s1_maps}

\end{figure*}

Recovered maps with joint calibration and imaging for the estimation of DDEs ($S=5\times5$, $\tau=8$) and DIEs ($S=1\times 1$, $\tau=8$) along with the ones obtained with {Adaptive~PPD} imaging are displayed in Figure~\ref{fig:s1_maps}. The reconstructed model images confirm the high sensitivity achieved via DDE estimation. Residual maps and their statistics\footnote{When inspecting the residual visibilities of the different methods, important outliers have been noticed. In order to have meaningful statistics, in particular the non-Gaussianity metrics, computations have been made after removing the residual visibilities whose amplitude is above 5 standard deviations.} displayed in Table~\ref{tab:S_vis} also confirm the higher fidelity to data obtained with DDE calibration, in comparison with DIE calibration and {Adaptive~PPD} imaging, with the SNRs reaching 75.76~dB in the image space and 53.8~dB in the data space. However, non-Gaussianity metrics, in particular the excess kurtosis of the residual image obtained with DDE calibration, suggest that the image departs from a Gaussian distribution. {At this low observation frequency, numerous degrading effects such as ground reflections, and antenna-antenna reflections and coupling, can explain the non Gaussian nature of the residuals}. On a further note, the STD value of the residual data that is above one, suggests that the instrumental noise is not yet reached.

\begin{figure*}
\begin{minipage}[]{0.3\linewidth}
\includegraphics[width=\linewidth]{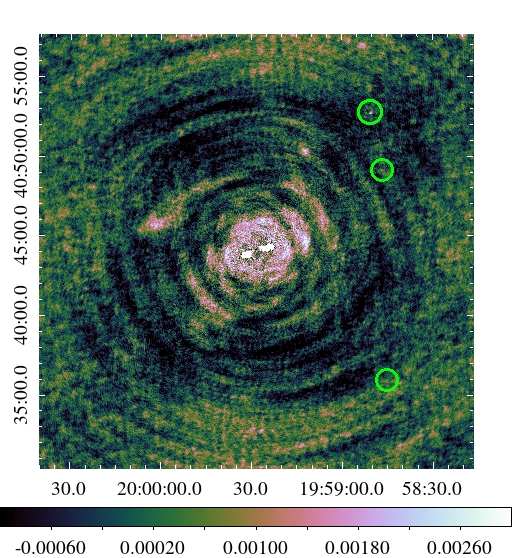}
\end{minipage}
~
\begin{minipage}[]{0.67\linewidth}
     \begin{subfigure}[a]{0.32\textwidth}
        \label{fig:src1}
   \includegraphics[width=\linewidth]{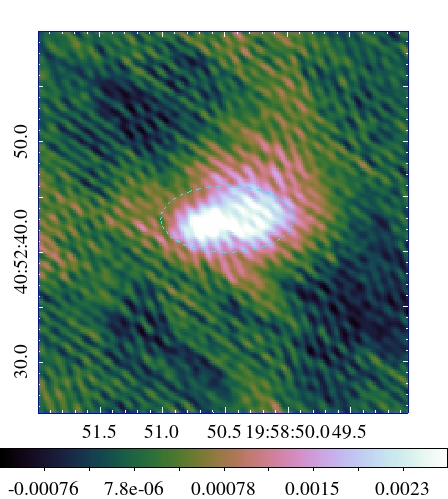}
   \caption{SRC1}
   \end{subfigure}
   \begin{subfigure}[b]{0.32\textwidth}
   \label{fig:src2}
   \includegraphics[width=\linewidth]{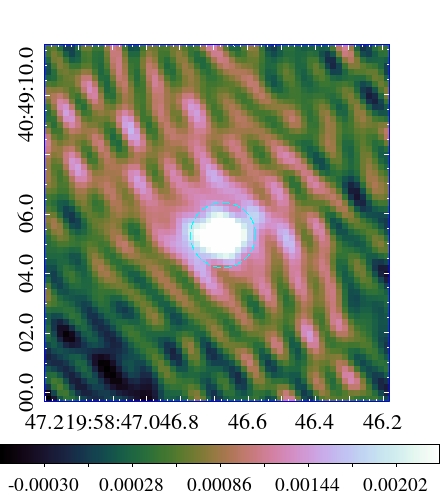}
   \caption{SRC2}
   \end{subfigure}
   \begin{subfigure}[c]{0.32\textwidth}
   \label{fig:src3}
\includegraphics[width=\linewidth]{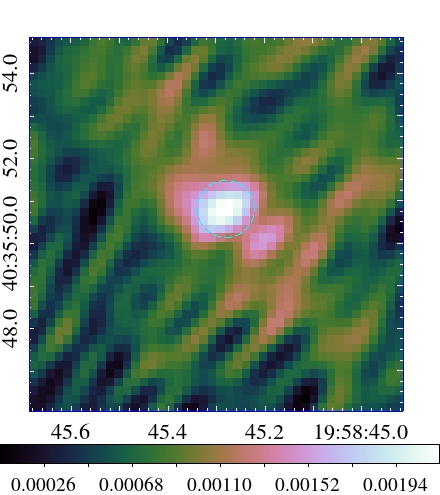}

\caption{SRC3}
   \end{subfigure}
   
    \end{minipage}%
\caption{S band: from left to right, residual image computed over the FoV $\Omega =0.45512\degr \times 0.45512\degr $, centred at the core of Cyg~A and zooms on the detected background sources (highlighted with green circles on the left panel). {{Apparent flux values of the detected sources are computed over the regions highlighted in cyan, dashed line, of their associated panels.
}}}
\label{fig:s1_srcs}
\end{figure*}

In light of these observations, estimated residual visibilities via DDE calibration are imaged via NUFFT over a large FoV, $\Omega =0.45512\degr \times 0.45512\degr$ that is eight times larger than the jointly calibrated and imaged FoV. As an encouraging illustration of the fidelity of the image reconstruction, we detect three faint background sources in the vicinity of Cyg~A. To our knowledge, this is the first published image of Cyg~A that is deep enough to detect background sources. The sources are displayed in Figure~\ref{fig:s1_srcs}. Details of their positions and their apparent fluxes are listed in Table~\ref{tab:S_srcs}.
\begin{table}
    \caption{{{S band: details of the detected background sources}}.}
      \centering
     \hspace{-0.5cm}
    \resizebox{0.5\textwidth}{!}{    
    \begin{tabular}{llll} 
	\hline
       \textbf{Background sources} 
       & RA~(J2000) & DEC  & Flux~(mJy) \\
     	\hline
     	SRC1  & $19\rm{h}58\rm{mn}50.481\rm{s}$  & $+40\degr52\arcmin42.93\arcsec $  &  13.42  \\
     	SRC2   &$19\rm{h}58\rm{mn}46.672\rm{s}$ &  $+40\degr49\arcmin05.28\arcsec$ &  9.5 \\
     	SRC3    &$19\rm{h}58\rm{mn}45.279\rm{s}$  & $+40\degr35\arcmin50.80\arcsec$  &  2.8 \\
	\hline
    \end{tabular}
   } 
   \label{tab:S_srcs} 
\end{table}


\subsection{Computational efficiency assessment}
The joint calibration and imaging approach is implemented in  \mbox{MATLAB}. All experiments reported in this section are performed on Cirrus, a UK Tier2 HPC service\footnote{{https://epsrc.ukri.org/research/facilities/hpc/tier2/}}. Cirrus is an SGI ICE XA system consisting of 280 computing nodes, each having two processors (2.1~GHz, 18-core, Intel Xeon E5-2695 (Broadwell) series) and 256~GB memory shared between the two processors. More specifically, each experiment is conducted on a single node of Cirrus, using its 36 cores as follows. The updates of the antenna gains in Steps \ref{algo:step:startU}-\ref{algo:step:stopDDEs} are performed in parallel with respect to the antennas. The number of cores used is therefore equal to the number of the antennas. All 36 cores are then exploited to update the mapping operator prior to the imaging cycle in Step~\ref{algo:step:updateG}. Finally, in the imaging step, we leverage a faceted implementation of the wavelet bases proposed by \citet{Prusa2012}. In this context, 16 facets are adopted, and consequently 16 cores are deployed in Steps~\ref{algo:step:startim}-\ref{algo:step:endim}. 
\begin{table}

    \caption{Computing cost of the joint calibration and imaging approach when varying the spatial bandwidth of the antenna gains.}
      \centering
           \hspace{-0.5cm}
    \resizebox{0.5\textwidth}{!}{    
    \begin{tabular}{lllll} 
	\hline
      \textbf{Computing Cost} 
       & $S=1\times1$  & $S=3\times3$ &  $S=5\times5 $  &  $S=7\times7$\\
      \hline
      Calibration steps (CPU~hour)  & 54 & 306 & 1909 & 4059 \\
      Mapping operator update (CPU~hour) & 53 & 193 & 416 & 650 \\
      Imaging steps (CPU~hour) & 88 &  165 & 216  & 220 \\
       Total Time & 8h  & 24h & 81h  & 144h\\
       
	\hline
    \end{tabular}
   } 
   \label{tab:cost} 
\end{table}

In general, we have observed that the overall computational cost of our approach is highly correlated with the spatial {bandwidth} of the estimated antenna gains. This is showcased in Table~\ref{tab:cost}, where we report the computational cost of the method applied to C band data when varying the spatial bandwidth of the antenna gains (see Section~\ref{sec:cyga-c} for details of these experiments). In fact, we have noticed that the main computational bottleneck lies in the calibration step, more precisely in the update of the operators $\left(\Hc_{\alpha,1},\Hc_{\alpha,2}\right)_{\alpha \in \{1,\dots,n_a\}}$ (Steps~\ref{algo:step:h1} and \ref{algo:step:h2} of Algorithm~\ref{algo:VMBCFB}), which involves convolutions of the NUFFT kernels with the spatial Fourier transform of the image estimate and the antenna gains. These operations are performed several times within each global iteration of Algorithm~\ref{algo:VMBCFB}. Yet, they can benefit from a massive parallelisation with respect to the antennas and time instances, which is not achieved with the current \mbox{MATLAB} implementation.

The overall computing time of the method spans hours to days depending on the spatial bandwidth of the estimated antenna gains. In comparison, a traditional \alg{CLEAN}-based deconvolution takes few minutes only when applied to the same data, and utilising the same computing resources. Given the complexity of the joint DDE calibration and imaging problem, the total computing time of the method in its current implementation is indeed considerably long. Nevertheless, a fully parallelised implementation of the approach is expected to achieve important gains in the computing time. 

\section{Conclusions}\label{sec:cc}
In this paper, we provide a first application of the joint calibration and imaging approach initially proposed by \citet{Repetti2017} to highly sensitive RI data. The data considered herein are {{VLA}} observations of the radio galaxy Cyg~A at three bands. The non-convex optimisation approach calibrates for the unknown DDEs and provides high quality estimates of the radio sky, thanks to its suitable DDEs prior, in particular, the smoothness in time and space, and its powerful SARA image prior. The estimated radio maps exhibit high fidelity where faint features of the galaxy, previously buried in the errors induced by the inaccurate measurement operator, have emerged. In addition, efficient DDEs estimation results in residual visibilities reaching instrumental noise levels on some data sets (e.g. Cyg~A~reconstructions at bands X and C). From the computational perspective, DDEs estimation is highly parallelisable with respect to the antennas, making the approach computationally attractive. Moreover, we have noticed that highly smooth DDEs whose dimension in the spatial Fourier domain is limited to $S=3\times3$ can already achieve good quality reconstructions, with important fidelity to the data and limited artefacts in the model images, thus providing a good compromise between accuracy and computational cost. Interestingly, the approach succeeds in {{detecting}} background sources in the vicinity of Cyg~A at S band, illustrating the level of accuracy of the estimated mapping operator and Cyg~A radio map when imaging the highly sensitive {{VLA}} data. Future developments consist in extending the current work to the context of wide-field wide-band polarimetric imaging for real RI data by leveraging Faceted HyperSARA \citep{Thouvenin20}, a recently proposed wide-band imaging algorithm involving sophisticated facet-specific prior models to efficiently handle the sheer volume of wide-band RI image cubes and the theoretical developments of \citet{Birdi2019} proposing a joint polarimetric calibration and imaging framework, dubbed Polca~SARA.

\section{Data availability}
 The data underlying this article were provided by the National Radio Astronomy Observatory (NRAO) (Program code: 14B-336). The self-calibrated data can be shared upon request to RP. The MATLAB implementation of the proposed approach is available at https://basp-group.github.io/SARA-CALIB-realdata/.
Reconstructed images underlying this article are available at {https://doi.org/10.17861/529cdcbc-7c18-47a6-970f-755a5da19071}.
\section*{Acknowledgements}
The National Radio Astronomy Observatory is a facility of the National Science Foundation operated under cooperative agreement by Associated Universities, Inc. The research of AD, AR and YW is supported by the UK Engineering and Physical Sciences Research Council (EPSRC, grant EP/M008843/1) and is conducted using Cirrus UK National Tier-2 HPC Service at EPCC ({http://www.cirrus.ac.uk}) funded by the University of Edinburgh and EPSRC (grant EP/P020267/1). The research of OS is supported by the South African Research Chairs Initiative of the Department of Science and Technology and National Research Foundation. 




\bibliographystyle{mnras}

\bibliography{biblio} 



\appendix
\section{Plots of the estimated DDEs at C band, spanning the imaged Field of view for selected antennas and time slots.}
\begin{figure*}
\begin{minipage}[t]{1\linewidth}
\centering
\includegraphics[width=0.33\linewidth]{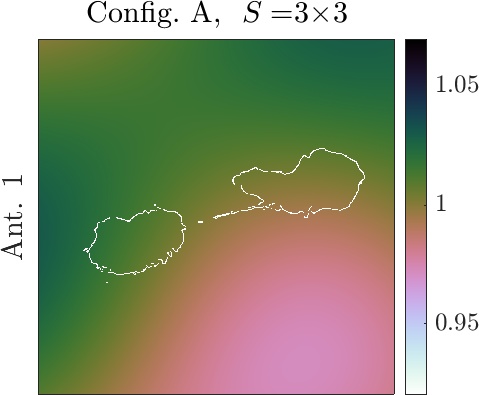}
\includegraphics[width=0.31\linewidth]{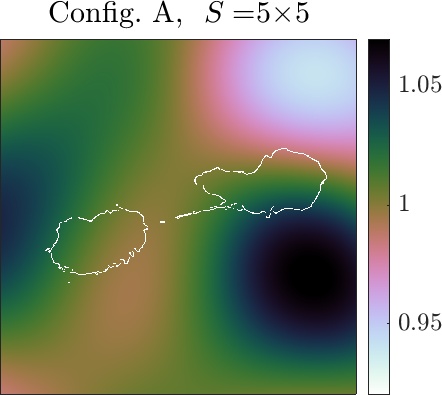}
\includegraphics[width=0.31\linewidth]{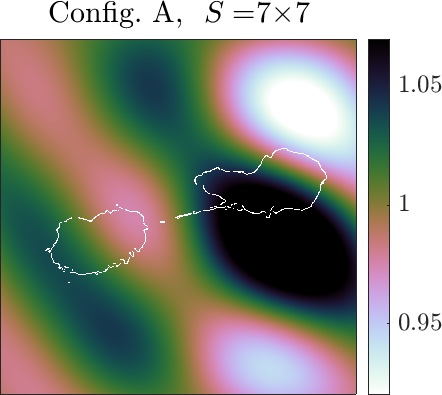}
\includegraphics[width=0.33\linewidth]{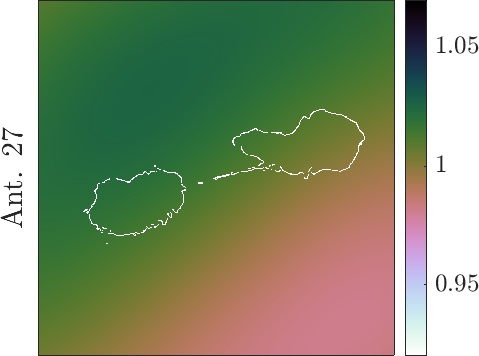}
\includegraphics[width=0.31\linewidth]{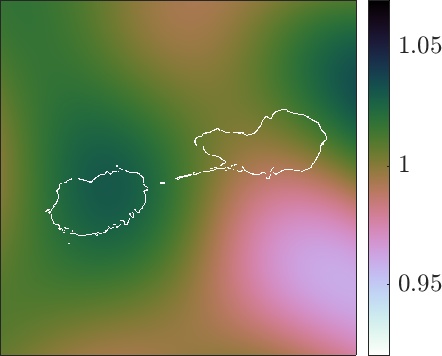}
\includegraphics[width=0.31\linewidth]{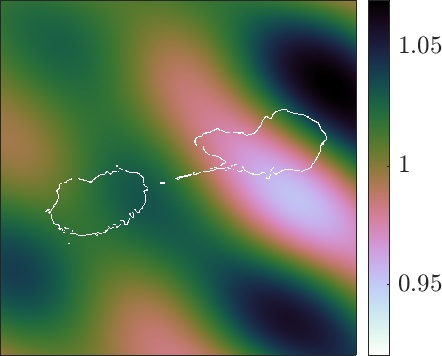}
\end{minipage}
~\\
\begin{minipage}[t]{1\linewidth}\centering
\includegraphics[width=0.33\linewidth]{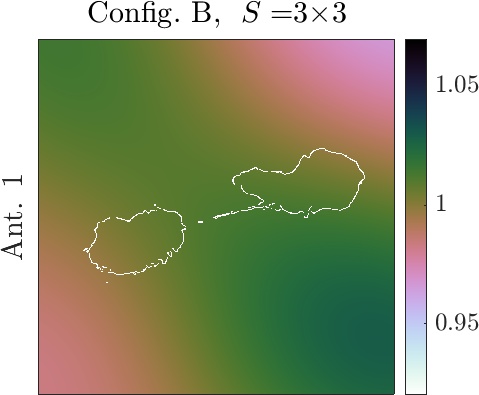}
\includegraphics[width=0.31\linewidth]{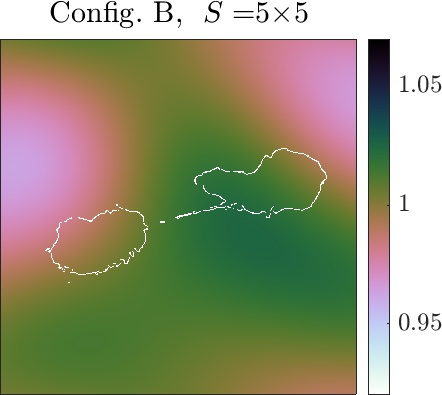}
\includegraphics[width=0.31\linewidth]{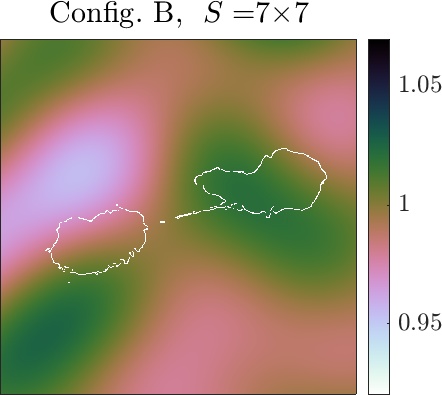}
\includegraphics[width=0.33\linewidth]{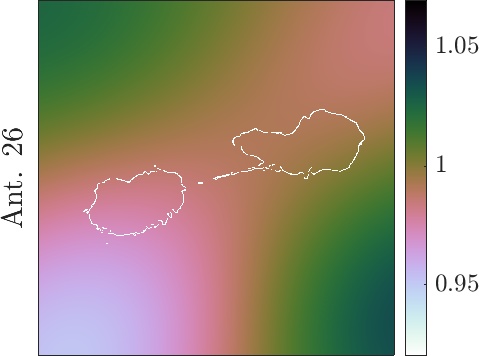}
\includegraphics[width=0.31\linewidth]{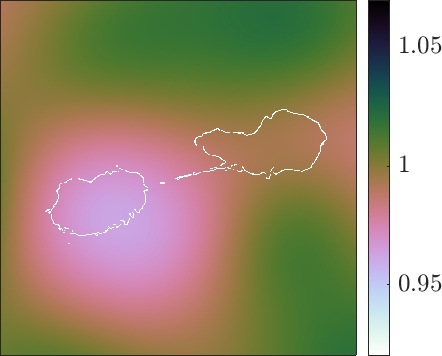}
\includegraphics[width=0.31\linewidth]{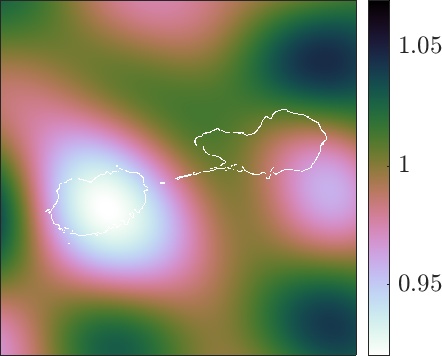}
\end{minipage}
\caption{{C band: amplitudes of DDE solutions in the image domain obtained with the joint calibration and imaging approach ($\tau =8$). Estimated DDEs are displayed over the imaged FoV $\Omega= 0.0455\degr \times 0.0455\degr$. Cyg~A is demarcated by the white contours. From left to right, results obtained for DDEs spatial Fourier dimension $S$ set to $3\times 3,~5\times 5,~ 7\times 7$, respectively. From top to bottom, DDE estimates of two selected antennas at VLA configurations A and B, at {the $10^{\textrm{th}}$} time slot. Rows 1-2, configuration A, antennas 1 and 27.  Rows 3-4, configuration B, antennas 1 and 26. }}
\label{fig:c_dde_solutions_full1}

\end{figure*}
\begin{figure*}
\begin{minipage}[t]{1\linewidth}\centering
\includegraphics[width=0.33\linewidth]{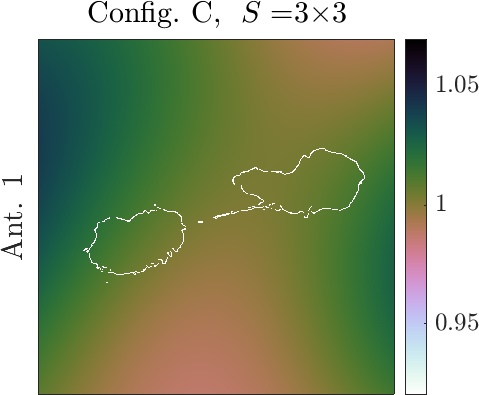}
\includegraphics[width=0.31\linewidth]{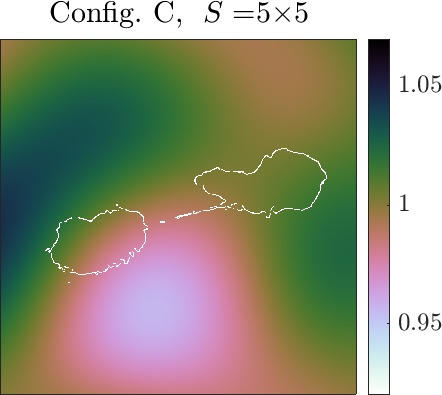}
\includegraphics[width=0.31\linewidth]{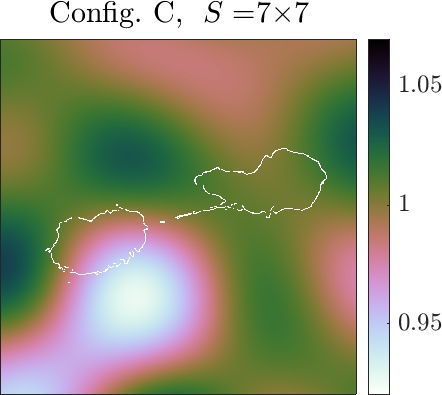}
\includegraphics[width=0.33\linewidth]{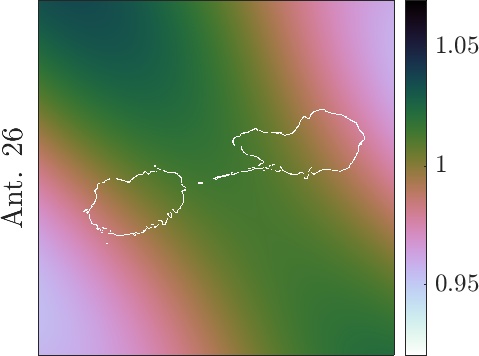}
\includegraphics[width=0.31\linewidth]{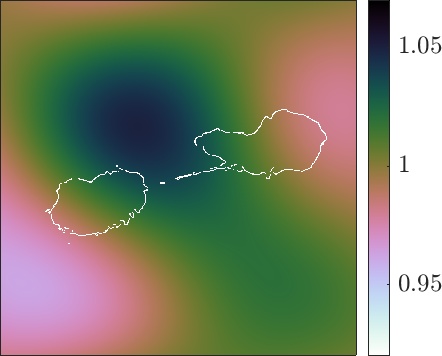}
\includegraphics[width=0.31\linewidth]{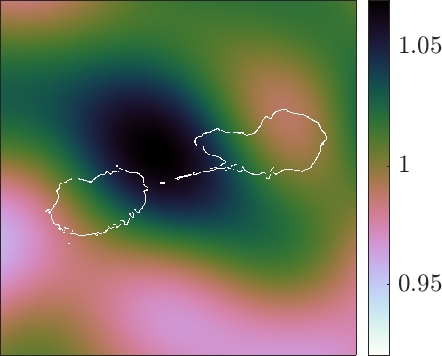}
\end{minipage}
~\\
\begin{minipage}[t]{1\linewidth}\centering
\includegraphics[width=0.33\linewidth]{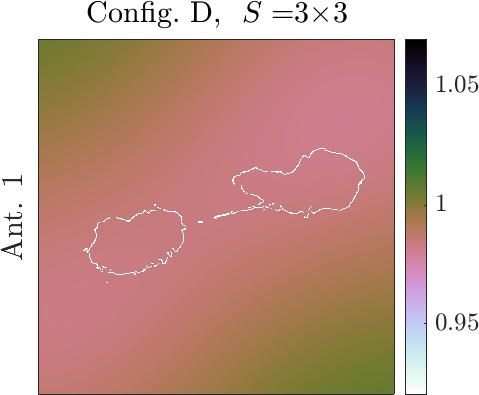}
\includegraphics[width=0.31\linewidth]{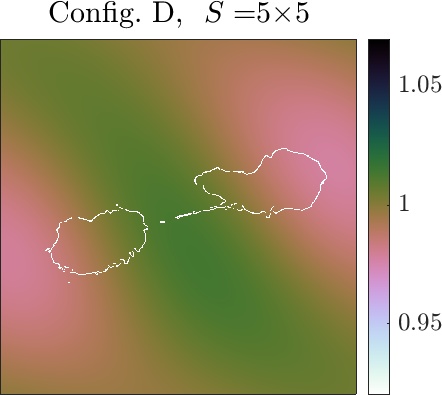}
\includegraphics[width=0.31\linewidth]{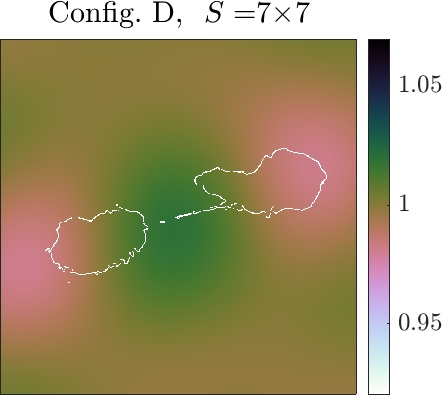}
\includegraphics[width=0.33\linewidth]{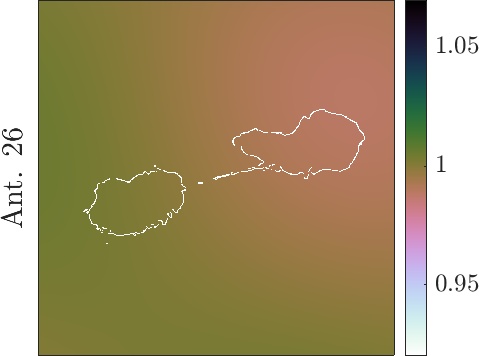}
\includegraphics[width=0.31\linewidth]{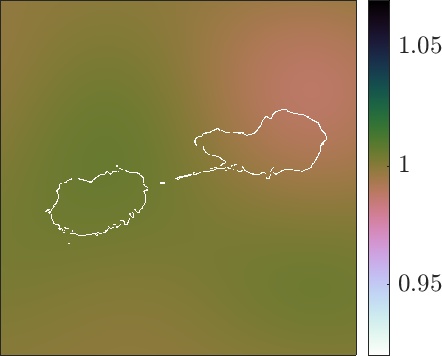}
\includegraphics[width=0.31\linewidth]{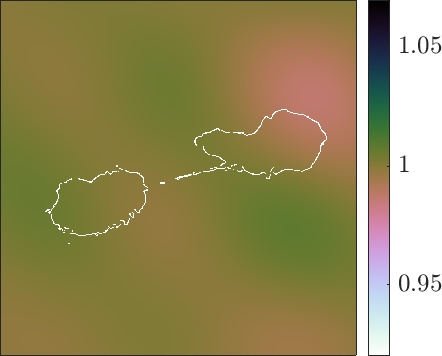}
\end{minipage}
\caption{{C band: amplitudes of DDE solutions in the image domain obtained with the joint calibration and imaging approach ($\tau =8$) over the imaged FoV $\Omega= 0.0455\degr \times 0.0455\degr$. Cyg~A is demarcated by the white contours.  From left to right, results obtained for DDEs spatial Fourier dimension $S$ set to $3\times 3,~5\times 5,~ 7\times 7$, respectively. From top to bottom, DDE estimates of two selected antennas at VLA configurations C and D at {the $10^{\textrm{th}}$} time slot. Rows 1-2, configuration C, antennas 1 and 26. Rows 3-4, configuration D, antennas 1 and 26.}}
\label{fig:c_dde_solutions_full2}
\end{figure*}


\bsp	
\label{lastpage}
\end{document}